\documentclass[11pt]{article}

\usepackage{amsthm}
\usepackage{amsmath}
\usepackage{amssymb}
\usepackage{bm} 
\usepackage{bbm}
\usepackage{mathrsfs}
\usepackage{mathtools}
\usepackage{dsfont}
\usepackage{float}
\usepackage{bm}
\usepackage{natbib}
\setcitestyle{authoryear,open={(},close={)}}

\RequirePackage{crop}
\RequirePackage{graphicx}
\RequirePackage{caption}
\RequirePackage{amsmath}
\RequirePackage{array}
\RequirePackage{color}
\RequirePackage{xcolor}
\RequirePackage{amssymb}
\RequirePackage{flushend}
\RequirePackage{stfloats}
\RequirePackage[figuresright]{rotating}
\RequirePackage{chngpage}
\RequirePackage{totcount}
\RequirePackage{fix-cm}

\RequirePackage{colortbl}
\RequirePackage{bbm}
\RequirePackage{subcaption}
\RequirePackage{multirow}
\RequirePackage[normalem]{ulem}

\usepackage{enumitem}
\usepackage{graphicx}
\usepackage{adjustbox}
\usepackage{booktabs}
\usepackage{multirow}
\usepackage{caption}
\usepackage{comment}
\usepackage{nicefrac}
\usepackage{enumitem}
\usepackage{apptools}
\usepackage{color}
\usepackage{colortbl}
\usepackage{adjustbox}

\usepackage{graphicx} 
\usepackage{latexsym}
\usepackage{amsmath}
\usepackage{amsfonts}
\usepackage{amssymb}
\usepackage{amsthm}
\usepackage{mathtools}
\usepackage{mathrsfs}
\usepackage{bbm}
\usepackage[utf8]{inputenc}
\usepackage{algorithm}
\usepackage{algpseudocode}
 \usepackage{caption}
\usepackage{tikz}
\usepackage{tikzscale}
\usetikzlibrary{positioning}
\usepackage{enumerate}
\usepackage{empheq}
\usepackage{bbm}
\usepackage{dsfont}
\usepackage{hyperref}

\theoremstyle{remark}

\newcommand{\X}{\mathcal{X}}

\newcommand{\G}{\mathcal{G}}

\newcommand{\ci}{\!\perp \! \! \! \perp\!}

\newcommand{\new}[1]{{\color{black}{#1}}} 

\def\*#1{\bm{#1}} 

\def\@fnsymbol#1{\ensuremath{\ifcase#1\or *\or \dagger\or \ddagger\or
   \mathsection\or \mathparagraph\or \|\or **\or \dagger\dagger
   \or \ddagger\ddagger \else\@ctrerr\fi}}
\newcommand{\ssymbol}[1]{^{\@fnsymbol{#1}}}

\let\OLDthebibliography\thebibliography
\renewcommand\thebibliography[1]{
  \OLDthebibliography{#1}
  \setlength{\parskip}{0pt}
  \setlength{\itemsep}{3pt plus 0.3ex}
}

\def\*#1{\bm{#1}} 

\usepackage[left=2.3cm, right=2.3cm, top=3cm]{geometry}

\usepackage{authblk}
\title{\bf  {Bayesian Inference of Multiple Ising Models for Heterogeneous
Public Opinion Survey Networks}}
\author[1,2,*]{Alejandra Avalos-Pacheco}
\author[3]{Andrea Lazzerini}
\author[5,6]{Monia Lupparelli}
\author[3,*]{Francesco C. Stingo}
\affil[1]{Institute of Applied Statistics, Johannes Kepler University Linz}
\affil[2]{Harvard-MIT Center for Regulatory Science, Harvard Medical School}
\affil[3]{Depart.\ of Statistics, Computer Science, Applications ``G. Parenti'', University of Florence}
\affil[*]{\textit {\textcolor{blue}{alejandra.avalos\textunderscore pacheco@jku.at}}}
\date{October 2024}

\begin{document}




\def\spacingset#1{\renewcommand{\baselinestretch}%
{#1}\small\normalsize} \spacingset{1}

\setcounter{Maxaffil}{0}
\renewcommand\Affilfont{\itshape\small}

\spacingset{1.42} 

\maketitle
\begin{abstract}
In public opinion studies, the relationships between opinions on different topics are likely to shift based on the characteristics of the respondents.
Thus, understanding the complexities of public opinion requires methods that can account for the heterogeneity in responses across different groups. 
Multiple graphs are used to study how external factors—such as time spent online or generational differences—shape the joint dependence relationships between opinions on various topics. Specifically, we propose a class of multiple Ising models where a set of graphs across different groups are able to capture these variations and to model the heterogeneity induced in a set of binary variables by external factors.
The proposed Bayesian methodology is based on a Markov Random Field prior for the multiple graph setting.
Such prior enables the borrowing of strength across the different groups to encourage common edges when supported by the data. Sparse inducing spike-and-slab priors are employed on the parameters that measure graph similarities to learn which subgroups have a shared graph structure.
Two Bayesian approaches are developed for the inference of multiple Ising models with a special focus on model selection: (i) a Fully Bayesian method for low-dimensional graphs based on conjugate priors specified with respect to the exact likelihood, and (ii) an Approximate Bayesian method based on a quasi-likelihood approach for high-dimensional graphs where the normalization constant required in the exact method is computationally intractable.
These methods are employed for the analysis of data from two public opinion studies in US.
The first one  analyzes the heterogeneity of the confidence in a network of political institutions induced by different time users spent on web. 
The second one studies the relationships between opinions on relevant public spending areas in diverse inter-generational groups.
The obtained results 
display a good trade-off between identifying significant edges (both shared and group-specific) and having sparse networks, all while quantifying the uncertainty of the graph structure and the graphs' similarity, ultimately shedding light on how external factors shape public opinion.
\end{abstract}

\noindent %
{\it Keywords: binary data, log-linear model, MRF prior, public institution performance, quasi-likelihood, undirected graphs}

\spacingset{1.45}

\section{Introduction}
\label{sec:Intro}

\new{The US General Social Survey (GSS) provides a comprehensive resource for understanding the diverse and evolving attitudes of the American public on matters ranging from government spending to religious believes \cite{gss}. 
This paper leverages two distinct datasets from the GSS, each composed of heterogeneous groups, to examine pivotal societal issues. 
By analyzing these datasets, we aim to uncover significant patterns and insights, common and sub-group specific, that contribute to a deeper understanding of public opinion in the United States.


Graphical models are a powerful tool for achieving the goals of this paper. 
They are effective tools to model complex relationships in multivariate distributions of a set of variables and to provide a  graphical representation of their conditional independence structure \citep{lauritzen1996graphical}. 
Thus, such tools enable us to visualize and analyze the relationships between various opinions within the datasets from the GSS. 
This approach is particularly useful in identifying how these connections vary across different subgroups. 
For example, in this paper we show that when modeling the independence structure of public opinions on government spending for key social issues, we discovered notable associations with the crime variable and spending on social programs such as education, military, and parks within the younger sub-group. 
These connections, however, tend to disappear among older respondents, suggesting that only young people see a link between spending on parks and recreation and crime control.

The first application studies confidence in government institutions. 
Recent studies have shown that opinions about confidence in institutions are strongly encouraged and influenced by the public perception of the institutional performances channeled by mass media; for a recent review on the effect of trust in mass media see \cite{fawzi-etal:2021}. 
Our goal is to compare the confidence network of different users, which are divided by their weekly time spent on web navigation. 
Such confidence network is composed of binary variables that reflect the opinion of distrust or at least some trust on people running government institutions such as organized religion, congress, and banks and financial institutions.

The second application analyzes the opinion on the public spending adequacy in the US. 
The effect of age on social policy preferences represents a longstanding debate.
Some studies showed that elderly people are less inclined to support public spending for family care than young people, as older sub-populations are more prone to support pension policy reforms \citep{jag-sch:2016, ponza-etal:1988},
Conversely, other authors argue that age has a negligible impact on public spending opinion, as the family transfers most of the time flow from the elderly to young generations \citep{koli-kun:2003, preston:1984}.
We study how the opinion on public spending varies across different ages of the respondents. 
In particular, we analyze the network structure of binary variables that reflect the opinion of underfunded or not underfunded public spending for different social aspects such as welfare, mass transportation and improving the conditions of African Americans.

In order to model such heterogeneous binary networks, we introduce novel Multiple Ising models and propose new Bayesian methods for their inference.
}

Collections of graphical models have been increasingly used, as they can capture the heterogeneity of the data involved in more realistic settings. 
In such models, the dependence structure of the variables may differ according to one or more discrete factors, used to define subgroups or subpopulations. 
This approach has led to the development of the multiple graphical methodology. 
Multiple graphical models for Gaussian random variables have been widely studied in the recent literature \citep{guo2011joint, danaher2014joint, peterson2015bayesian, HaMinJin2021BSLi}. 
Conversely, there have been only a few proposals for multinomial sampling models. 
\citep{hojsgaard2003split, corander2003labelled, nyman2014stratified, nyman2016context} proposed multinomial graphical models for context-specific independences, which allow conditional independences to hold only for a subset of one or more variables that are conditioned upon. 
Multiple graphical models are typically more general than the above-mentioned methods since they allow context-specific independences to vary not only with respect to adjacent vertices. 
\citep{ChengJie2014ASIM}  proposed Ising graphical models with covariates to simultaneously study the conditional dependency within the binary data and its relationship with the additional covariates. 
More recently, \citep{ballout2019structure}  used multiple Ising models to study associations among the injuries suffered by victims of road accidents according to the road user type and proposed frequentist methods based on graphical lasso approaches for model selection. 
The aforementioned techniques are not particularly suited for the analysis of public opinion studies because, given the uncertain role of external factors, such analyses require methods that encourages similarity across sub-groups only if supported by the data and that clearly characterize the inherent uncertainty present in this type of data.

%
We discuss inference for multiple graphical models under the assumption of an Ising model for the joint distribution of multivariate binary variables. 
This model was originally introduced by \citep{f1925} to study the behaviour of atomic variables representing solid magnetic materials. 
Ising models are now widely used in several contexts as tools for the analysis of pairwise associations of binary data in an undirected graph setting, where the probability of the status for each variable depends only on the status of its neighbours \citep{banerjee2008model,ravik-etal:2010}.
%
Inference for these models is particularly challenging due to the presence of intractable normalizing functions, even for a moderate number of variables \citep{ParkJaewoo2018BIit}. 
Strategies for single graphs with a small variable set include the use of conjugate priors for log-linear parameters with Laplace approximations for inference \citep{massam2009a}, and MCMC methods for a stochastic search of the best model \citep{MollerEtAlBKA2006, dobra2010the, FangZaili2016BIGM}. 
These methods, while useful, do not scale well with the number of variables and require to perform numerical approximations of the normalizing constant. 
The lack of a closed form of the normalizing constant implies that maximum likelihood estimation generally cannot be performed. 
Various solutions have arisen in both the frequentist and Bayesian literature. 
In the frequentist literature, the use of (quasi/pseudo)-likelihood methods \citep{besag1974spatial}, instead of the exact-likelihood, for discrete graphical models is widely used. 
Several examples can be found in the frequentist literature when dealing with large graphical models \citep{meinshausen2006high, ravikumar2010high, guo2015estimating}. 
In the Bayesian framework, the use of a non-likelihood function for inference is currently of growing popularity \citep{jiang2008gibbs,kato2013quasi,bhattcharya2019a,HaMinJin2021BSLi}. 
Another Bayesian inference strategy is the use of a latent variable representation for low-rank Ising networks \citep{MarsmanM2015Bifl}. 
However, it is not recommended to use this strategy if every variable can be correlated with nearly all other variables, like in some social sciences applications \citep{MarsmanM2015Bifl}. 
More recently, it has been proposed the use of generalised Bayesian inference leveraging a discrete Fisher divergence \citep{MatsubaraTakuo2022GBIf} and piecewise linear approximations for Approximate Bayesian Computation \citep{MooresMatthewT.2015PfaB}.

One of our main methodological contributions is the development of two Bayesian approaches for the selection of  multiple Ising  models: an exact-likelihood inference for low-dimensional binary response data; and a quasi-likelihood approach for high-dimensional data. 
The exact-likelihood approach builds on the work by \citep{massam2009a}, based on conjugate priors for log-linear parameters. 
We implement a computational strategy that uses Laplace approximations and a Metropolis-Hastings algorithm that allows us to perform a stochastic model search. 
The quasi-likelihood Bayesian approach is inspired by the work of \citep{bhattcharya2019a}, where the normalization constant results computationally intractable.
We set spike-and-slab priors to encode sparsity and provide MCMC algorithms for sampling from the quasi-posterior distribution, which, simultaneously, enables variable selection and estimation. 
Another of our main contributions is the use of
a Markov Random Field prior on the graph structures to encourage the selection of the same edges in related graphs \citep{peterson2015bayesian} in both inference methods. 
To our knowledge, this is the first adaptation of such priors to Ising models. 
These inferential strategies have the twofold aim  (i) to learn the sub-groups network structures by borrowing strength across heterogeneous binary datasets\new{, i.e.\ binary data originating from different groups or subpopulations,} and encouraging the selection of the same edges in related graphs; and (ii) to provide a measure of uncertainty for model selection and parameter inference.

\new{We remark that in this manuscript, we refer to our methods as `Approximate Bayesian' and `Fully Bayesian' in a non-traditional sense. Here, `approximate' and `fully' refer to how the likelihood of the model is handled. 
Our approach involves either setting up the full likelihood or approximating it with a quasi-likelihood. 
This is in contrast to traditional Bayesian approximation methods for inference, such as Laplace approximations, INLA: integrated nested Laplace approximations \cite{Rue2009ABif}, ALA: approximate Laplace approximation \cite{Rossell2021ALaf}, variational inference, and ABC: Approximate Bayesian Computation \cite{Pritchard1999Pgoh}, which primarily focus on directly approximating the posterior through computational strategies.}

The utility of these methods is studied via an extensive simulation study and  their performance is compared against the competing methods Indep-SepLogit \citep{meinshausen2006high} and DataShared-SepLogit \citep{ollier}, as well as with the same competing methods using identical and independent Bernoulli distributions as a prior distribution for the model, as in \citep{bhattcharya2019a}. 
Then, both inferential methodologies are applied to the analysis of two data sets provided by the US GSS. 

Our results showed that our approaches perform comparatively well \new{to the state-of-the-art methods used here as competitors} for both of the public opinion surveys. 
Exhibiting two exclusive features: learning which groups are related and providing measures of uncertainty for model selection and parameter inference.  
The selected multiple Ising models provide a good balance between network sparsity and edge selection.

\new{Through rigorous analysis of these GSS datasets, this paper seeks to contribute to the broader understanding of how diverse groups within the American populace perceive and engage with critical societal and governmental issues.}

The article is organized as follows. 
\new{Sections~\ref{sec:governmentData} and \ref{sec:public} motivates our new approaches by describing the GSS datasets and all the variables considered in our analysis.} 
Section~\ref{sampling} provides background on multiple Ising  models and reviews them.
Section~\ref{sec:prior} proposes prior formulations, including the priors for the canonical parameter: 
(i) conjugate priors for low-dimensional data, leading to an exact-likelihood, and 
(ii) spike-and-slab priors for high dimensional-data leading to a quasi-likelihood. 
Section~\ref{sec:posterior} describes the MCMC procedure for posterior inference. 
Sections~\ref{sec.simulation} and ~\ref{sec:CS} present, respectively, simulations and applications  on two public opinion case studies. 
Section~\ref{sec:conclusion} concludes.

\new{\section{Confidence in government institutions}
\label{sec:governmentData}
We consider a GSS dataset (\url{https://gssdataexplorer.norc.org/})  with 450 individuals, 10 categorical response variables about the confidence on government institutions for the only year 2018 and one continuous variable for the hours spent on the Web. 
In order to have a more interpretable and parsimonious model and to make the implementation of the FB approach feasible, we grouped the original variables. 

The original response variables which answer  the question \textit{``I am going to name some institutions in this country. As far as the people running these institutions are concerned, would you say you have a great deal of confidence, only some confidence, or hardly any confidence at all in them?''} have been properly dichotomized in dummies that take value $1$ if the original respondent answer was \textit{``Hardly any''} while take value $0$ if the original answer was \textit{``Only some''} or \textit{``A great deal''}. Therefore the $10$ binary variables reflect the opinion of distrust versus at least some trust on people running government institutions. We report in Table \ref{tab10resp} the response variables with description.

We create the grouping variable $\textit{Web}= \{0,1,2\}$, by categorizing the original continuous variable \textit{wwwhr} that answers to the following question \textit{``Not counting e-mail, about how many hours per week do you use the Web? (Include time you spend visiting regular web sites and time spent using interactive Internet services like chat rooms, Usenet groups, discussion forums, bulletin boards, and the like.)''}. Specifically, $\textit{Web}=0$ if the respondent uses the web at most 5 hours per week, $\textit{Web}=1$ if the respondent uses the web more than 5 hours and at most 15 hours per week, $\textit{Web}=2$ if the respondent uses the web more than 15 hours  per week.
These threshold values are set equal to the empirical quantiles of the observed  variable \textit{wwwhr} so to get a well-balanced sample size within each group and to make inference comparable; the resulting size of the three sub-groups is 147, 153 and 150, respectively. 
\begin{table}[h!]
	\caption{\label{tab10resp} Response variables for the institution confidence data}
		\begin{center}
		\scalebox{0.8}{
	\begin{tabular}[t]{lll}
		\hline
		Name&Confidence in \\
		\hline\hline
		\textit{tv}&TV\\
		\textit{press}&press\\
		\textit{lab}&organized labor\\
		\textit{exec}&executive branch of federal government\\
		\textit{edu}&education\\
		\textit{rel}&organized religion\\
		\textit{comp}&major companies\\
		\textit{bank}&banks and financial institutions\\
		\textit{court}&in United States supreme court\\
		\textit{congr}&congress\\
		\hline
	\end{tabular}}
	\end{center}
\end{table}

\section{Public spending opinion}
\label{sec:public}

The  data set we consider was collected at \url{https://gssdataexplorer.norc.org/} and it includes 768 observations for 18 categorical response variables about a survey on opinions, for the only year 2018, on the public spending amount on social problems and one continuous variable for the respondent age. Given the large set of variables, only the approximate Bayesian procedure can be considered for model selection. Furthermore, to have a more parsimonious model and to reduce the model complexity,  the  original response variables which answer to the question \textit{``We are faced with many problems in this country, none of which can be solved easily or inexpensively. I'm going to name some of these problems, and for each one I'd like you to tell me whether you think we're spending too much money on it, too little money, or about the right amount.''} have been dichotomized in dummies that take value $1$ if the original respondent answer was \textit{``Too little''} while take value $0$ if the original answer was \textit{"About right"} or \textit{``Too much''}. Therefore the resulting $18$ binary variables reflect the opinion of not underfunded  public spending for each social aspect. 
Table \ref{tab18} includes the response variables with description. We create the grouping variable $\textit{Age}$ taking three values $\{0,1,2\}$, by categorizing the original continuous variable:
\[
Age = 
\begin{cases}
0, & \text{if } \text{ \textit{ age}} \leq 36 \\
1, & \text{if } 36 <  \text{ \textit{age}} \leq 55  \\
2, & \text{if }  \text{ \textit{age} } >  55.
\end{cases}
\]
The threshold values are set equal to empirical quantiles of the observed continuous variable  so to have balanced sample sizes, equal to 263, 250 and 255, respectively. 
\begin{table}[h!]
	\caption{\label{tab18} Response variables for  public spending opinion data.}
		\begin{center}
		\scalebox{0.8}{
	\begin{tabular}[t]{lll}
		\hline
		Name&Public spending opinion on \\
		\hline\hline
		\textit{welf}&welfare\\
		\textit{high}& highways and bridges\\
		\textit{sec}& social security\\
		\textit{trans}& mass transportation\\
		\textit{park}& parks and recreation\\
		\textit{chi}& assistance for childcare\\
		\textit{sci}& supporting scientific research\\
		\textit{ern}& developing alternative energy sources\\
		\textit{for}& foreign aid\\
		\textit{mil}& military, armaments and defense\\
		\textit{bla}& improving the conditions of African
Americans\\
		\textit{spa}& space exploration program\\
		\textit{env}& improving and protecting environment\\
		\textit{hea}& improving and protecting nations health\\
		\textit{cit}& solving problems of big cities\\
		\textit{crime}& halting rising crime rate\\
		\textit{drug}& dealing with drug addiction\\
		\textit{edu}& improving nations education system\\
		\hline
	\end{tabular}}	
		\end{center}
\end{table}}

\section{Sampling model}
\label{sampling}



A graphical Markov model is a statistical model for a joint probability distribution defined over a graph whose vertices  correspond to random variables \citep{lauritzen1996graphical, edwards1990introduction}. 
Under suitable Markov properties, missing edges of the graph are translated into conditional independence restrictions that the model imposes on the joint distribution of the variables. 
Here, we focus on a collection of undirected graphical models where the independence structure of the variables may differ with respect to one or more factors, for instance different subpopulations or subgroups under study \citep{guo2011joint,peterson2015bayesian}.

Let $G=(V,E)$ be a graph defined by a set $V$ of $p$ vertices/nodes  and a set $E$ of edges, drawn by  undirected lines, joining pairs of nodes. 
Let $Z_{V}$ be the random vector indexed by the vertex set $V$, which includes a set of $p$ binary variables, and let $X$ be a random variable corresponding to a categorical factor not included in $V$, taking the value $x \in \mathcal{X}$, with $|\mathcal{X}|=q$.

We consider a collection of multiple undirected graphs $G_{V|\X}=[G(x)]_{ x \in \mathcal{X} }$, where each graph $G(x)=(V,E(x))$ is associated to the random vector $Z_{V}|\{X=x\},$ with a node set $V$ and an edge set $E(x)$ which depends on $x$, $x \in \mathcal{X}$. 
For any couple $r,j \in V$ and $x \in \X$, if $(r,j) \in E(x)$ there is an edge between $r$ and $j$ in the corresponding graph $G(x)$ while if $(r,j) \notin E(x)$ then the two nodes are disjointed in $G(x).$ 
Missing edges in these graphs correspond to conditional independences for the associated conditional probability distribution \citep{lauritzen1996graphical}. 
For the pairwise Markov property, if $(r,j) \notin E(x)$ then $Z_{r} \ci  Z_{j} |\{ Z_{V \setminus \{r,j\}}, X=x \}$. 
In case of strictly positive probability distributions, the pairwise Markov property is able to capture all the independences encoded by a graph  to specify and undirected graphical model for any $Z_V|\{X=x\}$, with $x \in \mathcal{X}$.  

\begin{figure}[h!]
	\begin{center}
\includegraphics[scale=0.25]{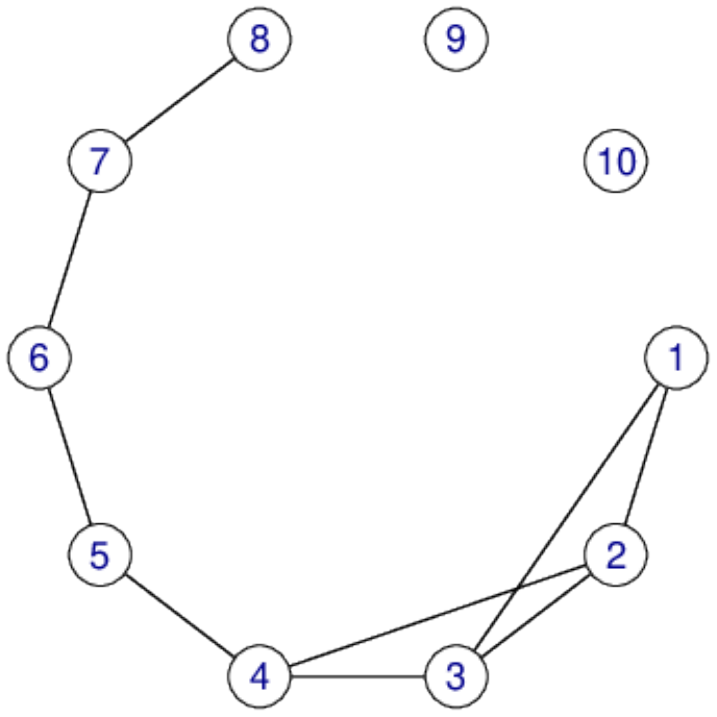}\tiny{G(0)} \hspace*{0.3cm}
\includegraphics[scale=0.25]{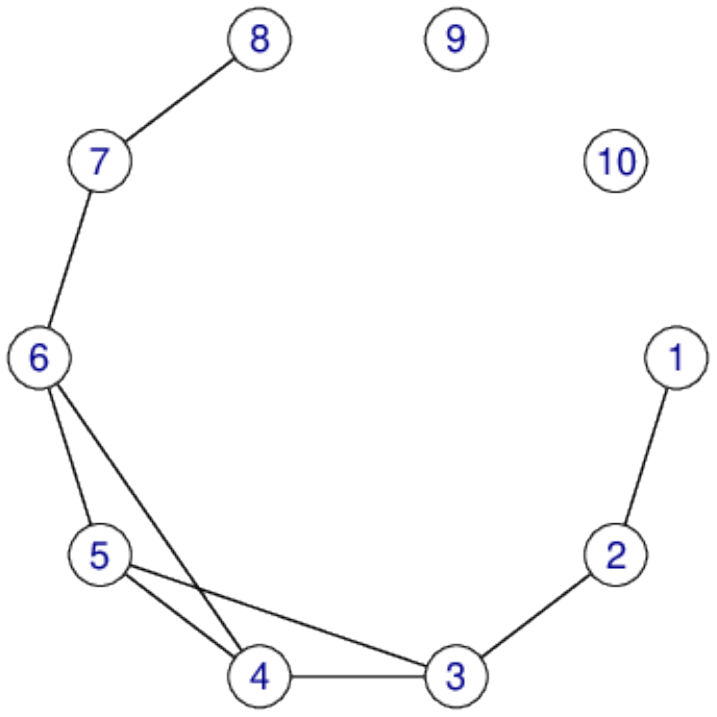} G(1) 
\\ \vspace{0.3cm}
\includegraphics[scale=0.25]{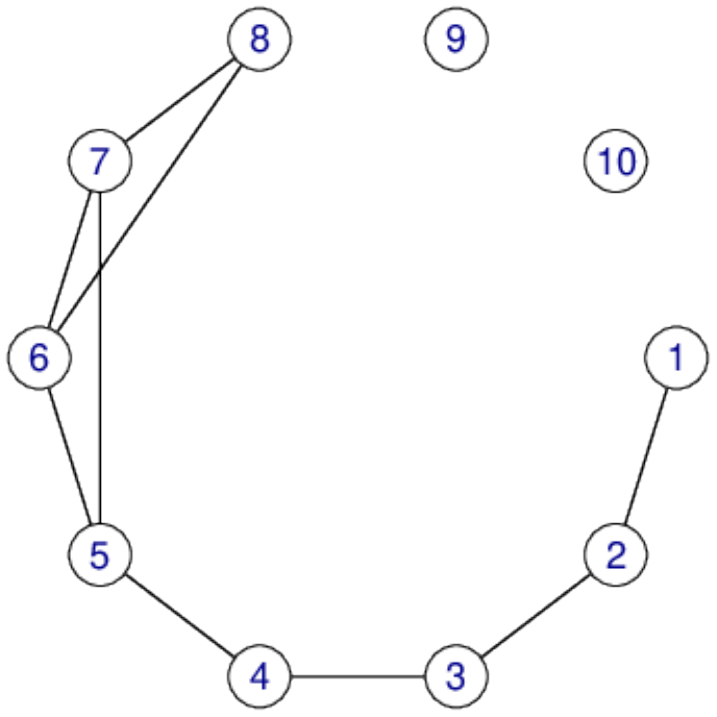} G(2) 
\hspace*{0.3cm} 
\includegraphics[scale=0.25]{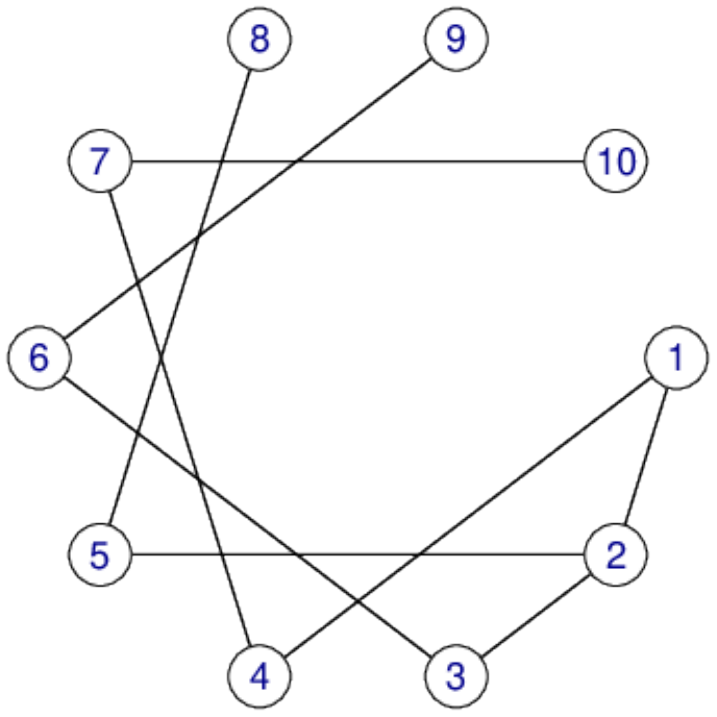} G(3) 

	\end{center}
	\caption{\label{fig} Illustration of a collection of multiple undirected graphs with $p=10$ variables and $q=4$ levels of $X$, $\X=\{0,1,2,3\}$.}
\end{figure}

Figure \ref{fig} provides an example of multiple undirected graphs with $p=10$ nodes and $q=4$ levels of $X$, $\X=\{0,1,2,3\}$, where, for instance, $Z_{1} \ci Z_{10} |\{ Z_{V \setminus \{1,10\}}, X=x \}$ for all $x \in \X$ while $Z_{1} \ci Z_{3} |\{ Z_{V \setminus \{1,3\}}, X=x \}$ for any $x \in {\{1,2,3\}}$. 
In this setting, $X$ is the external factor with respect to which the network structure may change, i.e. the weekly time that people use to spend for web navigation and the age group.

\subsection{Multiple Ising  models}
\label{ising}

We consider the case of binary random variables for response data  $Z_V$. 
Modelling binary data can be cumbersome and challenging because as the number of the variables increases, the number of parameters can become so large and thus, intractable. 
To address this issue and make inference more scalable, we assume an Ising model \citep{besag1974spatial} for the joint probability distribution of $Z_V$ so that, in the canonical log-linear parameter, all higher than two-way interaction terms vanish \citep{ballout2019structure}. 
The Ising model assumption is realistic when the probability of the status of each variable depends only of the status of its neighbours in the graphical representation. 

For each $x \in \mathcal{X}$, let $n_x$ be the observed realizations of $Z_{V}|\{X=x\} \sim \text{Ising}(\lambda_x)$ where $\lambda_x=[\lambda_{rj,x}]_{r,j \in V} \in \mathbb{R}^{p+(p \times (p-1))/2}$ is the log-linear parameter, also known as canonical parameter in the exponential family theory \citep{barndorff:1978}. 
Let $\lambda_{r,x}=[\lambda_{rj,x}]_{j \in V}$ denote the $r$-th vector of log-linear parameters, for any $r \in V$. 
Missing edges in each undirected graph $G(x)$ correspond to zero pairwise log-linear interactions, since $Z_r \ci Z_j |\{Z_{V \setminus r,j}, X=x\}$ if and only if $\lambda_{rj,x} = 0$ in the related model Ising($\lambda_x$), for any $x \in \mathcal{X}$ \citep{whittaker:1990}.
Let $Z(x)$ be the corresponding $n_x \times p$ observed binary matrix, with $i$-th row $z_x^{i} \in \{0,1\}^{p}$ and entries $z_{r,x}^{i} \in \{0,1\}$; $i=1,\ldots, n_x$ indexes units for which $X=x$. The likelihood function for $\lambda_x$ can be expressed, for any $x \in \mathcal{X}$, as

\begin{equation}
\label{lik}
p(Z(x)|\lambda_x) =\prod_{i=1}^{n_x} \dfrac{1}{\Psi(\lambda_x)} \exp \left\{ \sum_{r=1}^{p} \lambda_{rr,x} z^{i}_{r,x} + \sum_{r=1}^{p} \sum_{j < r} \lambda_{rj,x} z^{i}_{r,x} z^{i}_{j,x}  \right\}, 
\end{equation}
where 
\begin{equation}
\label{cnlik}
\Psi(\lambda_x)= \hspace{-0.4cm} \sum_{z_x^{i} \in \{0,1\}^{p}}  \hspace{-0.3cm}  \exp \left\{ \sum_{r=1}^{p} \lambda_{rr,x} z^{i}_{r,x} + \sum_{r=1}^{p} \sum_{j < r} \lambda_{rj,x} z^{i}_{r,x} z^{i}_{j,x} \right\}
\end{equation}
is the normalization constant. 
Likelihood based inference on $\lambda_x$ is computationally tractable only for moderate values of $p$, because to calculate $\Psi(\lambda_x)$ is required to compute a sum that grows exponentially in $p$. 
We develop an approach based on the proper likelihood \eqref{lik} that can be applied to a moderate number of variables ($\leq 10$), referred to as Fully Bayesian (FB), as well as a second approach based on \textit{quasi-likelihoods} \citep{bhattcharya2019a} for intractable high-dimensional settings ($> 10$), referred to as Approximate Bayesian (AB). 

In high-dimensional settings, we express the r-th node conditional likelihood for $\lambda_{r,x} \in \mathbb{R}^{p}$, $r \in V$ and for any $x \in \mathcal{X}$, as
\begin{equation}
\label{rlik}
p_{r}(Z(x)|\lambda_{r,x}) = \prod_{i=1}^{n_x} \dfrac{1}{\Psi_{r,x}^{i}(\lambda_{r,x})}  \exp \left\{\lambda_{rr,x} z^{i}_{r,x} + \sum_{j < r} \lambda_{rj,x} z^{i}_{r,x} z^{i}_{j,x} \right\},
\end{equation}
with a normalization constant equal to
\begin{equation}
\label{cnqlik}
\Psi^{i}_{r,x}(\lambda_{r,x})=1+ \exp \left\{  \lambda_{rr,x} +  \sum_{j < r} \lambda_{rj,x}  z^{i}_{j,x} \right\}, \hspace{5mm} x \in \mathcal{X}.
\end{equation}
We then approximate the likelihood in \eqref{lik} with a quasi-likelihood obtained as the product of $p$ conditional likelihoods, i.e.
\begin{equation}
\label{eqquasilikc}
p_{q}(Z(x) \mid \lambda_x) = \prod_{r=1}^{p} p_{r}(Z(x) \mid \lambda_{r,x}), \hspace{5mm} x \in \mathcal{X},
\end{equation}
so the inference problem on $\lambda_{x} \in \mathbb{R}^{p+(p \times (p-1))/2}$ simplifies into $p$ separable sub-problems on $\mathbb{R}^{p}$.

\section{Prior formulation}
\label{sec:prior}
\new{
The proposed methodology aims to infer the set of graphs $G_{V|\X}$, accounting for their potential similarities, by using priors that encourage shared edge selection in related graphs when supported by data; these priors, described in Sections \ref{sec:linking} and \ref{sec:theta}, apply to both the FB and AB approaches.}


\subsection{Markov Random Field prior, $\delta_x$}
\label{sec:linking}
%
The parameter vector $\lambda_x$ is estimated for all levels $x \in \mathcal{X}$ through the specification of a Markov Random Field (MRF) prior on the graph structures as in \citep{peterson2015bayesian} to encourage similarities in related graphs. 
Specifically, the MRF prior 
is set on binary indicators of edge inclusion $\delta_{rj,x} \in \{0,1\}$. 
These latent indicators serve as a proxy for which $\lambda_{rj,x}$ are significantly non-zero ($\delta_{rj,x}=1$) and which are zero ($\delta_{rj,x}=0$). 

The conditional probability of $\delta_{rj,x}$, the inclusion of edge $(r,j)$ in $G(x)$, given $\delta_{rj,-x}$, the inclusion of edge $(r,j)$ in the remaining graphs $[G(h)]_{ h \in \{ \mathcal{X} \setminus x \}}$, is 
\begin{equation}
\label{eq:MRF}
p(\delta_{rj,x} \mid \delta_{rj,-x},\nu_{rj},\theta_x)=\dfrac{\exp[\delta_{rj,x}(\nu_{rj}+ \bm{1}^\top \theta_x \delta_{rj,x})]}{1+\exp[\delta_{rj,x}(\nu_{rj}+ \bm{1}^\top\theta_x \delta_{rj,-x})]}, \hspace{5mm} x \in \mathcal{X},
\end{equation} 
where $\nu_{rj} \in \mathbb{R}$ is a sparsity parameter specific for the edge $(r,j)$ for all levels $x \in \mathcal{X}$, and $\theta_x=[\theta_{ xh }]_{h \in \{ \mathcal{X} \setminus x \}}$, where $\theta_{xh } \in \mathbb{R}$ represents the relatedness between the graphs $G(x)$ and $G(h)$,  for all $x,h \in \mathcal{X}, x \neq h$. 
The parameters $\theta_{ xh }$ and $\nu_{rj}$ are an important aspect of the formulation and will be discussed in Section~\ref{sec:theta}.

For Equation~\eqref{eq:MRF}, the conditional probability of the $r$-vector $\delta_{r,x}=[\delta_{rj,x}]_{j \in V}$, for any $r \in V$, is
\begin{equation}
\label{depriorr}
p(\delta_{r,x} \mid \delta_{r,-x},\nu_{r},\theta_x)=\prod_{j=1}^{p} p(\delta_{rj,x} \mid \delta_{rj,-x},\nu_{rj},\theta_x), \hspace{5mm} x \in \mathcal{X},
\end{equation}
where $\delta_{r,-x}=[\delta_{r,h}]_{h \in \{ \mathcal{X} \setminus x \} }$ and $\nu_{r}=[\nu_{rj}]_{j \in V}$; and the conditional probability of the entire vector $\delta_x=[\delta_{rj,x}]_{r,j \in V,j<r}$ is
\begin{equation}
\label{depriorr}
p(\delta_x \mid \delta_{-x},\nu,\theta_x)=\prod_{r=1}^{p}\prod_{j<r} p(\delta_{rj,x} \mid \delta_{rj,-x},\nu_{rj},\theta_x), \hspace{5mm} x \in \mathcal{X},
\end{equation}
where $\delta_{-x}=[\delta_h]_{h \in \{ \mathcal{X} \setminus x \} }$ and $\nu=[\nu_{rj}]_{r,j \in V,j<r}$. 

Note that since this MRF prior is an Ising model, the joint distribution of the vector $\delta_{ij}$ has the same functional form as equation \eqref{lik}, this greatly simplifies the computations.

\subsection{Priors for the graph similarity $\theta_{xh}$ and the edge-specific parameter $\nu_{rj}$}
\label{sec:theta}
Let us consider the prior specification for  parameter $\theta_{xh}$, for any $x,h \in \mathcal{X}$. 
If $\theta_{xh}=0$,  the two graphs $G(x)$ and $G(h)$ are independent a priori.
Following \citep{peterson2015bayesian}, we set an spike-and-slab prior on $\theta_{xh}$ \citep{george1997approaches}  of the form:
\begin{equation}
\label{eqpriort}
\begin{split}
p(\theta_{xh} \mid \epsilon_{xh}) &= (1-\epsilon_{xh}) \: \delta_{0}(\theta_{xh}) + \epsilon_{xh} \: p(\theta_{xh} \mid \alpha, \beta),
\\
p(\theta_{xh} \mid \alpha, \beta) & = \text{Gamma}(\theta_{xh}; \alpha, \beta),
\end{split}
\end{equation}
where $\delta_{0}(\cdot)$ is a Dirac delta in 0, $\alpha$ and $\beta$ are fixed hyperparameters for the Gamma distribution and $\epsilon_{xh} \in \{0,1\}$ are latent indicators that determines if profiles $x$ and $h$ have a similar graph structure, i.e. if $\epsilon_{xh} = 0$, profiles $x$ and $h$ have different graph structure, 
if $\epsilon_{xh} = 1$, $\theta_{xk}$ is generated from the slab Gamma distribution, which encourages borrowing strength between profiles $x$ and $h$ leading to similar graph structures.

Note that the probability density function Gamma$(x;\alpha, \beta)$ is equal to zero at the point $x = 0$. 
Thus,  the sparse inducing prior in Equation~\eqref{eqpriort} is a discrete mixture of a non-local prior \citep{JohnsonValenE.2010Otuo} and a point mass at zero, i.e. a non-local spike-and-slab prior. 
This type of priors have shown to substantially improve the parameter inference, as they discard spurious parameters faster as the sample size grows, but preserve exponential rates to detect important coefficients and can lead to improved parameter estimation shrinkage \citep{JohnsonValenE.2010Otuo, RossellDavid2017NPfH, Avalos-PachecoAlejandra2022HLDI}. 

The joint prior for $\theta=[\theta_{xh}]_{x < h}$ given $\epsilon=[\epsilon_{xh}]_{x < h}$ can be written as the product of the marginal densities of any $\theta_{xh}$ because the parameters are variation independent and there are no constraints on the structure of $\theta$, such that
\begin{equation}
\label{eqpriortc}
\begin{split}
p(\theta\mid \epsilon) = \prod_{x<h} p(\theta_{xh} \mid \epsilon_{xh}).
\end{split}
\end{equation}

We complete the model specification with independent Bernoulli priors over the latent indicators $\epsilon_{xh}$ as follows,
\begin{equation}
\label{eqpriore}
\epsilon_{xh}\mid \omega \sim \text{Bernoulli}(\epsilon_{xh}; \omega), 
\end{equation}
where $w \in [0,1]$ is a fixed prior probability.


We use the edge-specific parameter $\nu_{rj} \in \mathbb{R}$ to further encourage sparsity in the graph structures. 
Specifically, the probability of inclusion of edge $(r,j)$, for all $r,j \in V$, in $G(x)$, for all $x \in X$, can be written as 
\begin{equation}
p(\delta_{rj,x} \mid \nu_{rj}) =\dfrac{e^{\nu_{rj}}}{1+e^{\nu_{rj}}} =q_{rj}.
\end{equation}
If no prior knowledge on the graph structure is available, a prior that favors lower values, such as $q_{rj} \sim \text{Beta}(a,b)$ with $a < b$ can be chosen to encourage overall sparsity. 
After applying a univariate transformation of variables to the $\text{Beta}(a, b)$ prior on $q_{rj}$, the prior on $\nu_{rj}$, for all $r,j \in V$, can be written as
\begin{equation}
\label{eqpriornu}
p(\nu_{rj}) =\dfrac{1}{B(a,b)} \hspace*{0.1cm} \dfrac{e^{a\nu_{rj}}}{(1+e^{\nu_{rj}})^{a+b}},
\end{equation}
where $B(\cdot)$ represents the beta function.

\subsection{Priors for the canonical parameter $\lambda_x$}

We define different prior distributions on the $\lambda_x$ parameters for the low-dimensional (proper likelihood) and the high-dimensional (conditional likelihood) cases corresponding to the FB and AB approaches, respectively.

\subsubsection{Fully Bayesian (FB) approach}

In the low-dimensional settings we set the Diaconis and Ylvisaker prior distribution \citep{diaconis1979conjugate}, as this is a conjugate prior for $\lambda_x$ \citep{massam2009a}, leading to a fully Bayesian inference. 
For convenience, we rewrite the likelihood \eqref{lik} as function of the marginal counts, for any $x \in \mathcal{X}$, 
\begin{equation}
\label{lik2}
\begin{split}
p(y_x \mid \lambda_x,\delta_x) = & \exp \left\{ \sum_{r} {\lambda_{rr,x} y_{rr,x}} + \sum_{r<j} {\delta_{rj,x} \lambda_{rj,x} y_{rj,x}} \right.  \\
& \left. - n_x \log \left[ \sum_{\{ \mathcal{I}_x \setminus z_{\emptyset,x} \}} \exp \left( \sum_{r} \lambda_{rr,x} + \sum_{r<j} {\delta_{rj,x} \lambda_{rj,x}} \right) \right]  \right\},
\end{split}
\end{equation}
for any $x \in \mathcal{X}$, with $\delta_{rj,x} \in \{0,1\}$, $x \in \mathcal{X}$ an indicator parameter for the inclusion of parameter $\lambda_{rj,x}$ (See Section~\ref{sec:linking}), $\mathcal{I}_x= \{0,1\}^p$ the probability table for $Z_V|\{X=x\}$ with a generic cell $z_x \in \mathcal{I}_x$, $z_{\emptyset,x}$ the baseline cell where all variables take level $0$, and $y_x =[y_{rj,x}]_{r,j \in V}$ the vector of of marginal observed counts, compute as 
\begin{equation}
y_{rj,x}=\sum_{i=1}^{n_x} \mathbbm{1}_{\left\{{z^{i}_{r,x}=1,z^{i}_{j,x}=1}\right\}},
\end{equation}
We then set a Diaconis and Ylvisaker prior for \eqref{lik2} as
\begin{equation}
\label{dyp}
\begin{split}
p(\lambda_x \mid \delta_x) = &C(s_x,\alpha_x)^{-1} 
\\& \times   \exp \left\{\sum_{r} {\lambda_{rr,x} s_{rr,x}} + \sum_{r<j} {\delta_{rj,x} \lambda_{rj,x} s_{rj,x}} \right.
\\&\left. - g_x \log \left[ \sum_{\{\mathcal{I}_x \setminus z_{\emptyset,x}\}} \exp \left( \sum_{r} {\lambda_{rr,x}} + \sum_{r<j} {\delta_{rj,x} \lambda_{rj,x}} \right) \right]  \right\},\\
&x \in \mathcal{X}, 
\end{split}
\end{equation}
where $C(s_x,\alpha_x)$ is an unknown normalization constant that depends on the hyperparameters $g_x \in \mathbb{R}$ and $s_x=[s_{rj,x}]_{r,j \in V}$, $s_x \in \mathbb{R}^{p+(p \times (p-1))/2}$.

\subsubsection{Approximate Bayesian (AB) approach}
In the high-dimensional case ($p > 10$), we set a continuous Normal spike-and-slab prior \citep{GeorgeEdwardI.1993VSvG} for the $r$-th vector of log-linear parameters $\lambda_{r,x}$ in \eqref{rlik}, defined as:
\begin{equation}
\label{priorlahigh}
p(\lambda_{r,x} \mid \delta_{r,x}) =
 \prod^p_{j=1} \left\{ \delta_{rj,x} N(\lambda_{rj,x}; 0,\rho ) +(1- \delta_{rj,x}) N(\lambda_{rj,x}; 0,\gamma) \right\},
\end{equation}
with $\rho>>\gamma>0$ and  $\delta_{rj,x} \in \{0,1\}$ an indicator parameter for the inclusion of $\lambda_{rj,x}$.
The indicator $\delta_{rj,x}$ (defined more in detail in Section~\ref{sec:linking}) signals which $\lambda_{rj,x}$ were generated by each component (spike or slab).

The spike-and-slab priors provide sparse canonical parameters $\lambda_{rj,x}$, effectively performing model selection on the number of edges included in $(r,j) \in G(x)$, as $\lambda_{rj,x}=0$ corresponds to the missing edge $(r,j)$. 
Furthermore, the continuity of the spike distribution provides computationally efficient updates for the MCMC algorithm derived in Section~\ref{sec:posterior}.

\section{Posterior inference}
\label{sec:posterior}

We conduct posterior inference and model selection of the graph structures $G_{V|\X}$ using  MCMC methods, 
summarized in Algorithm~\ref{alg:MCMC} and described in detail in Appendix~\ref{app:post},
by iteratively sampling the parameters of interest. 
Firstly, we update the graphs $G(x)$ for each profile according to the dimension of the $p$ nodes: 
(i) For low-dimensional $p$ (FB), we perform a stochastic search of the graph space, leveraging a Laplace approximation \citep{tierney1986accurate} of the marginal likelihood (see Appendices~\ref{app:exinf} and \ref{app:BELg} for details). 
(ii) For high-dimensional $p$ (AB), we sample the graph and the canonical parameter $\lambda_x$ for each group
from the conditional quasi-posterior distribution \eqref{eqposteriorc} as in \cite{bhattcharya2019a} (see Appendices~\ref{app:inf} and \ref{app:BALg}). 
We then use a Metropolis-Hastings algorithm to sample the graph similarity and the latent indicators from their conditional posterior distribution, for both the FB and AB. 
Here, we perform two steps: a between-model and a within-model move as in \cite{gottardo2008markov}. 
This strategy is carried out as it usually improves the mixing of the chains \citep{gottardo2008markov} (see Appendix~\ref{app:theta}). 
Finally, we update the edge-specific parameters $\nu$, for both FB and AB, using a traditional Metropolis-Hastings approach to sample from their conditional posterior distribution (see Appendix~\ref{app:nu}).

\begin{algorithm}[h!]
\textbf{while }{$t<T$}{

  \:\:\textbf{FB}:\\
  \begin{tabular}{rl}
& Update the graph$^{\dagger}$ ${G}(x)$ for each profile $x \in \mathcal{X}$ \\
&  $^{\dagger}$ see Appendix~\ref{app:BELg}\\
\end{tabular}\\

\:\: \textbf{AB}:\\
  \begin{tabular}{rl}
&Update the graph$^{\ddagger}$ ${G}(x)$ and the canonical parameter$^{\ddagger}$ ${\lambda}_{x}$\\
&for each profile $x \in \mathcal{X}$ \\
& $^{\ddagger}$ see Appendix~\ref{app:BALg}
\end{tabular}

\:\: \textbf{FB \& AB}:\\
  \begin{tabular}{rl}
&Update the graph similarity$^{\S}$ ${\theta}_{xh}$ and the latent \\
&indicators$^{\S}$ ${\epsilon}_{xh}$ for $1\leq x < h \leq q$ \\
&  $^{\S}$ see Appendix~\ref{app:theta}\\
&Update the edge-specific parameter$^{\P}$ ${\nu}_{rj}$ for $1\leq r < j \leq p$.\\
& $^{\P}$ see Appendix~\ref{app:nu}
\end{tabular}

\textbf{set}  $t=t+1$ 
 }
 \textbf{end  while }
\caption{MCMC algorithms for multiple Ising models under the FB and AB approaches}
 \label{alg:MCMC}
\end{algorithm}

The MCMC procedures provide a list of visited models and their corresponding posterior probabilities, based on the sampled values of ${\delta}_x$, $x \in \X$. 
Thus, graph and edge selection can then be achieved either by looking at the ${\delta}_x$ vectors with largest joint posterior probabilities among the visited models or, marginally, by calculating frequencies of inclusion for each $\delta_{rj,x}$ and then choosing those $\delta_{rj,x}$'s with frequencies exceeding a given cut-off value. 

\section{Simulation studies}
\label{sec.simulation}
We first asses the performance of both the FB and AB approaches on simulated data. 
We consider simulated scenarios that mimic the characteristics of the real data that motivated the development of the models.
In order to evaluate the effect of the MRF prior, we compare both approaches with methods that replace this prior with independent Bernoulli distributions, such methods are equivalent otherwise. 
We refer to these as Fully Bayesian Separate (FBS) and Approximate Bayesian Separate (ABS). 
We also compare both methods with the frequentists Indep-Seplogit (SL) \citep{meinshausen2006high} and  DataShared-SepLogit (DSSL) \citep{ollier}.
The \textbf{\textsf{R}}-code for our methods is available at \url{https://github.com/kinglaz90/phd}. 
The SL and DSSL methods were implemented using the \texttt{glmnet} \textbf{\textsf{R}}-package \citep{glmnet}. 
We simulated data from two different data-generating mechanisms.

\new{Important details about the parameter settings for the simulations and for the model fitting, prior sensitivity of the model, assessment of MCMC convergence, uncertainty quantification of the graph structure, frequentist coverage and computational cost,  can be found in Appendices~\ref{app:LDSS}, \ref{app:sensitivity}, \ref{app:MCMCconver} , \ref{app:UQ}, \ref{app:FC} and \ref{app:CompCost} respectively}.

\subsection{Low-dimensional simulation study}
\label{sec:reslow}
We simulated $n_x=100$ observations from $Z_{V}|\{X=x\} \sim \text{Ising}(\lambda_x)$ with associated undirected graph $\G$ with $p=10$ nodes and $q=4$ levels of $X$, $x \in \X = \{0,1,2,3\}$. 
Note that $\G$ includes at most $p(p-1)/2=45$ edges that are identified by the selection parameter $\delta_x$, $x \in \X$. 

We considered four different profile undirected graphs $\G(A)$, $\G(B)$, $\G(C)$ and $\G(D)$, displayed in  in Figure \ref{mulgraphs}. 
\new{The simulated graphs are scale-free networks using a preferential attachment mechanism, leveraging the Barab\'asi-Albert model to grow the network \citep{Albert1999EoSi}. Graphs
structures were generated using the \texttt{igraph} \textbf{\textsf{R}}-package \cite{csardi2006}.} 
We first studied scenarios in which all four graphs were identical, as in Scenario (A), or completely different, as in Scenario (B). 
The aim of Scenario (B) was to investigate if the joint inference performed by our approaches FB and AB provided a poor estimation when the graphs were totally different.
We then considered the arguably more interesting cases were some groups, but not all, share the same graph structure.  
In Scenario (C) the pairs $G(0), G(1)$, and $G(2), G(3)$ are identical between each other, but each pair is completely different to the other one. 
In Scenario (D) $G(0)$, $G(1)$ and $G(2)$ are identical and completely different to $G(3)$.


\small{\begin{figure*}[h!]
	\begin{center}
		\footnotesize Scenario (A)\\
		 	\includegraphics[scale=0.2]{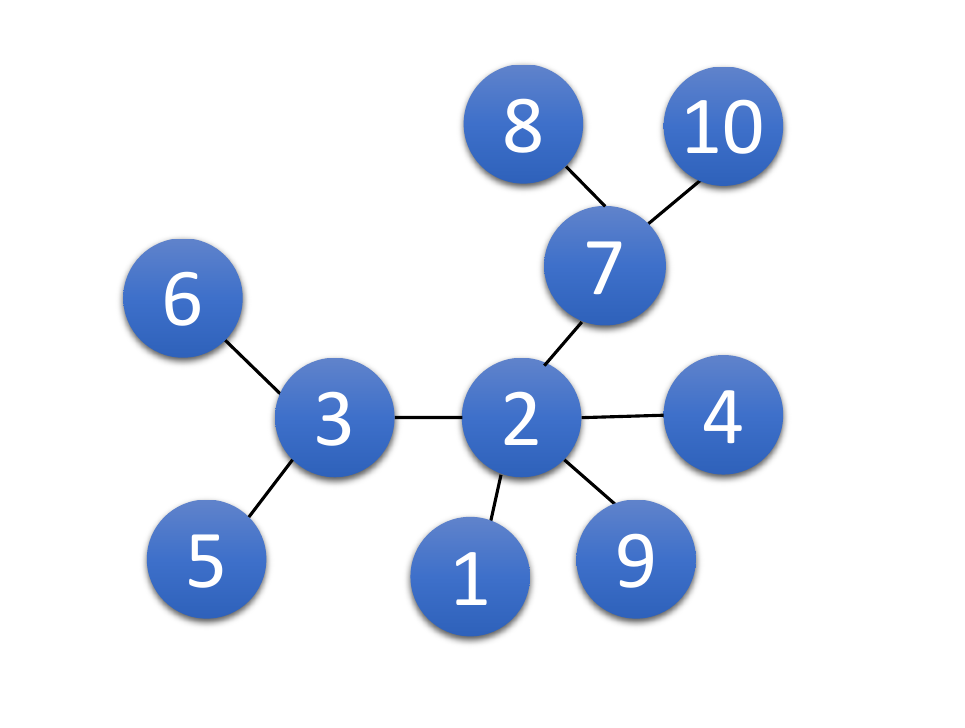}\tiny{$G(0)$}
		\includegraphics[scale=0.2]{figures/G1SFN.pdf} $G(1)$ 
		\includegraphics[scale=0.2]{figures/G1SFN.pdf} $G(2)$ 
		\includegraphics[scale=0.2]{figures/G1SFN.pdf} $G(3)$ 
		\\ \vspace*{0.3cm} \footnotesize Scenario (B) \\\includegraphics[scale=0.2]{figures/G1SFN.pdf}\tiny{$G(0)$}
		\includegraphics[scale=0.2]{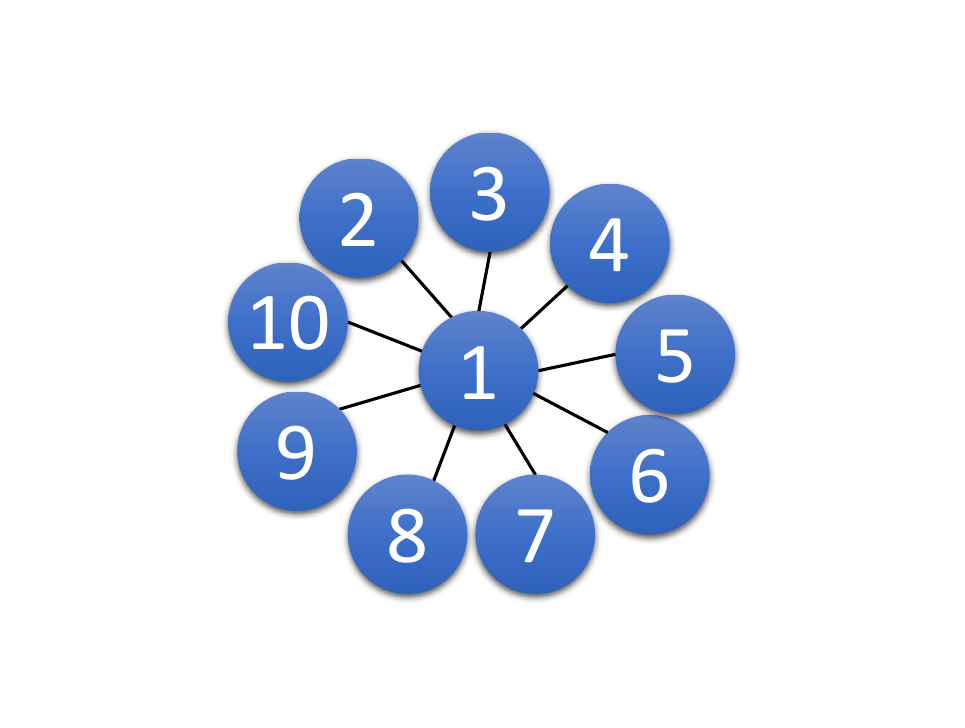}$G(1)$
		\includegraphics[scale=0.2]{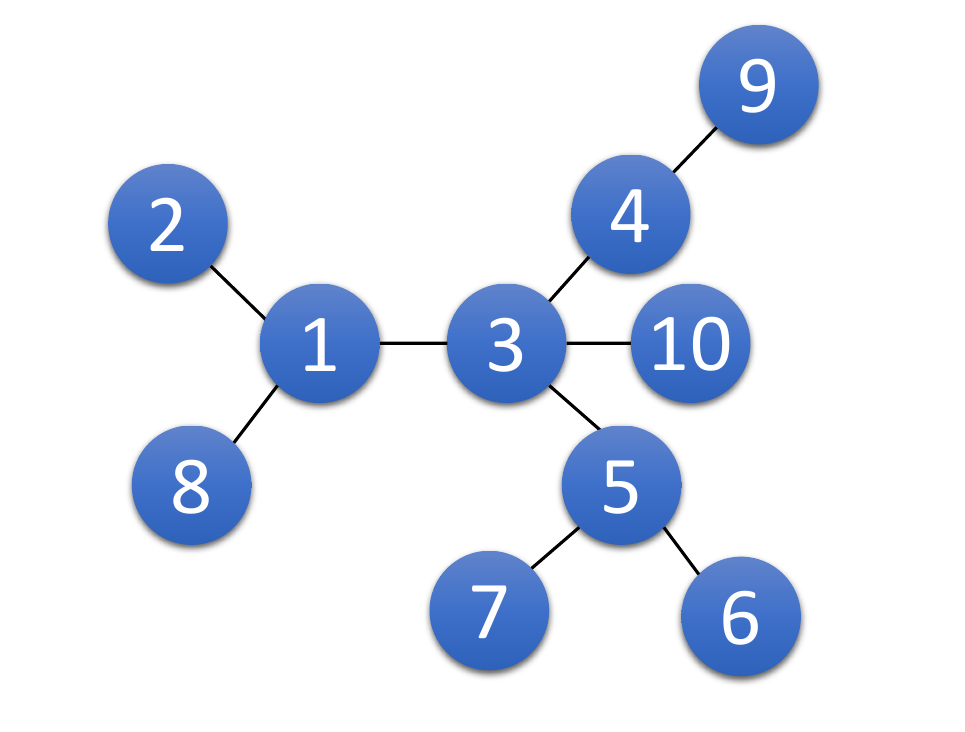}$G(2)$ 
		\includegraphics[scale=0.2]{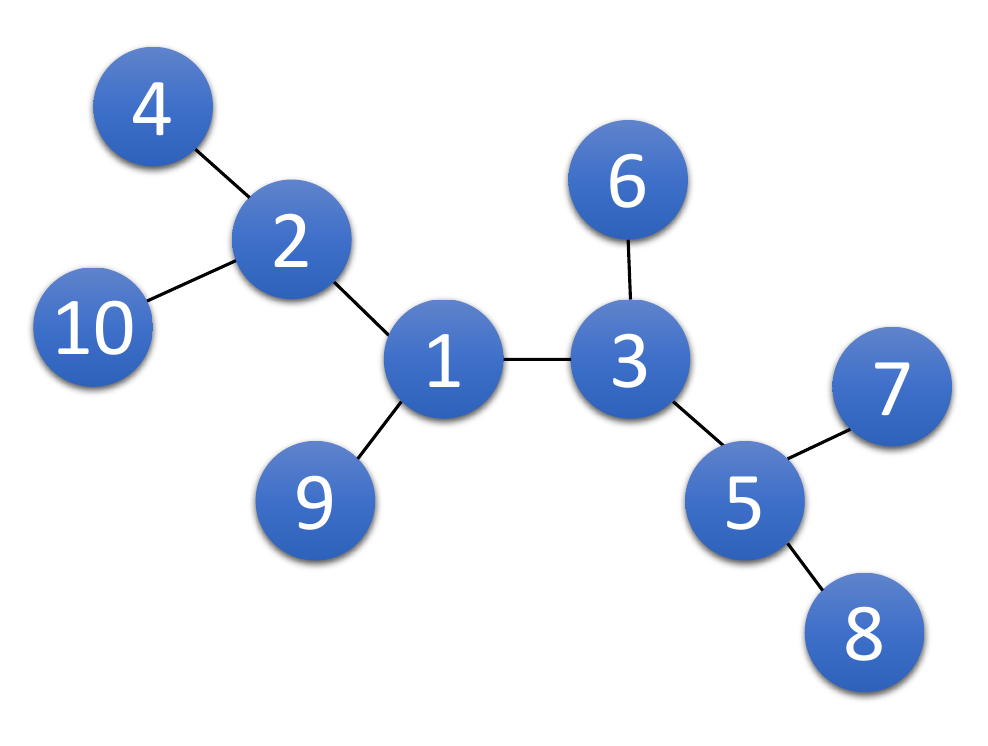}$G(3)$
		\\ \vspace*{0.3cm} \footnotesize Scenario (C) \\\includegraphics[scale=0.2]{figures/G1SFN.pdf}\tiny{$G(0)$}
		\includegraphics[scale=0.2]{figures/G1SFN.pdf}$G(1)$
		\includegraphics[scale=0.2]{figures/G2SFN.pdf}$G(2)$ 
		\includegraphics[scale=0.2]{figures/G2SFN.pdf}$G(3)$
			\\ \vspace*{0.3cm} \footnotesize Scenario (D) \\\includegraphics[scale=0.2]{figures/G1SFN.pdf}\tiny{$G(0)$}
		\includegraphics[scale=0.2]{figures/G1SFN.pdf}$G(1)$
		\includegraphics[scale=0.2]{figures/G1SFN.pdf}$G(2)$ 
		\includegraphics[scale=0.2]{figures/G2SFN.pdf}$G(3)$
  \caption{Multiple undirected graphs in the four different scenarios of the low-dimensional simulation study.}
  \label{mulgraphs}
	\end{center}
\end{figure*}}

\normalsize
To assess the accuracy of the graph structure estimation, we compute the Matthews correlation coefficient (MCC) and the F1 score (F1) of the true non-zero elements of $\lambda_x$, for all $x \in \X$, along with their associated standard errors (SE). 
The MCC is a balanced measure of binary classification that takes values between -1 (total disagreement) and +1 (perfect classification). 
The F1 $\in [0,1]$ is the harmonic mean of the precision and recall.
A value of F1=1 indicates perfect precision and recall, and F1=0 indicates that either the precision or the recall is zero. 
Table~\ref{T10} shows the MCC, the F1 score and their SE across ten different simulations.

Results in Table~\ref{T10} showed that, overall, the SL and ABS methods tend to perform similarly, as both are based on approximate inference.
The FBS approach had a better performance than SL and ABS, as it is based on exact inference. 
In Scenario (A) the SL and ABS methods performed poorly, not taking into account the homogeneity among the graphs. 
Conversely, the DSSL method outperformed all the methods, as it considered the graph homogeneity, but showed the worst performance in the remaining 3 scenarios, resulting in very little flexibility. 
The FB approach obtained the second best performance in Scenario (A) and the best in Scenarios (C) and (D), showing the advantages of the exact method when there is strong homogeneity among the graphs. 
The AB method had, a competitive performance overall, showing better results in scenarios where there is strong homogeneity among the graphs (Scenarios (A) and (D)), the second best in Scenario (D). 
It is important to note that the performance of both FB and AB approaches did not worsen noticeably in scenarios where there is no homogeneity (Scenario (B)). 
We finally highlight that both  methods FB and AB have two unique features: (i) they learn which groups are related, and (ii) they provide a measure of uncertainty for model selection and parameter inference.

\begin{table*}[t]
\caption{MCC and F1 score and their standard error (SE) rates for all scenarios of the low-dimensional synthetic datasets.\label{T10}}
\tabcolsep=0pt
\begin{tabular*}{\textwidth}{@{\extracolsep{\fill}}ccccc@{\extracolsep{\fill}}}
\toprule%
\multicolumn{5}{@{}c@{}}{MCC} \\
\hline
Model & Scenario (A) & Scenario (B) & Scenario (C) & Scenario (D) \\ 	\hline
SL     & 0.773(0.048) & 0.811(0.058) & 0.739(0.062) & 0.762(0.067) \\ 
DSSL   & \textbf{0.983}(0.020) & 0.589(0.060) & 0.518(0.087) & 0.714(0.040) \\
ABS & 0.734(0.070) & 0.761(0.058) & 0.674(0.078) & 0.712(0.070) \\
\cellcolor[HTML]{d9e6f2} AB    & \cellcolor[HTML]{d9e6f2} 0.814(0.036) &\cellcolor[HTML]{d9e6f2} 0.808(0.060) & \cellcolor[HTML]{d9e6f2} 0.744(0.050) &  \cellcolor[HTML]{d9e6f2} 0.772(0.057)\\
FBS     & 0.796(0.038) & \textbf{0.822}(0.070) & 0.749(0.054) & 
0.785(0.049)	\\
  \cellcolor[HTML]{d9e6f2} FB    & \cellcolor[HTML]{d9e6f2} 0.858(0.042) & \cellcolor[HTML]{d9e6f2} 0.804(0.064) & \cellcolor[HTML]{d9e6f2}  \textbf{0.764}(0.074)& \cellcolor[HTML]{d9e6f2} \textbf{0.812}(0.041)
		\\
		\hline \hline
  \multicolumn{5}{@{}c@{}}{F1} \\
\hline
Model & Scenario (A) & Scenario (B) & Scenario (C) & Scenario (D) \\ 	\hline
SL     & 0.812(0.040) & 0.838(0.052) & 0.770(0.055) & 0.794(0.062) \\ 
DSSL   & \textbf{0.986}(0.017) & 0.665(0.057) & 0.614(0.074) & 0.774(0.033) \\ 
ABS & 0.753(0.068) & 0.778(0.051) & 0.685(0.074) & 0.728(0.068) \\ 
\cellcolor[HTML]{d9e6f2} AB    & \cellcolor[HTML]{d9e6f2} 0.839(0.035) & \cellcolor[HTML]{d9e6f2} 0.828(0.048) & \cellcolor[HTML]{d9e6f2} 0.769(0.048) & \cellcolor[HTML]{d9e6f2} 0.797(0.056) \\
FBS     & 0.825(0.035) & \textbf{0.844}(0.066) & 0.781(0.047) & 0.812(0.044) \\ 
\rowcolor[HTML]{d9e6f2}
		FB    & 0.880(0.036) & 0.830(0.055) & \textbf{0.792}(0.065) & \textbf{0.833}(0.043) \\ 
		\hline
\end{tabular*}
\end{table*}
\subsection{High-dimensional simulation study}
We \new{
simulated} $n_x=200$ observations from $Z_{V}|\{X=x\} \sim \text{Ising}(\lambda_x)$ with 
undirected graph $\G$ 
\new{consisted of} $p=50$ nodes and $q=4$ levels of $X$, 
$x \in \X = \{0,1,2,3\}$. 
\new{Here,} $\G$ includes\new{
up to} $p(p-1)/2=1,225$ edges,
identified by the selection parameter $\delta_x$, $x \in \X$. As in Section~\ref{sec:reslow}, we consider 4 different profile undirected graphs $\G(A)$, $\G(B)$, $\G(C)$ and $\G(D)$: in Scenario (A) the four graphs are identical, in Scenario (B) the four graphs are completely different, in Scenario (C) the graphs $G(0)$ and $G(1)$ are identical but completely different to $G(2)$ and $G(3)$, which are identical to each other, and in Scenario (D) the graphs $G(0)$, $G(1)$ and $G(2)$ are identical and completely different to $G(3)$.

Table~\ref{T50}  provides the MCC, F1 score and their SE across ten different simulations. 
Results in Table~\ref{T50} showed that all the methods studied here, displayed a similar behaviour than in the low-dimensional settings. 
The AB approach perform comparatively well overall; having the best performance in Scenarios (B), (C) and (D), with little to strong homogeneity, and the second best in Scenario (A), with no homogeneity. 

\begin{table*}[t]
\caption{MCC and F1 score and their standard error (SE) rates for all scenarios of the high-dimensional synthetic datasets. \label{T50}}
\tabcolsep=0pt
\begin{tabular*}{\textwidth}{@{\extracolsep{\fill}}ccccc@{\extracolsep{\fill}}}
\toprule%
\multicolumn{5}{@{}c@{}}{MCC} \\
\hline
Model & Scenario (A) & Scenario (B) & Scenario (C) & Scenario (D) \\ 	\hline
SL & 0.916(0.010) & 0.911(0.011) & 0.915(0.012) & 0.916(0.008) \\ 
DSSL & \textbf{0.988}(0.007) & 0.798(0.031) & 0.740(0.017) & 0.838(0.018) \\
ABS & 0.920(0.010) & 0.914(0.012) & 0.910(0.010)& 0.915(0.010)\\
\rowcolor[HTML]{d9e6f2}
AB &0.947(0.008) & \textbf{0.924}(0.011) & \textbf{0.931}(0.010) & \textbf{0.935}(0.009)\\
		\hline \hline
  \multicolumn{5}{@{}c@{}}{F1} \\
\hline
Model & Scenario (A) & Scenario (B) & Scenario (C) & Scenario (D) \\ 	\hline
SL & 0.919(0.010) & 0.914(0.010) & 0.917(0.012) & 0.919(0.007) \\ 
DSSL & \textbf{0.988}(0.007) & 0.798(0.032) & 0.731(0.021) & 0.838(0.018) \\
ABS & 0.922(0.010) & 0.915(0.012) &  0.911(0.010) & 0.917(0.010) \\
\rowcolor[HTML]{d9e6f2}
		AB & 0.949(0.008) & \textbf{0.927}(0.011) & \textbf{0.933}(0.010) & \textbf{0.937}(0.009)
		\\
		\hline
\end{tabular*}
\end{table*}

\new{
Finally, Appendix \ref{app:OtherSims} presents the results of our methods applied to four additional undirected profiles $\G(A)$, $\G(B)$, $\G(C)$ and $\G(D)$, shown in Figure \ref{mulgraphsOld}. 
This analysis aimed to assess the sensitivity of the various methods to other chain-network structures. 
Our findings indicated a robust performance across the two cases, using both the FB and AB approaches.
}

\section{Public opinion case studies}
\label{sec:CS}
\new{We applied our }
proposed FB and AB approaches to infer networks and to explore model selection across related sample groups in two different case studies in social sciences. 
The first one studies the confidence in political institutions in different web users groups.
The second one analyzes the opinion on public spending in the USA in diverse age groups.
We apply the proposed methods, setting the parameters as in Appendix~\ref{app:LDSS}. 
We select edges with marginal posterior probability of inclusion greater than 0.5. 

\subsection{Confidence in government institutions}

We aim to study whether confidence in government agencies can be influenced by the perception of the performance of the political institutions channeled by mass media and, specifically, by the user's time-spent on the web. 
We implement the FB and AB model selection procedures   to study and compare the network of confidence across Americans' subpopulations with different web-access patterns in terms of time surfing the net.  
We compare the proposed methods with ABS, FBS, SL and DSSL. 


The networks selected by using the AB and FB methods, and by their competitors are shown in Figure \ref{fig.appl10} . 
We display in red solid lines the edges shared across all the three subgroups and in black dashed lines the differential edges. 
We measure the network similarity of the three different groups  with shared edges count (SEC) matrices. 
SEC matrices display the resulting edge counts for the number of overlapping edges between each pair of graphs. 
For our both methods, AB and FB, we provided in matrices $\bm{\theta}$(PPI) the posterior probabilities of inclusion for the elements of $\bm{\theta}$. 
To carry out edge selection, we estimate the posterior marginal probability of $\delta_{rj,x}$ as the proportion of MCMC iterations after the burn-in in which edge $(r-j)$ was included in graph $\mathcal{G}(x)$. 
For each profile, we then select the set of edges that appear with marginal posterior probability of inclusion (PPI) $>0.5$. Following \citep{peterson2015bayesian} we compute the expected false discovery rate (FDR). 
Then the expected FDR with bound $\zeta = 0.5$ is
\begin{equation}
\text{FDR}=\dfrac{ \sum_{x} \sum_{j < r} (\varepsilon_{rj,x}) \bm{1}[\varepsilon_{rj,x} \leq \zeta] }{ \sum_{x} \sum_{j < r} \bm{1} [\varepsilon_{rj,x} \leq \zeta] }, \qquad x \in \mathcal{X}, r,j \in V
\end{equation}
with $\varepsilon_{rj,x}$  the marginal PPI
for edge $(r-j)$ in graph $\mathcal{G}(x)$, and $\bm{1}$ the indicator function. 
{\small{\begin{figure}[h!]
	\begin{center}
		\includegraphics[scale=0.45]{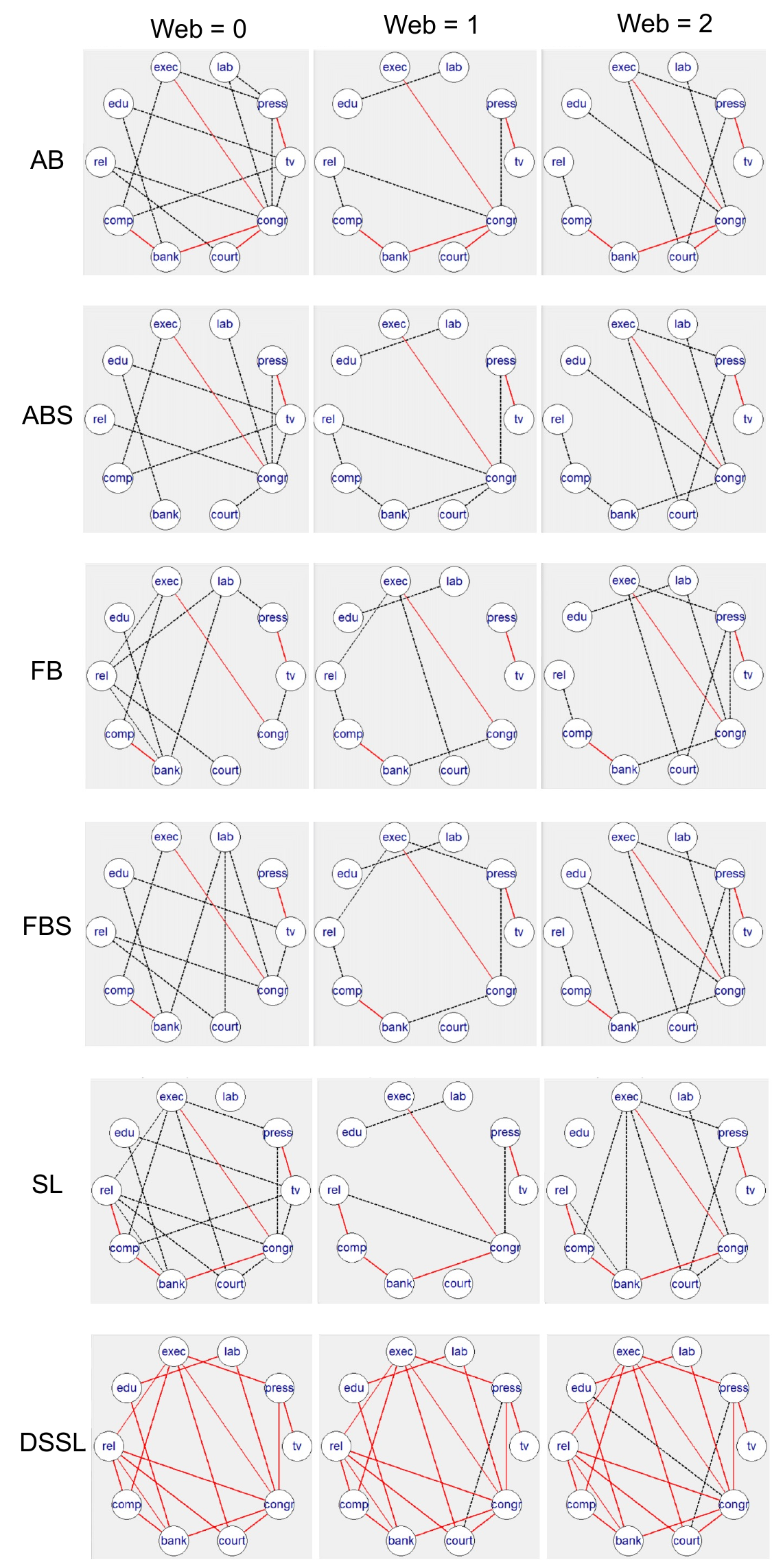}	
		\caption{\label{fig.appl10} Results for the case study of confidence in political institutions in US.  Edges shared across all groups are represented by red continuous lines.}
	\end{center}
\end{figure} }}

The obtained SEC and $\bm{\theta}$(PPI) matrices are the following:
\begin{center}
$\text{SEC}_{AB}$=$\begin{pmatrix} 
16 & 7 & 7 \\ 
    & 9 & 6 \\ 
     &  & 11  
\end{pmatrix} $
\hspace*{7mm}
$\text{SEC}_{ABS}$=$\begin{pmatrix} 
11 & 5 & 3 \\ 
    & 9 & 5 \\ 
     &  & 11    
\end{pmatrix} $
\\ \vspace*{5mm}
$\text{SEC}_{FB}$=$\begin{pmatrix} 
12 & 4 & 3 \\ 
    & 8 & 7 \\ 
     &  & 11   
\end{pmatrix} $
\hspace*{7mm}
$\text{SEC}_{FBS}$=$\begin{pmatrix} 
12 & 3 & 5 \\ 
    & 9 & 7 \\ 
     &  & 12   
\end{pmatrix} $
\\ \vspace*{5mm}
$\text{SEC}_{DSL}$=$\begin{pmatrix} 
17 & 17 & 17 \\ 
& 18 & 18 \\ 
&  & 19   
\end{pmatrix} $
\hspace*{7mm}
$\text{SEC}_{SL}$=$\begin{pmatrix} 
18 & 7 & 10 \\ 
& 8 & 5 \\ 
&  & 13   
\end{pmatrix} $
\\ \vspace*{5mm} \hspace*{6mm}
$\bm{\theta}\text{(PPI)}_{AB}$=$\begin{pmatrix} 
 1 & 1 \\ 
  & 1 \\ 
\end{pmatrix} $  \hspace*{10mm}
$\bm{\theta}\text{(PPI)}_{FB}$=$\begin{pmatrix} 
0.99 & 0.99 \\ 
 & 0.99 \\ 
\end{pmatrix}, $
\end{center}
and the expected FDR for edge selection obtained, averaged across the  graphs, is equal to 0.22 for ABS, 0.2 for AB, 0.14 for FBS and 0.12 for FB. 
We checked for MCMC chains convergence by running two different chains (with different starting points) getting two vectors containing the PPIs. 
We finally calculate the correlation of these two vectors obtaining the following results: 0.99 for ABS, 0.99 for AB, 0.96 for FBS and 0.81 for FB. 

From the diagonal elements of the $SEC_{AB}$ and $SEC_{FB}$ matrices we note that those who spend more hours on the web, i.e. \textit{Web}$=1,2$ display a greater sparsity than the ones who spend less time surfing, \textit{Web}$=0$. 
Looking at the elements outside the diagonals of the SEC matrices, the number of edges shared by the exact and approximate methods is quite comparable and, in general, the connections increase if we use our Bayesian approaches, in particular the approximate one. 
The FB displays a little sparser structures than the AB, although comparable between each other, providing more interpretable networks.

The red full edges to assess homogeneity in Figure~\ref{fig.appl10} show that the joint and separate approaches make no difference in the exact methods, while in the approximate method, the joint approach (AB) encourages homogeneity, providing three extra edges. 
The associations \textit{comp}-\textit{bank}, \textit{congr}-\textit{exec} and \textit{press}-\textit{tv} are present in all the groups for almost all the methods (except \textit{comp}-\textit{bank} for ABS in \textit{Web}$=0$).  
Overall, the node \textit{congr} is the one that shares more edges  for \textit{Web}$=2$, showing that the confidence in congress appears to be crucial, in particular the edges \textit{congr}-\textit{bank}, \textit{congr}-\textit{exec} and \textit{congr}-\textit{lab} are present in all the methods. 
In all the graphs for sub-group \textit{Web}$=1$ of the four Bayesian methods, the edge \textit{edu}-\textit{lab} is present while it is never present for the other sub-groups. 
The same behaviour is shown by the two associations \textit{bank}-\textit{edu} and \textit{congr}-\textit{tv} for \textit{Web} $=0$, where these two edges are not present in the other groups. 
Thus, these results suggest that, respondents who use the web from 6 to 15 hours a week (\textit{Web}$=1$), i.e. those who have an ``average'' web usage, tend to have a less structured confidence network than those who use the web little or a lot. 
In addition, only in this group there is an association between confidence in education and in organized labor. 
For respondents that use the web more than 15 hours, the confidence in congress seems to be quite relevant as it is associated with the confidence in organized labor, executive branch of federal government and banks. 
Respondents that use the web 5 hours or less a week, unlike the other two groups, show connections between confidence in bank and in education and also between confidence in congress and in TV. 
In all the methods, the variables \textit{edu}, \textit{rel} and \textit{lab} in sub-groups \textit{Web} $=1,2$ are poorly connected, this suggests that confidence in these areas is independent to confidence in other areas, with the exception of their connection to confidence in congress which, in general, is a variable with many connections. 
For those who spend a few hours on the web, the dependence structure of education, religions and work organizations increases with respect to remaining variables. 
It is important to highlight that the networks obtained with the FB and AB approaches, displayed all the previously discussed features but with an sparser representation than the other methods, suggesting a good balance between estimating important edges (both common and group specific), and sparse networks, all while providing uncertainty quantification of the \new{
graph structure} and the graph similarity.

\subsection{Public spending opinion}

We also employ the proposed selection strategies for modelling the independence structure of the opinions on public spending in US for some crucial social issues. 
We aim to study if these opinions are driven by self-interests which can be considerably divergent at different ages or if intergenerational conflict is actually mitigated by the family transfer from elderly to young people. 
Then, to account for an eventual heterogeneity, we study and compare the opinions for public spending networks for three age's groups using  multiple Ising model AB.


We compared our models with the same metrics than in the previous section. 
Figure \ref{fig.appl18} provides the networks estimated by all the considered methods, with edges shared across all subgroups in red and differential edges dashed as before. 
We also report the SECs and $\bm{\theta}$(PPI) matrices:
\begin{center}
	$\text{SEC}_{AB}$=$\begin{pmatrix} 
	25 & 7 & 12 \\ 
	& 21 & 12 \\ 
	&  & 25    
	\end{pmatrix} $
	\hspace*{7mm}
	$\text{SEC}_{ABS}$=$\begin{pmatrix} 
	17 & 4 & 6 \\ 
	& 9 & 6 \\ 
	&  & 19  
	\end{pmatrix} $
	\\ \vspace*{5mm}
	$\text{SEC}_{DSLL}$=$\begin{pmatrix} 
	48 & 47 & 47 \\ 
	& 48 & 47 \\ 
	&  & 47   
	\end{pmatrix} $
	\hspace*{7mm}
	$\text{SEC}_{SL}$=$\begin{pmatrix} 
	16 & 7 & 7 \\ 
	& 23 & 15 \\ 
	&  & 23   
	\end{pmatrix} $
	\\ \vspace*{5mm}
	$\bm{\theta}\text{(PPI)}_{AB}$=$\begin{pmatrix} 
	1 & 1 \\ 
	& 1 \\ 
	
\end{pmatrix}, $
\end{center}
the expected FDR, averaged over all the three graphs, getting a value equal to 0.22 for the ABS approach and a value equal to 0.18 for AB approach. 
We also checked for MCMC chains convergence, using the same strategy adopted in the previous case study getting a correlation of 0.97 for ABS and 0.98 for AB. 
{\small{\begin{figure}[h!]
	\begin{center}
		\includegraphics[scale=0.4]{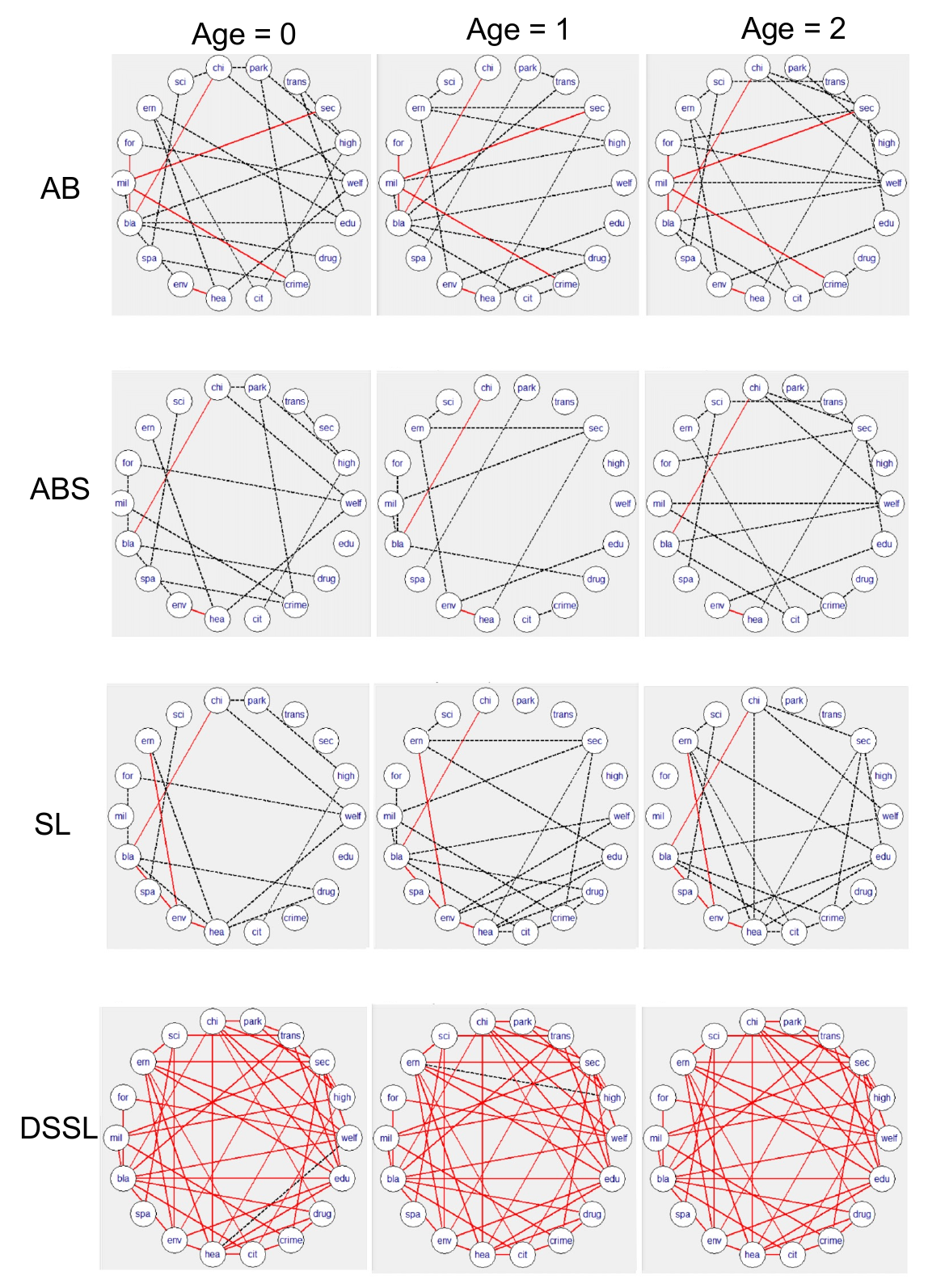} 	
	\caption{\label{fig.appl18} Results for the case study of the opinion on public spending in US.  Edges shared across all groups are represented by red continuous lines.}
	\end{center}
\end{figure}}}

We note from the diagonals elements of the SEC matrices that the number of connections of the three graphs in the AB approach is greater than in the ABS method.
The same behavior is observed for the off-diagonal elements outside of the SEC matrices, suggesting a noticeable effect of the MRF prior. 
The red full edges in Figure~\ref{fig.appl18}, which assess homogeneity, displayed that our Bayesian approach encourages homogeneity, showing 5 red full edges versus only two for the ABS method. 
We note that the association \textit{bla}-\textit{chi} is present in all the sub-groups independently of the method. 
The public spending on improving the conditions of African Americans appears to be an important variable as this node is the one that shares the highest number of edges in all the graphs and methods. 
For instance, respondents associate the conditions of African Americans with the assistance for childcare, drug addiction problem, nation education system and foreign aid. 
In the younger sub-group, \textit{Age}$=0$, the \textit{crime} variable is connected with \textit{spa}, \textit{mil} and \textit{park}. 
These connections tend to disappear for the older respondents, suggesting, for instance, that only young people associate the public spending on the care of parks and recreation with the public spending on crime control. 
The older sub-group, $Age=2$, place greater importance on welfare than other age sub-groups, in particular, there is association between the public spending on welfare with respect to the public spending on African Americans condition, military, foreign aid and childcare assistance.
Overall, AB achieved a good balance between sparsity and edge selection, displaying a sparser network than DSSL and a higher linked one than ABS.

\new{We remark that, in applications where the levels of $X$ have an inherent ordering, it is possible to incorporate this notion into the prior specification. 
Specifically, by adjusting the prior probability of graph relatedness $w$ within the MRF framework, one can influence the posterior probabilities to reflect expected similarities between consecutive levels of $X$. 
This allows for a more structured approach where graphs corresponding to adjacent levels of $X$ are more likely to be similar, aligning with practical expectations.}

\section{Discussion}
\label{sec:conclusion}
We have introduced two Bayesian approaches for the  analysis of multiple Ising  graphical: the FB and AB methods that can be used for low and, relatively, high-dimensional settings, respectively. 
The FB method is based on conjugate priors for log-linear parameters. 
This allow us to efficiently compute the marginal likelihood through the Laplace approximation. 
Inference for the FB approach is done via a stochastic search of the model with higher posterior probability, using MCMC methods. 
The AB method is based on a quasi-likelihood approach to  manage the normalization constant in a computational non-expensive manner. 
We set spike-and-slab priors for log-linear parameters to encode sparsity and we use MCMC algorithms to sample from the quasi-posterior distribution and update the model parameters.

Another key contribution 
is the use of a MRF prior on the binary indicator of edge inclusion \citep{peterson2015bayesian} incorporated to our both approaches to borrow strength across heterogeneous datasets and to encourage the selection of the same edges in related graphs. 
We remark that similarities between groups are not imposed a priori, but they  are learned from the data. 
This feature resulted particular useful in the two public opinion studies provided here, as the proposed methods resulted in much more plausible graphs than the competing penalized approaches (e.g., DSSL always resulted in very dense graphs). 
The analysis on the confidence in the USA government institutions allowed us to explore the changes in the confidence networks with respect to the weekly time that people use for web surfing. 
The public spending study, enable us to investigate the complex dependencies underlying such opinions across different ages of the respondents. 
Our inferential strategies show competitive performances overall, in comparison with other approaches for multiple Ising models.  
\new{In addition, the FB and AB approaches represents  flexible methodologies  to incorporate prior information about the network structure and to provide a measure of uncertainty for network selection.}
\new{Given the potential loss in frequentist coverage associated with the quasi-likelihood approach, we acknowledge that exploring curvature adjustments, such as those proposed by \cite{Shaby2014TOSA, Ribatet2012BIFC} and others, could be a valuable direction for future work. 
These adjustments could potentially improve the accuracy of the posterior distributions obtained under the AB model.}
It \new{also} remains as future work to generalize these approaches to different types of random variables, for instance, categorical response variables, to relax the assumption of the Ising model.

\section{Acknowledgements}
\new{The authors would like to thank Lassi Roininen and Blake Hansen for their assistance in deploying the code to the server.\\
The third and last authors were partially supported by the Italian Ministry of University and Research (MUR), Department of Excellence project 2023-2027 ReDS 'Rethinking Data Science' - Department of Statistics, Computer Science, Applications - University of Florence, the European Union - NextGenerationEU - National Recovery and Resilience Plan, Mission 4 Component 2 - Investment 1.5 - THE - Tuscany Health Ecosystem - ECS00000017 - CUP B83C22003920001, and the MUR-PRIN grant 2022 SMNNKY, CUP B53D23009470006, funded by the European Union Next Generation EU, Mission 4, Component 2.}

\appendix
\section*{Appendices}
\section{Posterior inference}
\label{app:post}
The main inferential goal is inference and selection of  the graph structures $G_{V|\X}$. For graphs of moderate dimensions, the proposed strategy for posterior inference is based on marginal likelihoods; this approach is introduced in Subsection \ref{app:exinf}. Algorithms that yields posterior samples of the graph structure parameters $G_{V|\X}$ are presented in Subsection \ref{app:mcmc}, for both the low and high-dimensional case.

\subsection{Marginal likelihoods and posterior distributions}
\label{app:marginal_lik} 
In subsection \ref{app:exinf} we follow a Bayesian exact-likelihood approach for low-dimensional cases, where we compute the marginal likelihood through the Laplace approximation. In subsection \ref{app:inf} we deal with high-dimensional cases, following a Bayesian approximate-likelihood approach that uses MCMC methods to sample from the quasi-posterior distribution.

\subsubsection{Low-dimensional case}
\label{app:exinf}
In the low-dimensional setting the posterior inference is based on the computation of the marginal likelihood.
Note that there is a one-to-one correspondence between the graph structure $G(x)$ and the binary indicator vector $\delta_x$; 
 we denote with $p(G(x)|G(-x))$ and $p(G(x)|Z{(x)},G(-x))$ the prior and the posterior probability of $G(x)$, $x \in \mathcal{X}$, conditional upon the graph structures of the other groups $G(-x)$, i.e., $p(G(x)|G(-x)) = p(\delta_x|\delta_{-x},\nu,\theta_x)$.

\noindent The posterior probability of $G(x)$ is proportional to the product of the prior distribution $\pi(G(x))$ and the marginal likelihood $m(Z|G(x))$, i.e.
\begin{equation}
\begin{split}
\label{postg}
p(G(x)|Z,G(-x)) \propto p(G(x)|G(-x)) ~ m(Z|G(x)), \hspace{5mm} x \in \mathcal{X},
\end{split}
\end{equation}
where 
\begin{equation}
\begin{split}
\label{marginal}
m(Z|G(x)) =
\int_{\Theta_{G(x)}} p(\lambda_x|s_x,g_x,\delta_x) ~ p(\lambda_x|y_x,\delta_x) ~ d \lambda_x.
\end{split}
\end{equation}
The posterior of $\lambda_x$, for any $x \in \mathcal{X}$, can be then written as:
\begin{equation}
\label{postlalow}
\begin{split}
p(\lambda_x|y_x,\delta_x) =& C(y_x+s_x,n_x+g_x)^{-1} \times \\
&\exp \Bigg\{ \sum_r {\lambda_{rr,x} (s_{rr,x}+y_{rr,x})} \\
&+ \sum_{r<j} { \delta_{rj,x} \lambda_{rj,x} (s_{rj,x}+y_{rj,x})}-
\\& - (g_x+n_x) \log \Bigg[ \sum_{\left\{ \mathcal{I}_x\setminus z_{\emptyset,x} \right\}} \exp \Big( \sum_r \lambda_{rr,x} \\
&+ \sum_{r<j} {\delta_{rj,x} \lambda_{rj,x}}   \Big) \Bigg]  \Bigg\},
\end{split}
\end{equation}
and in this way the integral in \eqref{marginal} is analytically derived as
\begin{equation}
\label{cnratio}
\dfrac{C(y_x+s_x,n_x+g_x)}{C(s_x,g_x)}, \hspace*{5mm} x \in \mathcal{X},
\end{equation}
i.e., the ratio between the normalizing constants of the posterior and prior distributions of $\lambda^{(x)}$ \citep{massam2009a}. We calculate both the normalizing constants through the Laplace approximation \citep{tierney1986accurate} such that, for instance 
\begin{equation}
\label{cnp}
C(s_x,g_x)=Ke(\lambda_x^{*})\dfrac{(2\pi)^{||\delta_x||/2}}{|A_x|^{1 / 2}},~~~~x \in \X,
\end{equation}
where $||\cdot||$ denotes the L1 norm, $Ke(\lambda_x^{*})$ is the kernel of the Diaconis and Ylvisaker prior, and $A_x$ is the Hessian matrix $(||\delta_x|| \times ||\delta_x||)$, both evaluated at a stationary point $\lambda_x^{*}$.

\subsubsection{High-dimensional case}
\label{app:inf}
In case of high-dimensional graphs, we devise a computational approach that builds upon \cite{bhattcharya2019a}. This approach is based on quasi-likelihoods; specifically, the $r$-th conditional posterior distribution of $(\delta_{r,x},\lambda_{r,x})$ given by
\begin{equation}
\label{eqposterior}
\begin{split}
p(\delta_{r,x},\lambda_{r,x} \mid Z{(x)},\delta_{r,-x},\nu_{r},\theta_x) & \propto p(\delta_{r,x} \mid \delta_{r,-x},\nu_{r},\theta_x)\times \\
&p_{r}(Z{(x)}|\lambda_x) p(\lambda_{r,x}|\delta_{r,x})
\end{split}
\end{equation}
such that the quasi-posterior of $(\delta_{r,x},\lambda_{r,x})$ for the graph $G(x)$, for all $x \in \X$, is given by
\begin{equation}
\label{eqposteriorc}
\begin{split}
p_{q}(\delta_x,\lambda_x \mid Z{(x)},\delta_{-x}\nu_{r},\theta_x)=\prod_{r=1}^{p} p(\delta_{r,x},\lambda_{r,x} \mid Z{(x)},\delta_{-x}\nu_{r},\theta_x)
\end{split}
\end{equation}

\subsection{Sampling algorithms}
\label{app:mcmc} 
In this section, we present the MCMC methods used for the posterior inference. In Step Ia, we perform a stochastic search of the graph space for the exact-likelihood method (low-dimensional case); in Step Ib we sample from the conditional quasi-posterior distribution (high-dimensional case) to update both $G(x)$ and $\lambda_x$. Step II and III are common to both algorithms.

\subsubsection{Low-dimensional case: Updating $G(x)$ - Step Ia}
\label{app:BELg}
Since the number of possible graphs grows exponentially with $p^2$, we rely on a stochastic model search. For $x \in \mathcal{X}$, we start from $G(x)$ the graph accepted at the previous iteration and propose a new graph $\widetilde{G}{(x)}$ by randomly sampling one element of $\delta_x$ and switching its value. We finally accept the new model $\widetilde{G}{(x)}$ with probability
\begin{equation}
\label{accmodel}
\begin{split}
r=\min \left(1,\dfrac{p(\widetilde{G}{(x)} \mid Z)}{p(G(x) \mid Z)} \right).
\end{split}
\end{equation}

\subsubsection{High-dimensional case: Updating $G(x)$ and $\lambda_x$ - Step Ib}
\label{app:BALg}
In the presence of high-dimensional data, our sampler relies on the quasi-posterior distribution \eqref{eqposteriorc}. 
In particular, we use a general Metropolis Adjusted Langevin Algorithm (MALA) which updates $\lambda_{r,x}, \delta_{r,s},\theta_x$ and $\nu_r$ respectively, for any $x \in \X$.
If $\lambda_{r,x}$ is considered the only random variable of interest, the full conditional \eqref{eqposterior} is proportional to
\begin{equation}
\label{h}
\begin{split}
h(\delta_{r,x},\lambda_{r,x} \mid Z{(x)}) = &p_{r}(Z{(x)} \mid \lambda_{r,x}) \\
&- \sum_{j} \left(\frac{\delta_{rj,x}\lambda_{rj,x}^2}{2\rho}  + \frac{ (1-\delta_{rj,x})\lambda_{rj,x}^2}{2\gamma} \right)
\end{split}
\end{equation}
which has a gradient given by
\begin{equation}
\label{grad}
\begin{split}
\mathbb{G}_j=&\nabla_{\hspace*{-1mm} \lambda_{rj,x}} ~ h(\delta_{rj,x},\lambda_{rj,x} \mid Z{(x)})=\nabla_{\hspace*{-1mm} \lambda_{rj,x}} ~ p_{r}(Z{(x)} \mid \lambda_{r,x}) \\
&-\delta_{rj,x} \frac{\lambda_{rj,x}}{\rho} - (1-\delta_{rj,x})\frac{\lambda_{rj,x}}{\gamma} .
\end{split}
\end{equation}
For any $j$-th component of $\lambda_{r,x}$ such that $\delta_{rj,x}=1$, we propose a new value
\begin{equation}
\label{eqprop}
\begin{split}
\widetilde{\lambda}_{rj,x}|\lambda_{r,x} \sim \text{N}\left(\lambda_{rj,x}+\dfrac{\sigma}{2} \mathbb{G}_{j} ,\sigma^{2}\right),
\end{split}
\end{equation}
where $\sigma$ is some constant step size and $\mathbb{G}_{j}$ represents the j-th component of the gradient. Let $f(\widetilde{\lambda}_{rj,x} \mid \lambda_{r,x})$ denote the density of the proposal distribution in \eqref{eqprop}. We also define $\widetilde{\lambda}_{r,x}=(\lambda_{r1,x}, \dots,\widetilde{\lambda}_{rj,x},\dots,\lambda_{rp,x})$ and the acceptance probability as
\begin{equation}
\label{acclambda}
\begin{split}
\zeta_{rj}=\min \left(1,\dfrac{f(\lambda_{rj,x} \mid \widetilde{\lambda}_{r,x})}{f(\widetilde{\lambda}_{rj,x} \mid \lambda_{r,x})} \times \dfrac{p(\delta_{r,x},\widetilde{\lambda}_{r,x}|Z{(x)})}{p(\delta_{r,x},\lambda_{r,x} \mid Z{(x)})} \right),
\end{split}
\end{equation}
such that we set $\lambda_{rj,x}=\widetilde{\lambda}_{rj,x}$ with probability $\zeta_{rj}$.

\noindent Conversely, for any $j$-th component of $\lambda_{r,x}$ such that $\delta_{rj,x}=0$, we update its value
\begin{equation}
\lambda_{rj,x} \sim \text{N}(0,\gamma).
\end{equation}
Finally, for each $j \in V$, we define $\bar{\delta}_{r,x}=(\delta_{r1,x}, \dots,(1-\delta_{rj,x}),\dots,\delta_{rp,x})$ and set
\begin{equation}
\label{acclambda}
\begin{split}
\tau_{rj}=\min \left(1,\dfrac{p(\bar{\delta}_{r,x},\lambda_{r,x} \mid Z{(x)})}{p(\delta_{r,x},\lambda_{r,x} \mid Z{(x)})} \right),
\end{split}
\end{equation}
the new value $\delta_{rj}=1-\delta_{rj}$ is accepted with probability $\tau_{rj}$. 

\subsubsection{Updating of $\theta_{xh}$  - Step II}
\label{app:theta}
In the second step of the proposed algorithms we sample from the full conditional of $(\theta_{xh},\epsilon_{xh})$,
\begin{equation}
\begin{split}
p{(\theta_{xh},\epsilon_{xh} \mid \nu_{rj},\delta)} & \propto  p(\theta_{xh} \mid \epsilon_{xh}) p(\epsilon_{xh}) \prod_{r < j} p(\delta_{rj}|\nu_{rj},\theta)
\end{split}
\end{equation}
where 
$\bm{\delta}_{rj}=[\delta_{rj,x}]_{x \in \X}$ and $\bm{\theta}$ the $(q \times q )$ symmetric matrix with entries $\theta_{xh}$, for all $x,h \in \X$. 
Since the normalizing constant is analytically intractable, we use Metropolis-Hastings steps to sample $\theta_{xh}$ and $\epsilon_{xh}$ from their joint posterior full conditional distribution for each pair $x,h \in \X$. 
At each iteration we perform two steps: a between-model and a within-model move (see \cite{gottardo2008markov} for more details). 
For the between-model move, if in the current state is $\epsilon_{xh}=1$, we propose $\widetilde{\epsilon}_{xh}=0$ and $\widetilde{\theta}_{xh}=0$. If in the current state is $\epsilon_{xh}=0$, we propose $\widetilde{\epsilon}_{xh}=1$ and sample $\widetilde{\theta}_{xh}$ from the proposal density $f(\widetilde{\theta}_{xh})= \text{Gamma}(\widetilde{\theta}_{xh}; \widetilde{\alpha},\widetilde{\beta})$.
When the proposed value includes $\widetilde{\epsilon}_{xh}=0$, the Metropolis-Hastings acceptance ratio is
\begin{equation}
\begin{split}
r=\dfrac{p{(\widetilde{\theta}_{xh},\widetilde{\epsilon}_{xh} \mid \nu_{rj},\delta)} f(\theta_{xh})}{p{(\theta_{xh},\epsilon_{xh} \mid \nu_{rj},\delta)}}, 
\end{split}
\end{equation}
whereas if we move from $\epsilon_{xh}=0$ to $\widetilde{\epsilon}_{xh}=1$, the 
acceptance ratio is
\begin{equation}
\begin{split}
r=\dfrac{p{(\widetilde{\theta}_{xh},\widetilde{\epsilon}_{xh} \mid \nu_{rj},\delta)} }{p{(\theta_{xh},\epsilon_{xh} \mid \nu_{rj},\delta)}f(\widetilde{\theta}_{xh})}. 
\end{split}
\end{equation}
We then perform a within-model move whenever the value of $\epsilon_{xh}$ sampled from the between-model move is 1. This step is not required to defined and ergodic Markov chain but it usually improves the mixing of the chains \citep{gottardo2008markov}. For this step, we propose a new value of $\theta_{xh}$ using the same proposal density as before. The Metropolis-Hastings ratio for this step is
\begin{equation}
\begin{split}
r=\dfrac{p{(\widetilde{\theta}_{xh},\widetilde{\epsilon}_{xh} \mid \nu_{rj},\delta)} f(\theta_{xh})}{p{(\theta_{xh},\epsilon_{xh} \mid \nu_{rj},\delta)}f(\widetilde{\theta}_{xh})}. 
\end{split}
\end{equation}

\subsubsection{Updating of $\nu_{rj}$  - Step III}
\label{app:nu}
In step III of the proposed algorithms we sample from the full conditional distribution of $\nu_{rj}$:
\begin{equation}
\begin{split}
p(\nu_{rj}|\delta) \propto \dfrac{\exp(a \nu_{rj})}{(1 + e^{\nu_{rj}} )^{a+b}} C(\nu_{rj},\theta)^{-1} \exp(\nu_{rj} \bm{1}^{T} \bm{\delta}_{rj} ).
\end{split}
\end{equation}
For each pair $r,j \in V,r \neq j$, we propose a value $\widetilde{q}_{rj}$  from the density Beta$(1,2)$, then set $\widetilde{\nu}_{rj}= \text{logit}(\widetilde{q}_{rj})$. The proposal density can be written in terms of $\widetilde{\nu}$ as
$f(\widetilde{\nu}_{rj}) =\dfrac{1}{B(\widetilde{a},\widetilde{b})} \hspace*{0.1cm} \dfrac{e^{\widetilde{a} \widetilde{\nu}_{rj}}}{(1+e^{\widetilde{\nu}_{rj}})^{\widetilde{a}+\widetilde{b}}}$
and the Metropolis-Hastings ratio is
$
r=\dfrac{p(\widetilde{\nu}_{rj} \mid \delta) f(\nu_{rj})}{p(\nu_{rj}|\cdot)f(\widetilde{\nu}_{rj})}.
$

\section{Parameter setting}
\label{app:LDSS}

\new{A crutial aspect in Ising models is the choice of priros and hyper-prior parameters ofr the model. 
Here we provide some default guidelines that can be followed in the absence of a priori knowledge.}

For our high-dimensional and low-dimensional simulations settings, we simulated observations from $Y_{V}|\{X=x\} \sim \text{Ising}(\lambda_x)$ with associated undirected graph $\G$ with $p=10$ nodes and $q=4$ levels of $X$, $x \in \X = \{0,1,2,3\}$. 
For all the levels $x \in \X$, we set $\lambda_{rr,x}=-1$, for all $r \in V$, and we set the non-zero interactions to $\lambda_{rj,x}=1.5$,  for all $r,j \in V, j <r$. 
We select the penalization parameter for the frequentists approaches SL and DSSL using BIC.
For the DSSL method we set the parameter that controls the degree of sharing between the levels of $X$ to $r=1/\sqrt{q}$, after having standardized the columns of the design matrix \citep{ollier}. 

\new{Two key aspects for our proposed methods are the prior elicitation for (i) the MRF and (ii) the sparse-inducing priors, given by the Diaconis and the spike-and-slab priors, for the FB and AB respectively. 
Further details on the sensitivity of the results to the choice of such parameters are given in Appendix \ref{app:sensitivity}.

Following the recommendations given in \cite{peterson2015bayesian},} 
we set the hyper-parameters of the MRF prior to $a=1$, $b=3$, $\alpha=1$, $\beta=2$ and $\omega=0.6$.

\new{For the sparse-inducing priors used here, it is common to estimate the hyper-parameters from the data, setting up a grid of values, and choosing the one that provides a better fit \cite{GeorgeEdwardI.1993VSvG}, leading to an increased computation cost. 
Thus, we propose to set such values in a way that results in moderate sparsity a-priori. 
}
For the FB approach, we set $g_x=0.02$ and $s_x$ in such a way that the prior probability of each cell of the contingency table $\mathcal{I}_x=\times (0,1)^{p}$ is equal to $g_x/|\mathcal{I}_x|$, for all $x \in \X$. 
For the AB approach, \new{the choice of $\rho$ and $\gamma$ has an impact of the posterior probability of edge inclusion, it is common practice to set $\gamma$ small to avoid over-shrinkage of large effects and $\rho$ large to shrink ignorable edges to zero. Here,} we set $\rho=2$ and $\gamma=0.5$ in the low dimensional simulations; these parameter values result in moderate sparsity a priori. 
For the high-dimensional scenarios, 
we set the spike-and-slab prior parameters $\rho=10$ and $\gamma=10^{-1}$, that result in an increased sparsity wrt the low-dimensional simulations. 
Finally, for the approaches FBS and ABS, we set the prior of the model as the product of $p$ Bernoulli$(0.2)$ for the low-dimensional simulations and a product of $p$ Bernoulli$(0.1)$ for the high-dimensional ones.

\new{For the real applications analysed here, we recommend setting priors based on the low-dimensional scenario, as our case studies have small number of $p$ edges (10 and 18), and it has been assumed to have a moderate sparsity graph structure a-priori.}

\section{Sensitivity analysis}
\label{app:sensitivity}
As mentioned earlier, two key aspects of our proposed methods are the prior elicitation for (i) the Markov Random Field (MRF) and (ii) the sparse-inducing priors, specifically the Diaconis and the spike-and-slab priors, for the Fully Bayesian (FB) and Adaptive Bayesian (AB) approaches respectively.

A sensitivity analysis for the impact of the parameters in the MRF prior is detailed in \cite{peterson2015bayesian}. Therefore, our focus here is on assessing the prior sensitivity of the model with respect to the sparse-inducing priors.

For the AB approach, the elicitation of parameters $\rho$ and $\gamma$ is crucial, with the condition $\rho \gg \gamma > 0$. These parameters represent the variance of the spike and slab densities respectively. Our goal is to identify values for $\rho$ and $\gamma$ that effectively distinguish practically relevant edges. In general, larger differences between $\rho$ and $\gamma$ result in more aggressive penalties.

\begin{figure*}
    \begin{center}
		\includegraphics[scale=0.45]{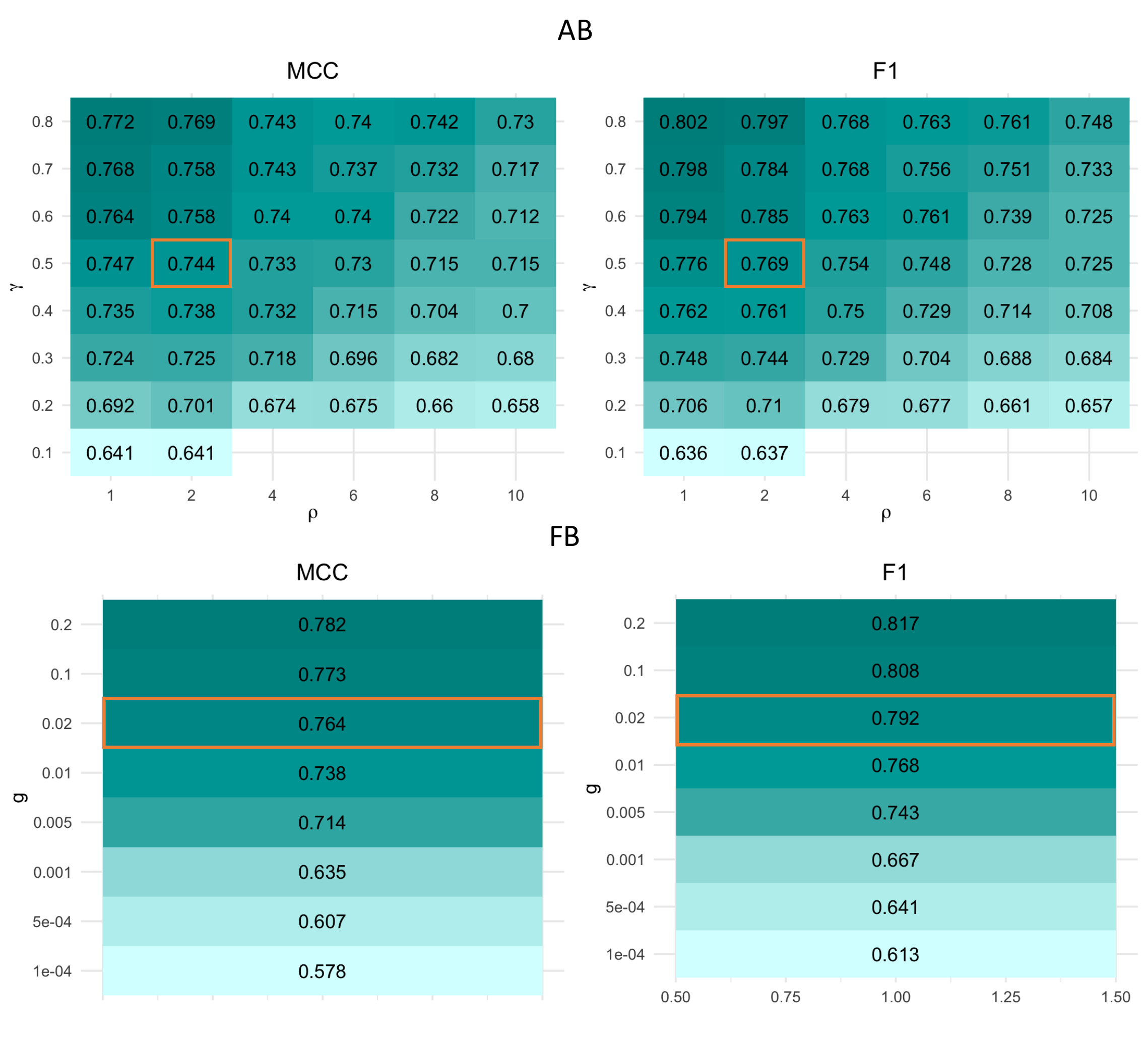}
  \caption{Sensitivity analysis. Average MCC and F1 score for different hyper-parameters for both the FB and AB. Reported values in this paper are in an orange square.}
  \label{fig:sensitivity}
    \end{center}
\end{figure*}

To evaluate the impact of $\rho$ and $\gamma$ on posterior inference, we applied our proposed FB approach across a range of values: $\rho = \{1, 2, 4, 6, 8, 10\}$ and $\gamma = \{0.1, 0.2, 0.3, 0.4, 0.5, 0.6, 0.7, 0.8\}$. These values were tested on ten fixed datasets generated following the setup of the low-dimensional simulation C, as described in the \ref{sec:reslow} section. 
The results presented in that section used $\rho = 2$ and $\gamma = 0.5$. The effect on the average MCC and average F1 score is summarized in Figure \ref{fig:sensitivity}. 
Our results show that as $\rho - \gamma$ increases, both average MCC and F1 score improve, ranging from 0.641 to 0.772 and from 0.636 to 0.802, respectively.

For the FB approach, the parameter $g_x$ is pivotal as it helps to distinguish practically relevant edges. Similarly, for the AB approach, we propose a range of $g_x$ values: $\{0.2, 0.1, 0.02, 0.01, 0.005, 0.001, 0.0005, 0.0001\}$, and applied this method to the same ten fixed datasets. 
The datasets were generated following the setup of the low-dimensional simulation C. The results reported in this paper were obtained using $g_x = 0.02$. 
Figure \ref{fig:sensitivity} demonstrates that as $g_x$ decreases, the average MCC and F1 score increase, ranging from 0.578 to 0.782 and from 0.613 to 0.817, respectively.

\section{MCMC convergence}
\label{app:MCMCconver}
\begin{figure*}[h!]
	\begin{center}
  \tiny{\textbf{AB}}\\
  \includegraphics[scale=0.4]{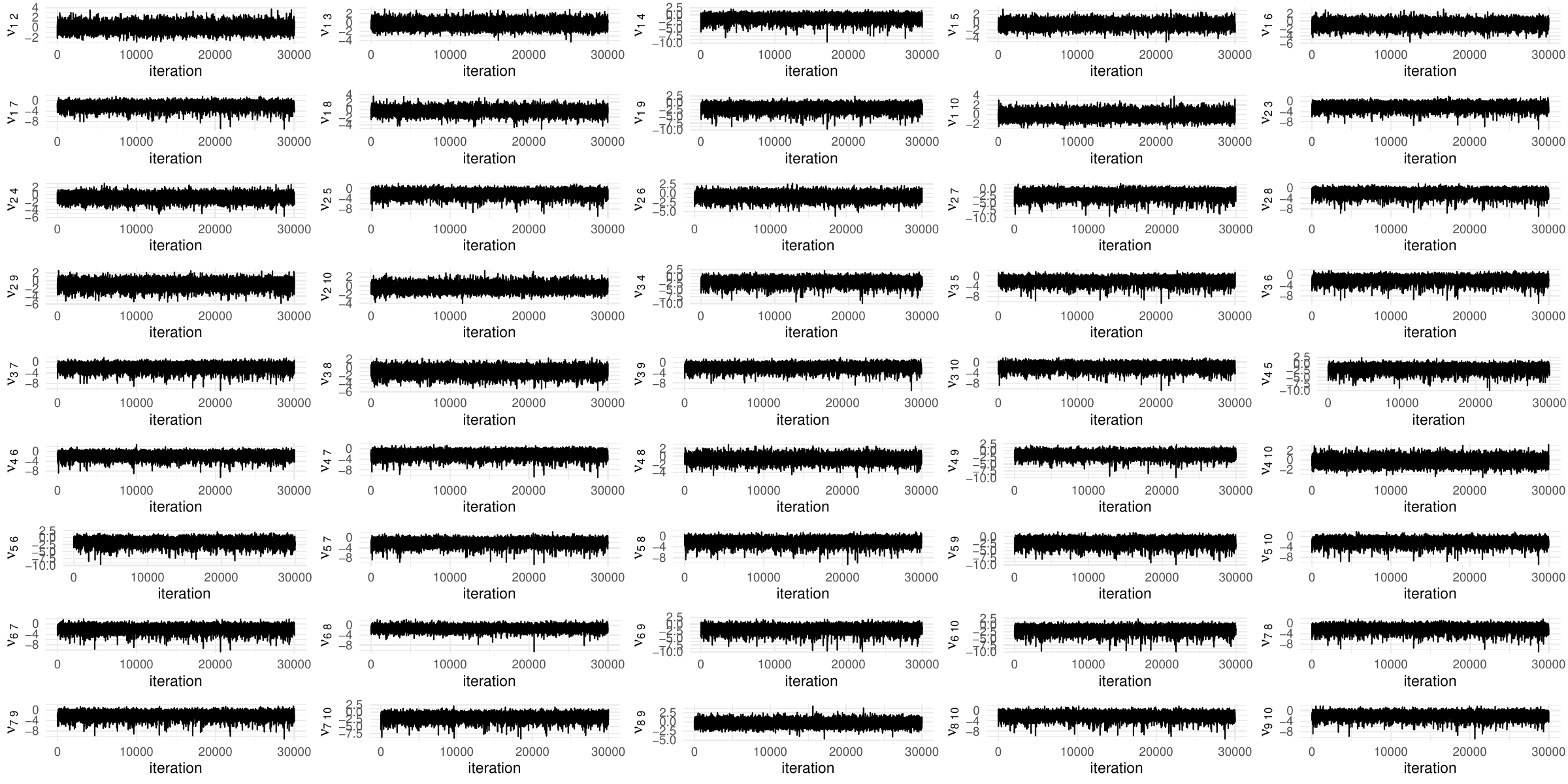}\\
  \vspace*{0.3cm}
  \tiny{\textbf{FB}}\\
  \includegraphics[scale=0.4]{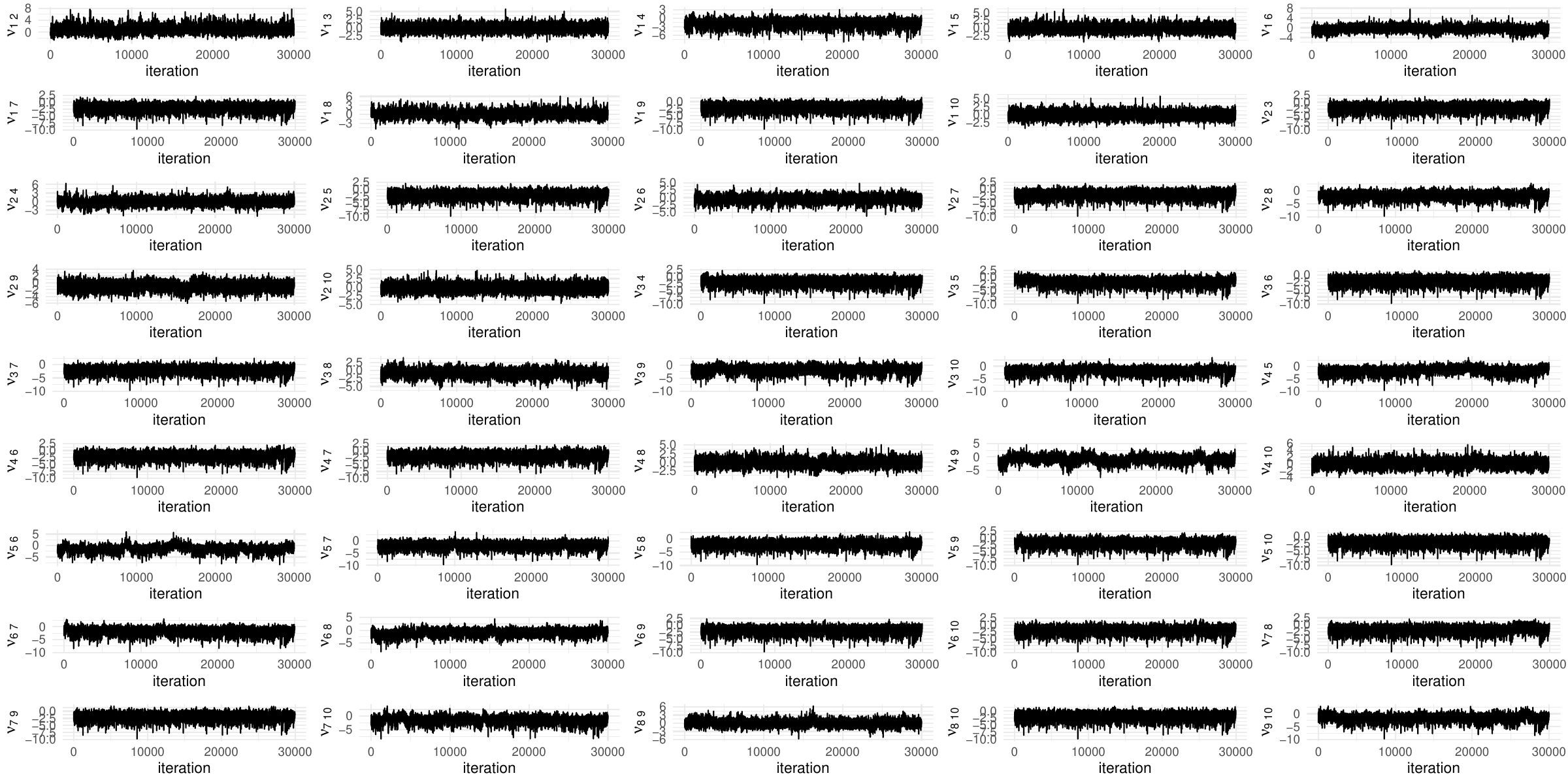}
  \caption{Traceplot for the posterior distribution of $\nu_{rj}$ for the AB (top) and FB (bottom) methods.}
		\label{NuMCMCconvergeAB}
	\end{center}
\end{figure*}

\new{We visually investigate convergence and
mixing, by plotting the traceplot of the posterior chain for the key parameters in our models $\theta_{ xh }$ and $\nu_{rj}$.
Figures \ref{ThetaMCMCconverge} and \ref{NuMCMCconvergeAB} show the trace plots obtained for one of the runs of the low-dimensional simulated Scenario (C).

Additionally, we investigated how robust
results are when running the algorithm multiple times. 
To this aim, we analysed each of the low-dimensional scenarios using distinct seeds for the random number generator (i.e. \texttt{set.seed} in \textbf{R}), and compared
results with a Pearson correlation coefficient, showed in 
Table \ref{T:MCMCconvergency}. 
Our results show a high correlation among all our runs, leading then to the same graph structure.
}

\small{\begin{figure}[h!]
	\begin{center}
  \tiny{\textbf{AB}}\\
		\includegraphics[scale=0.4]{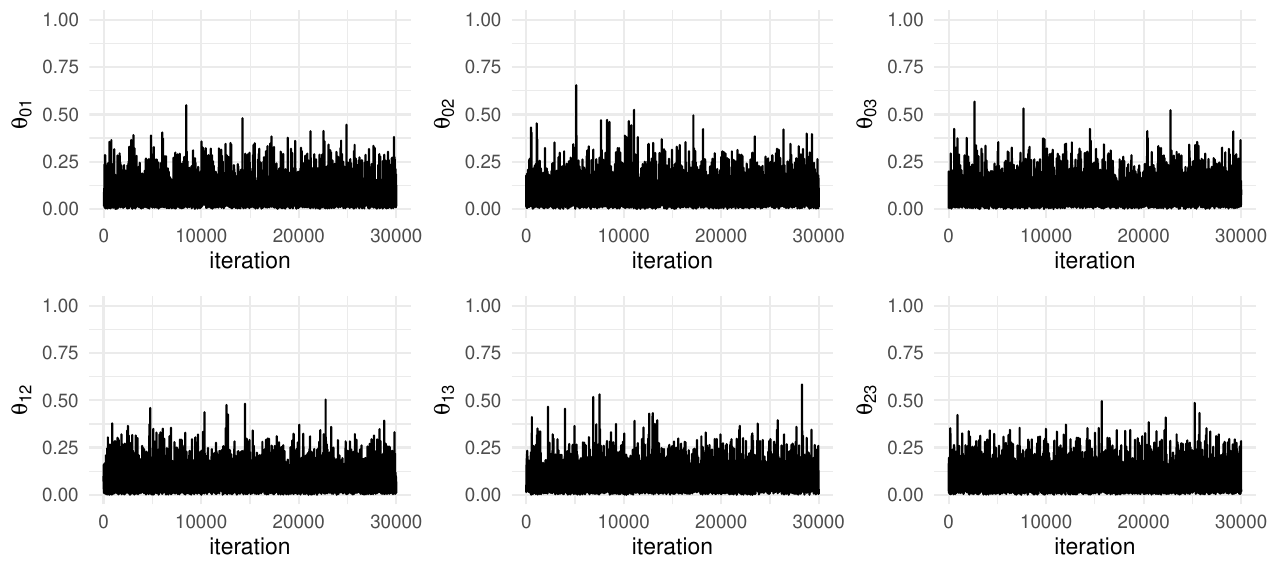}\\
  \vspace*{0.3cm}
  \tiny{\textbf{FB}}\\
		\includegraphics[scale=0.4]{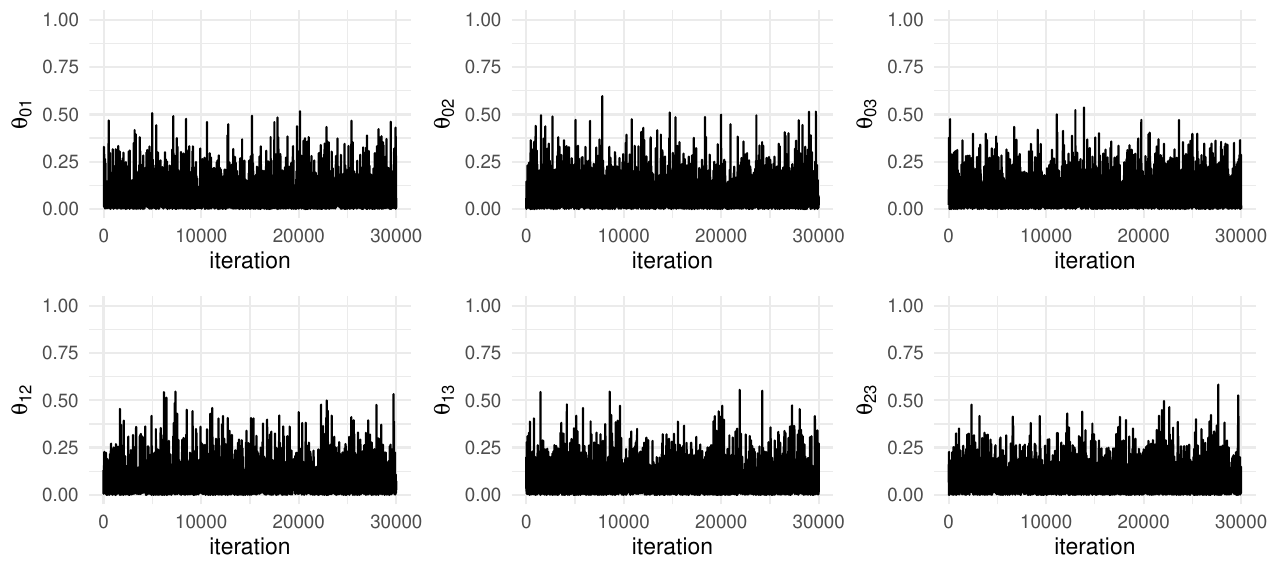}
  \caption{Traceplot for the posterior distribution of $\theta_{xk}$ for the AB (top) and the FB (bottom) methods.}
		\label{ThetaMCMCconverge}
	\end{center}
\end{figure}}

\begin{table}[ht]
\centering
\caption{Average Pearson correlation coefficient.\label{T:MCMCconvergency}}
\begin{tabular}{ccccc}
  \hline
Model &Scenario(A) & Scenario(B) & Scenario(C) & Scenario(D) \\ 
  \hline
AB & 1.000 & 1.000 & 1.000 & 1.000 \\ 
  FB & 1.000 & 0.922 & 0.959 & 1.000 \\ 
   \hline
\end{tabular}
\end{table}

\section{Uncertainty quantification of the graph structure}
\label{app:UQ}

\new{Both our proposed Fully Bayesian (FB) and Approximate Bayesian (AB) methods provide a framework for quantifying uncertainty in graph structures, a feature that cannot be achieved through state-of-the-art frequentist approaches. 
The Bayesian framework allows us to derive a posterior distribution for the edge inclusion probability, enabling a more nuanced understanding of the uncertainty associated with each edge.

Figures \ref{UQAB} and \ref{UQFB} visually represent the graph structures generated by the AB and FB methods, respectively, for Scenario (A) in the low-dimensional setting. 
These figures depict the graphs at the 25\% quantile, mean, and 75\% quantile of the posterior distribution of edge inclusion probability. 
Our findings indicate that the FB method exhibits smaller variance across this posterior distribution, as the network structures are highly consistent between the 25\% quantile, mean, and 75\% quantile.
In contrast, the AB method yields graph structures similar to FB when considering the mean of the posterior distribution. 
However, at the 25\% and 75\% quantiles, the AB method produces more variability, with additional spurious edges (depicted as dotted lines) appearing at the 75\% quantile. 
This increased variability indicates greater uncertainty in the AB approach at the extremes of the posterior distribution.

Figure \ref{UQEdges} illustrates the number of edges included in the graph at different quantiles, further emphasizing the smaller variance of the FB method. 
The FB approach consistently identifies the correct number of edges (9 per level) more quickly than the AB method, highlighting its effectiveness in uncertainty quantification and edge selection.}

\small{\begin{figure*}[h!]
	\begin{center}
		\footnotesize Ground Truth\\
		 	\includegraphics[scale=0.2]{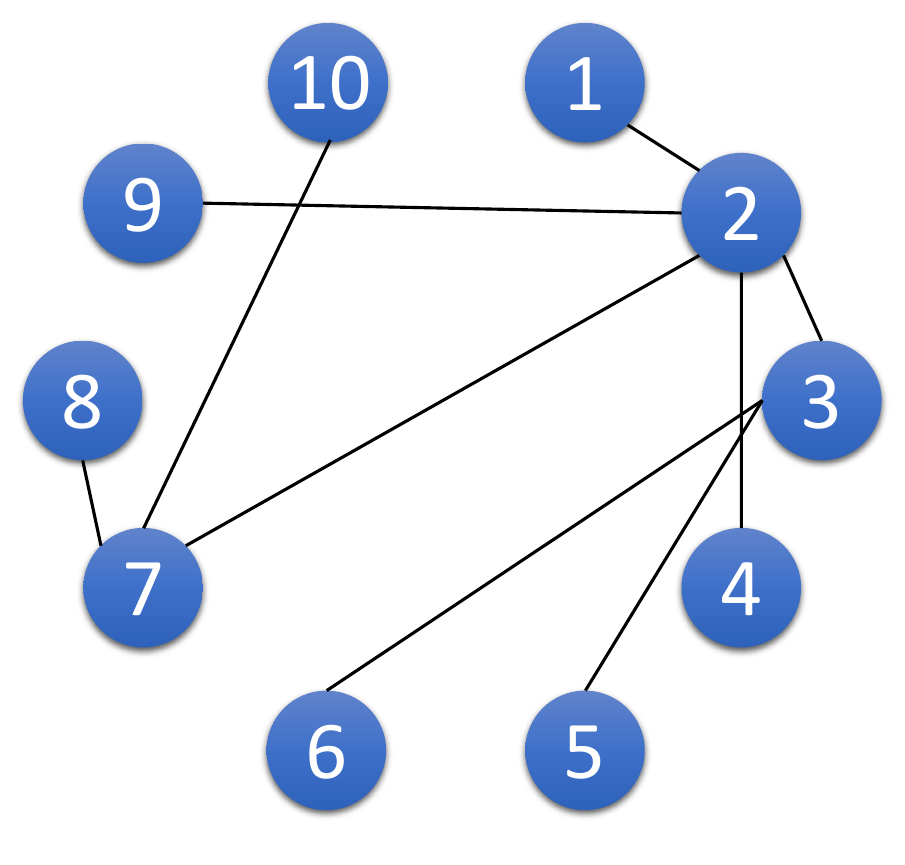}\tiny{$G(0)$}
		\includegraphics[scale=0.2]{figures/GT.pdf} $G(1)$ 
		\includegraphics[scale=0.2]{figures/GT.pdf} $G(2)$ 
		\includegraphics[scale=0.2]{figures/GT.pdf} $G(3)$ 
		\\ \vspace*{0.3cm} \footnotesize 75\% Quantile \\\includegraphics[scale=0.2]{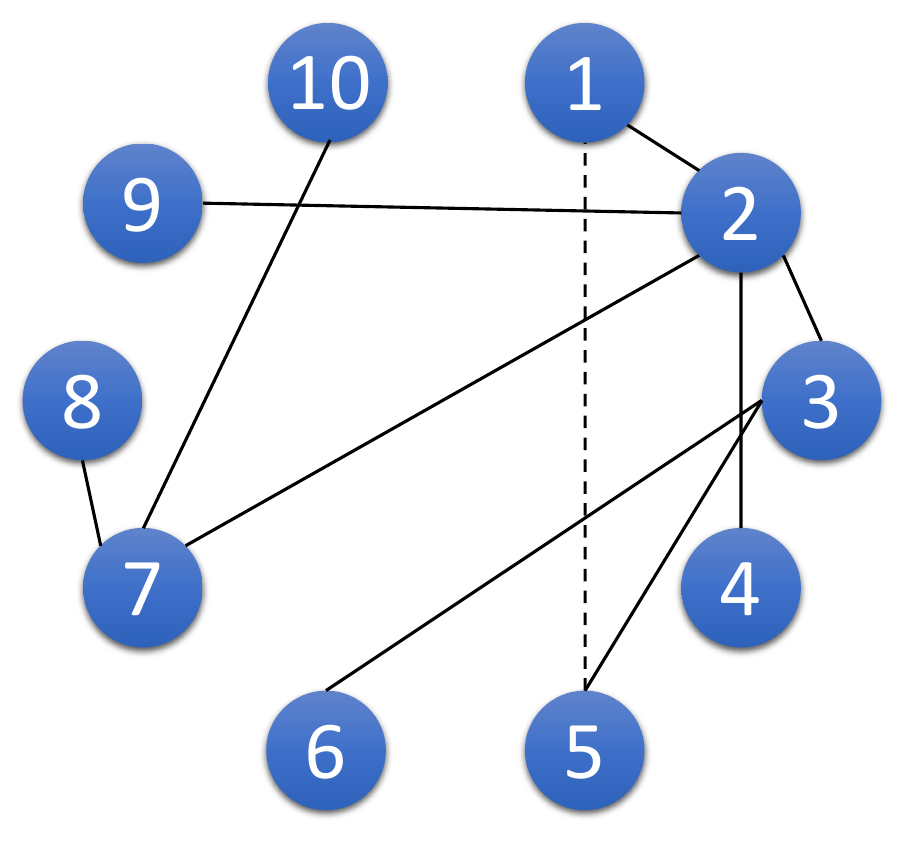}\tiny{$G(0)$}
		\includegraphics[scale=0.2]{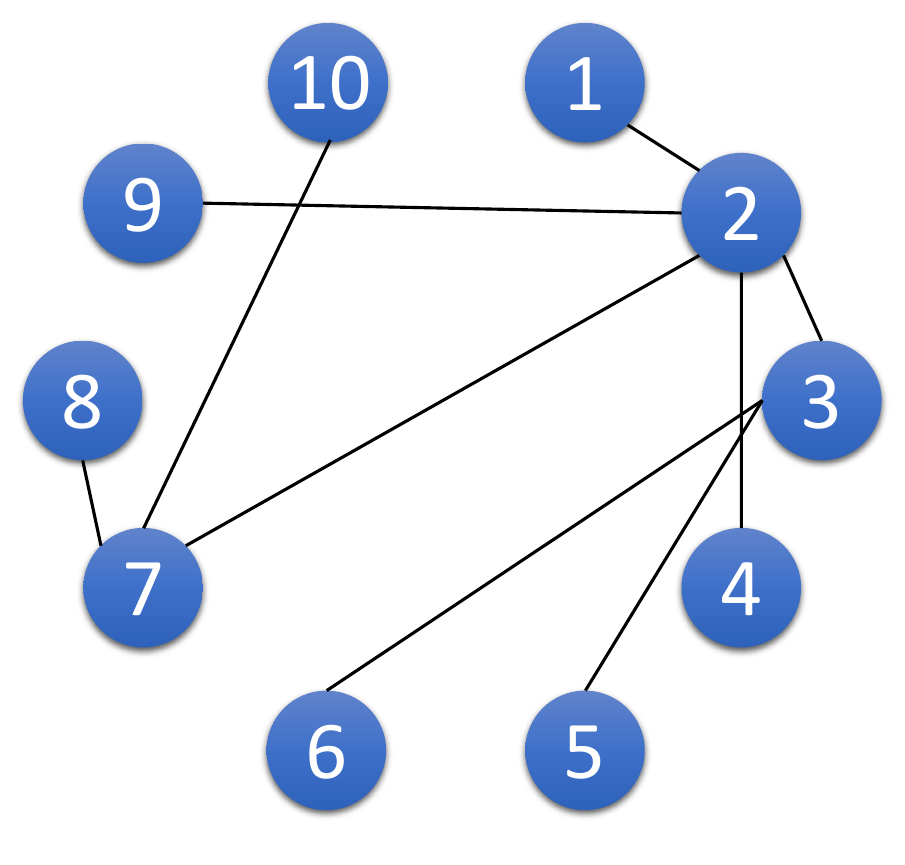}$G(1)$
		\includegraphics[scale=0.2]{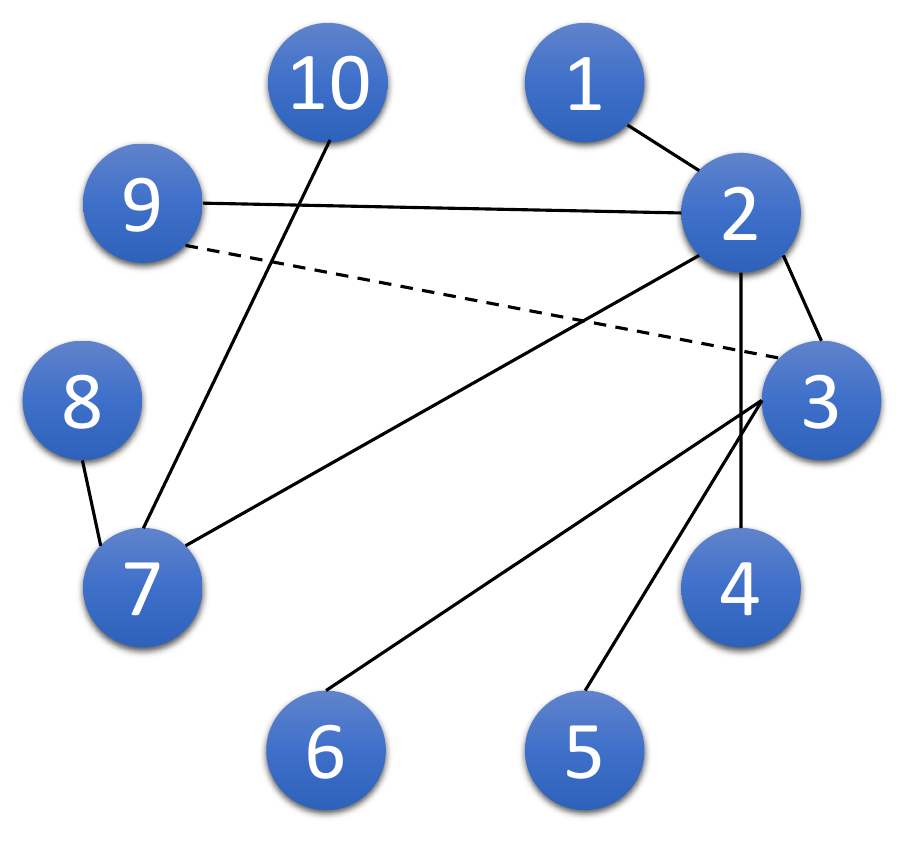}$G(2)$ 
		\includegraphics[scale=0.2]{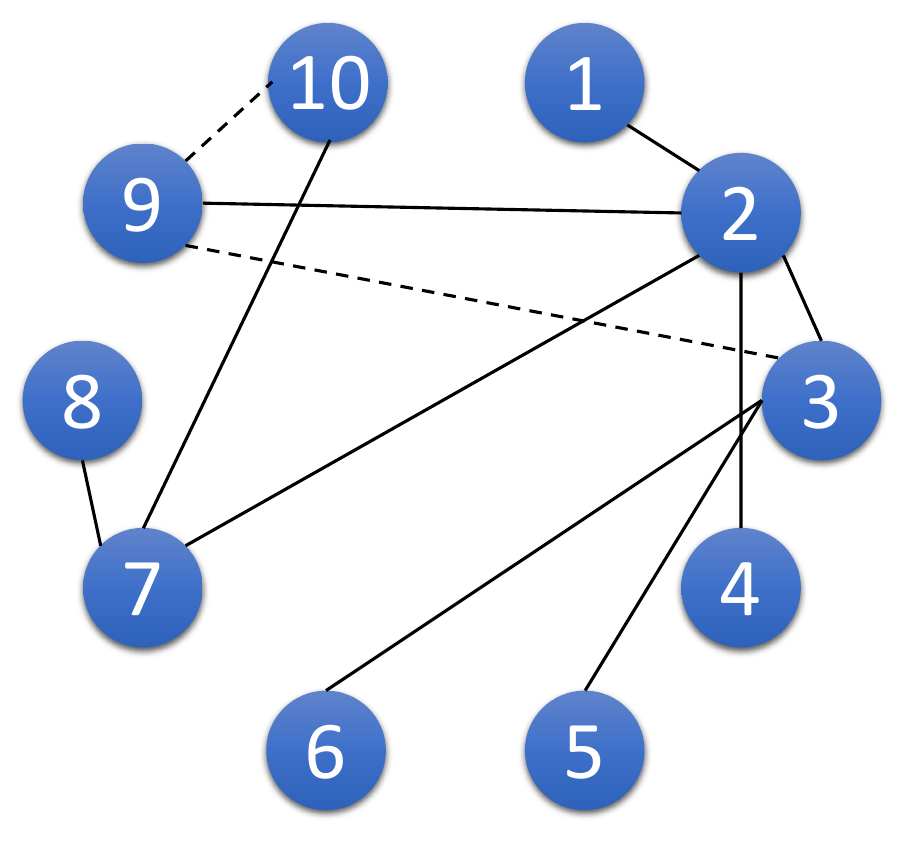}$G(3)$
		\\ \vspace*{0.3cm} \footnotesize Mean \\\includegraphics[scale=0.2]{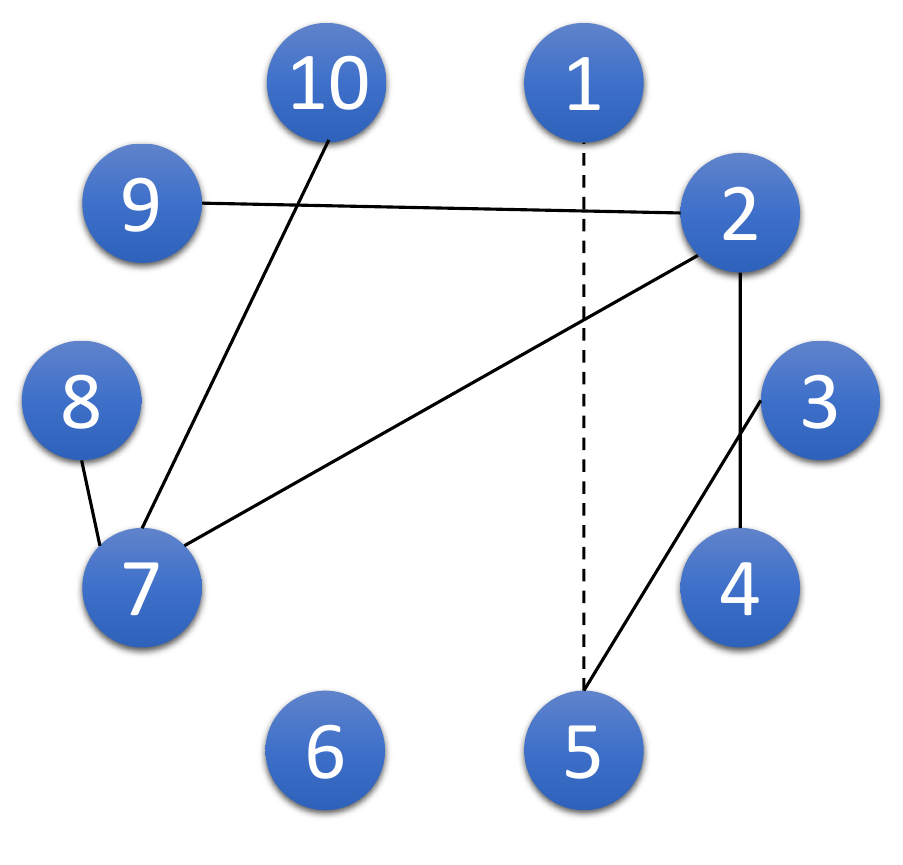}\tiny{$G(0)$}
		\includegraphics[scale=0.2]{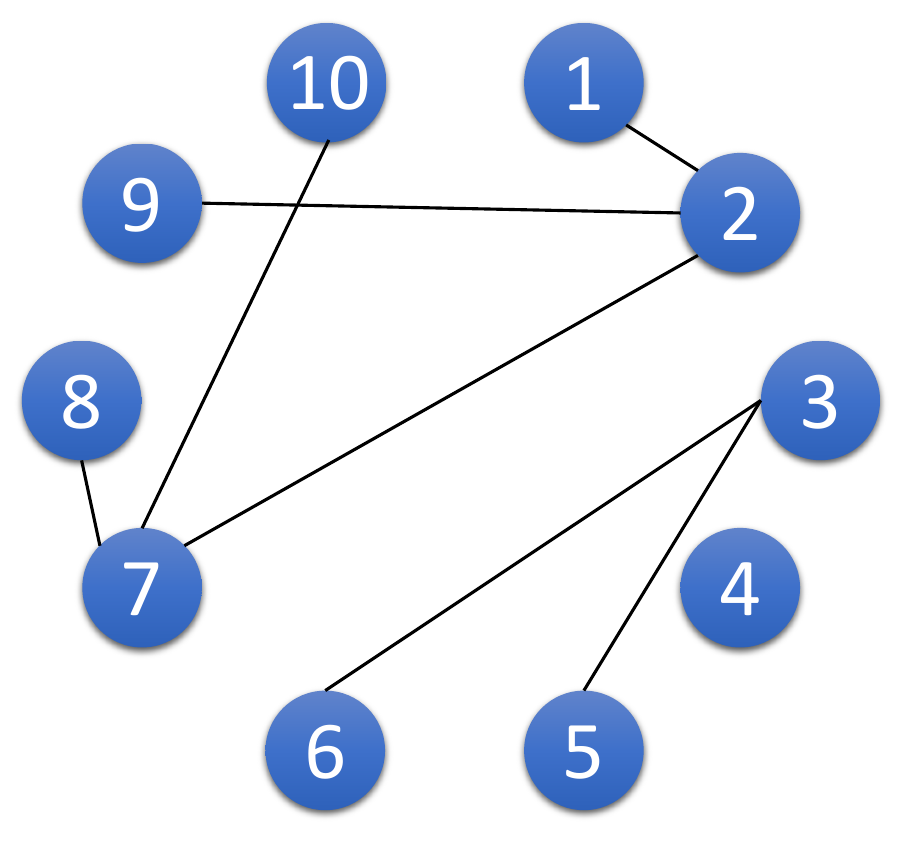}$G(1)$
		\includegraphics[scale=0.2]{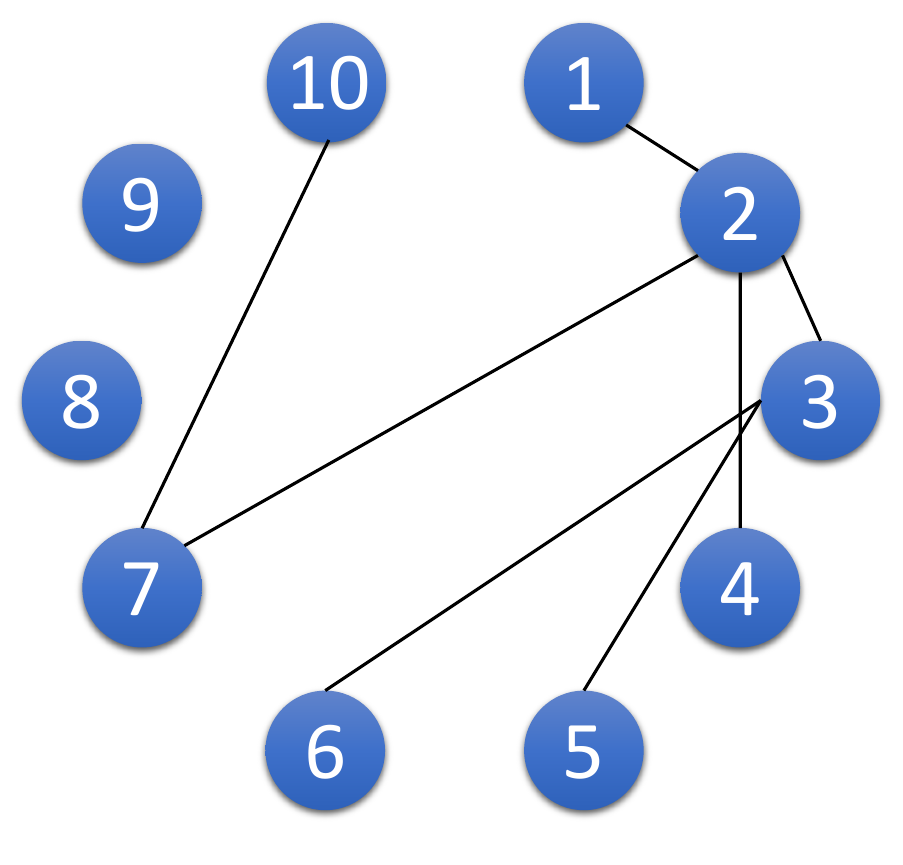}$G(2)$ 
		\includegraphics[scale=0.2]{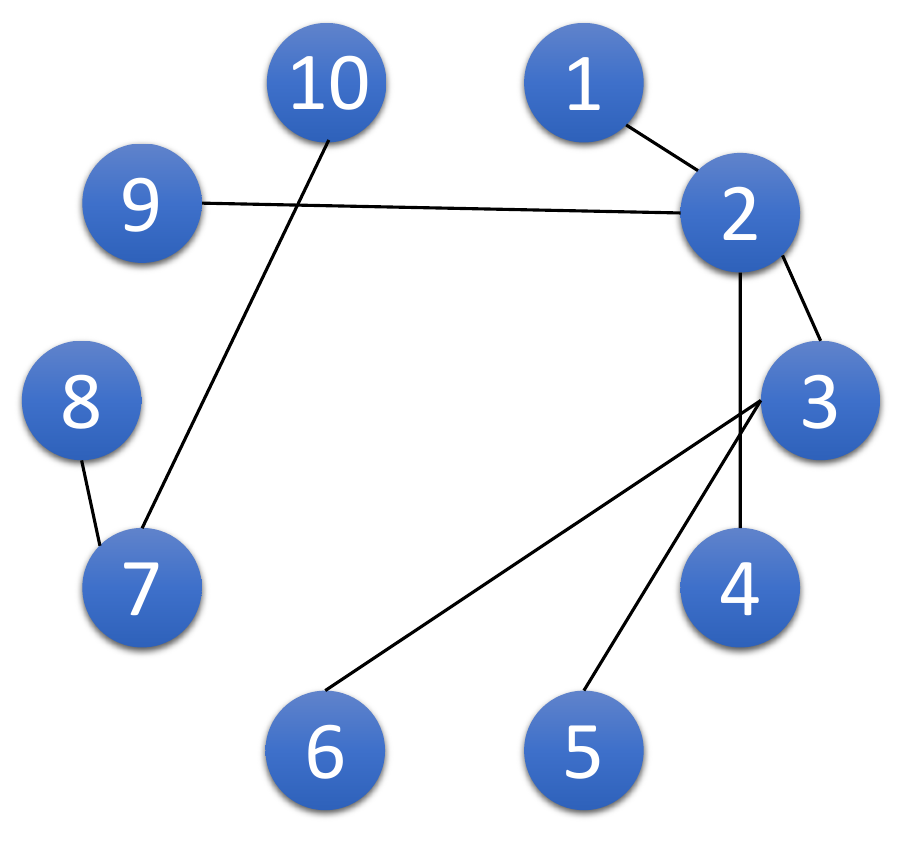}$G(3)$
			\\ \vspace*{0.3cm} 
   \footnotesize 25\% Quantile \\\includegraphics[scale=0.2]{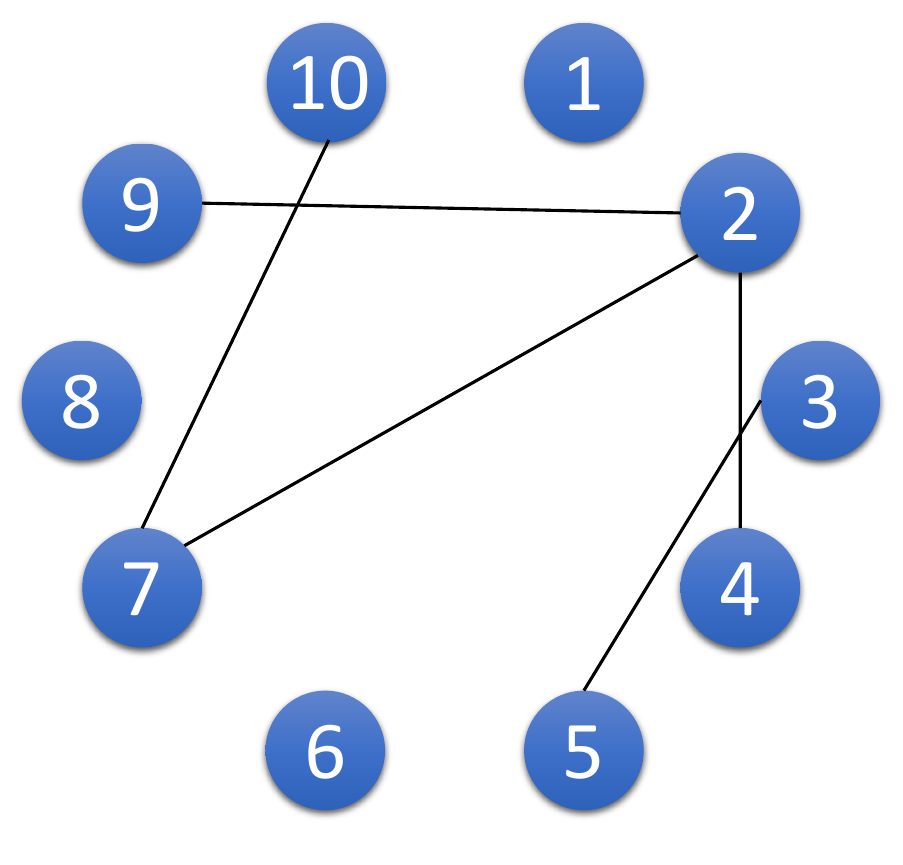}\tiny{$G(0)$}
		\includegraphics[scale=0.2]{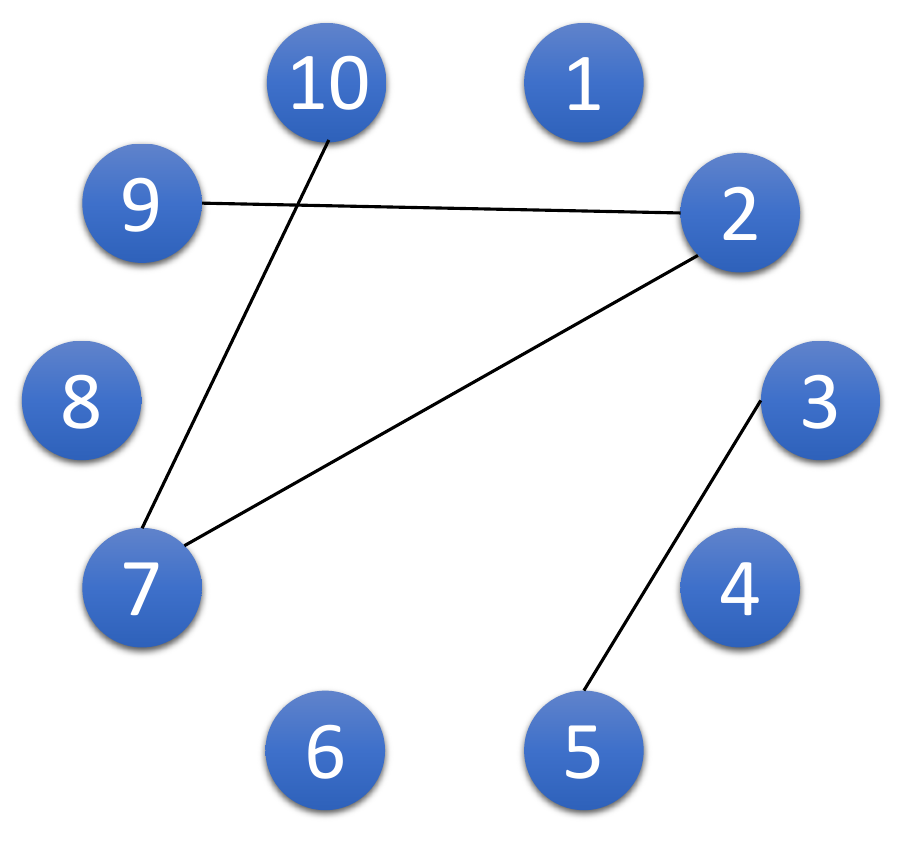}$G(1)$
		\includegraphics[scale=0.2]{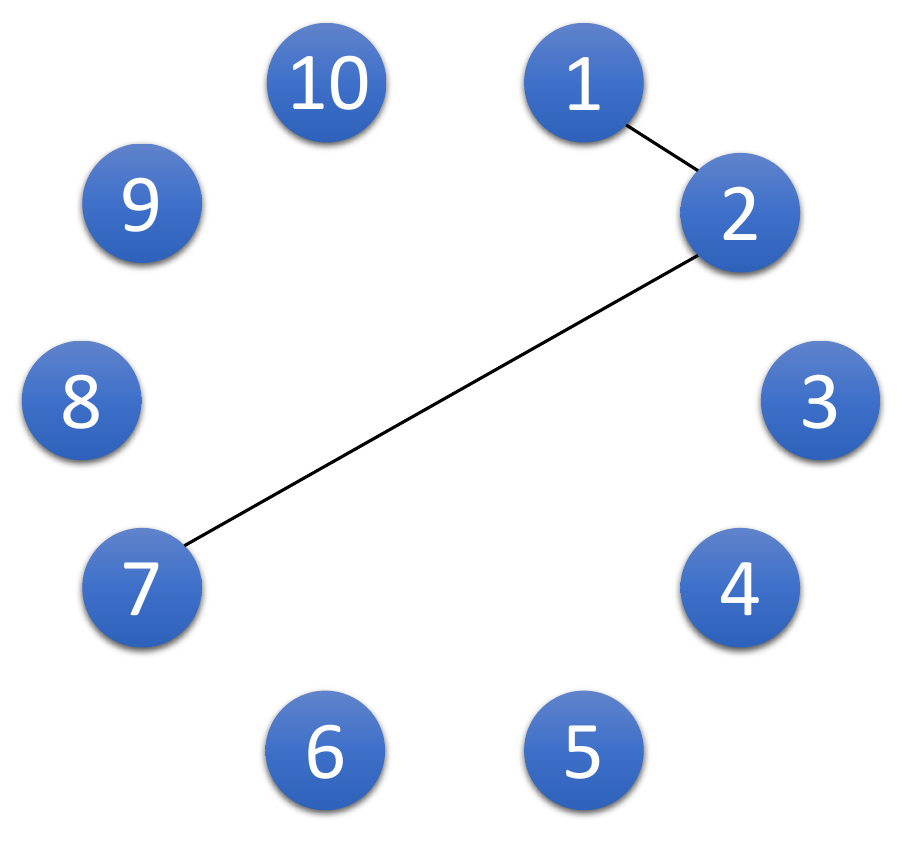}$G(2)$ 
		\includegraphics[scale=0.2]{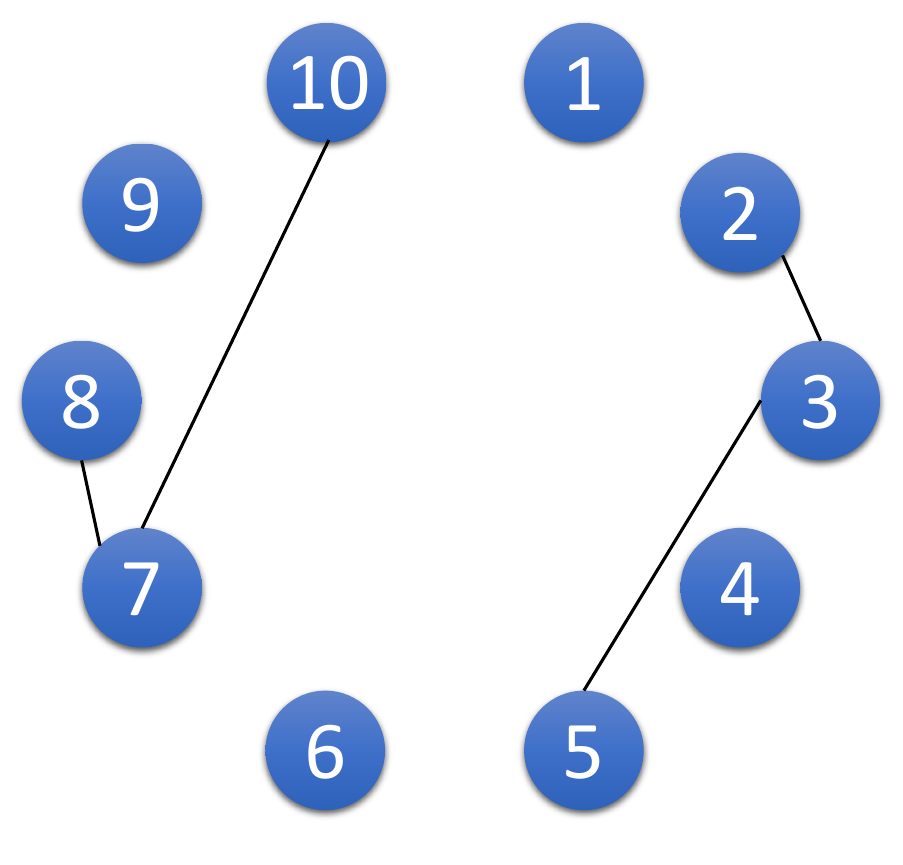}$G(3)$
  \caption{Graph structures obtained using the AB method in Scenario (A) in the low-dimensional setting at the 25\% quantile, mean, and 75\% quantile of the posterior distribution of edge inclusion probability, compared to the ground truth. Correctly identified edges are shown with solid lines, while spurious edges are represented by dotted lines.}
  \label{UQAB}
	\end{center}
\end{figure*}}

\small{\begin{figure*}[h!]
	\begin{center}
		\footnotesize Ground Truth\\
		 	\includegraphics[scale=0.2]{figures/GT.pdf}\tiny{$G(0)$}
		\includegraphics[scale=0.2]{figures/GT.pdf} $G(1)$ 
		\includegraphics[scale=0.2]{figures/GT.pdf} $G(2)$ 
		\includegraphics[scale=0.2]{figures/GT.pdf} $G(3)$ 
		\\ \vspace*{0.3cm} \footnotesize 75\% Quantile \\\includegraphics[scale=0.2]{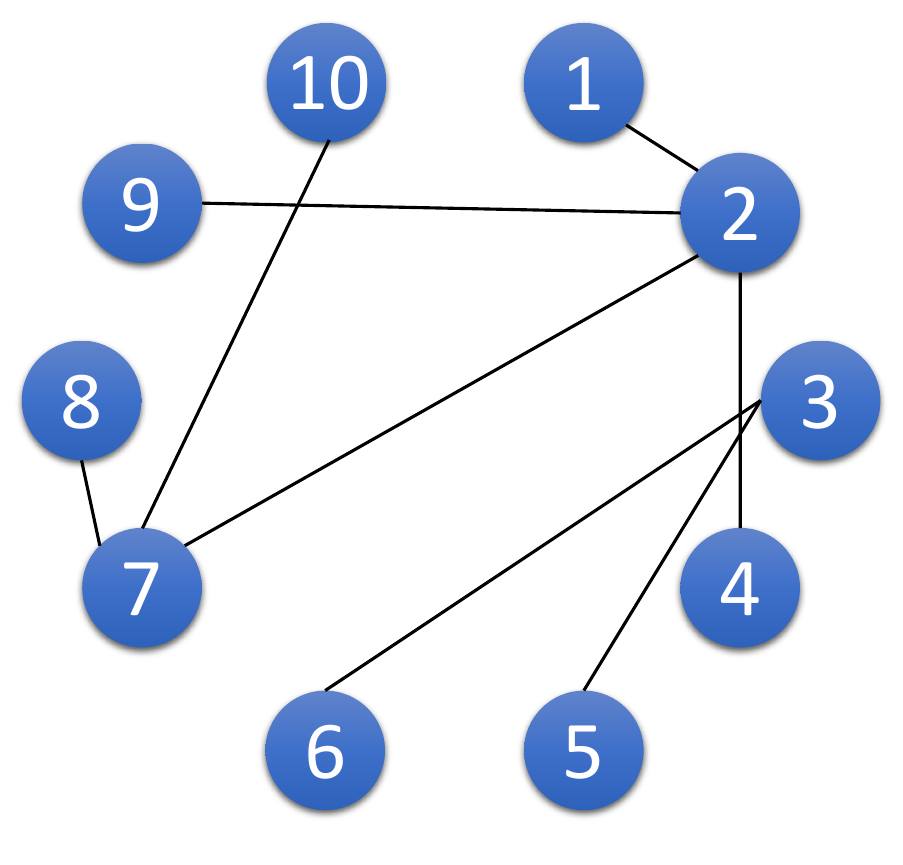}\tiny{$G(0)$}
		\includegraphics[scale=0.2]{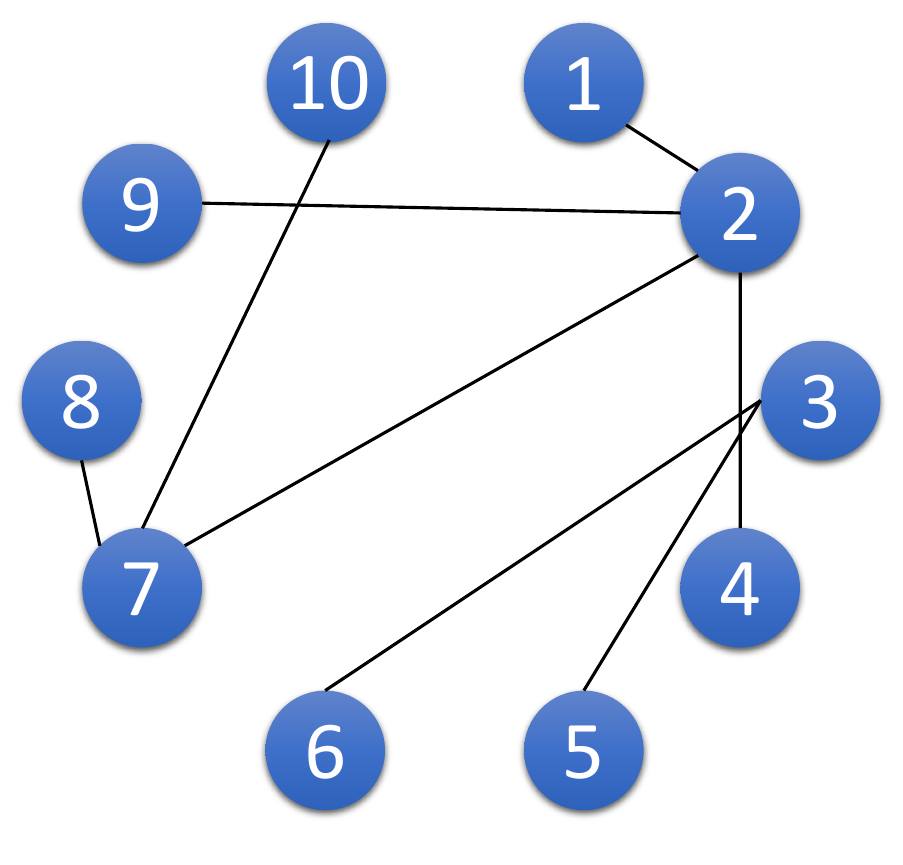}$G(1)$
		\includegraphics[scale=0.2]{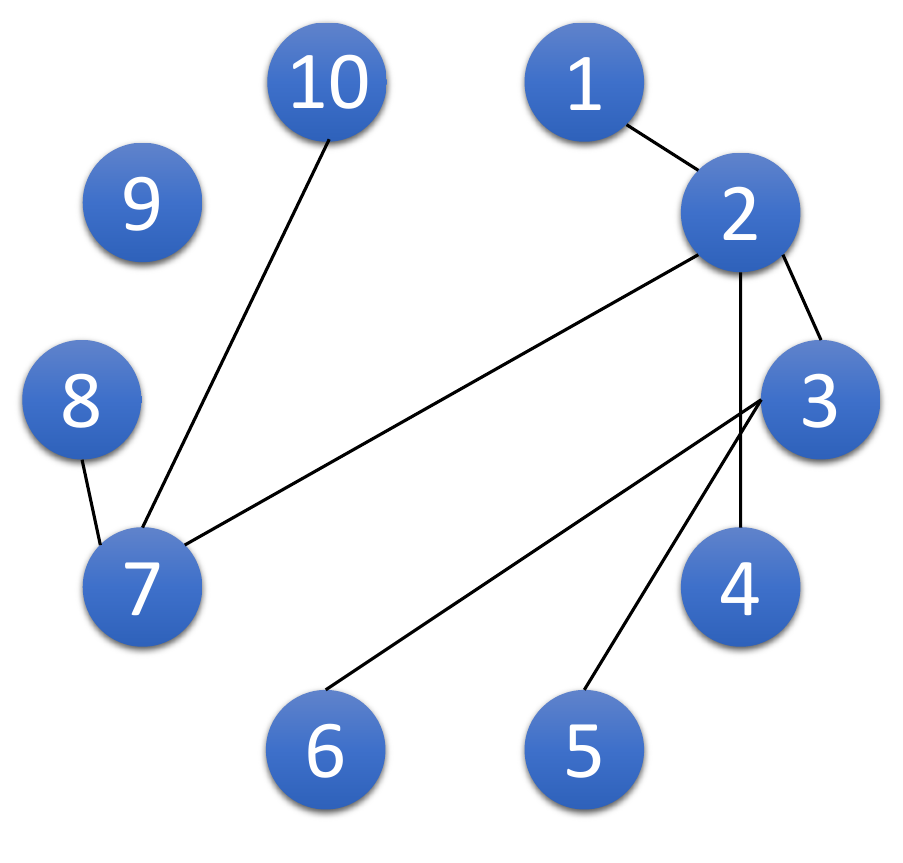}$G(2)$ 
		\includegraphics[scale=0.2]{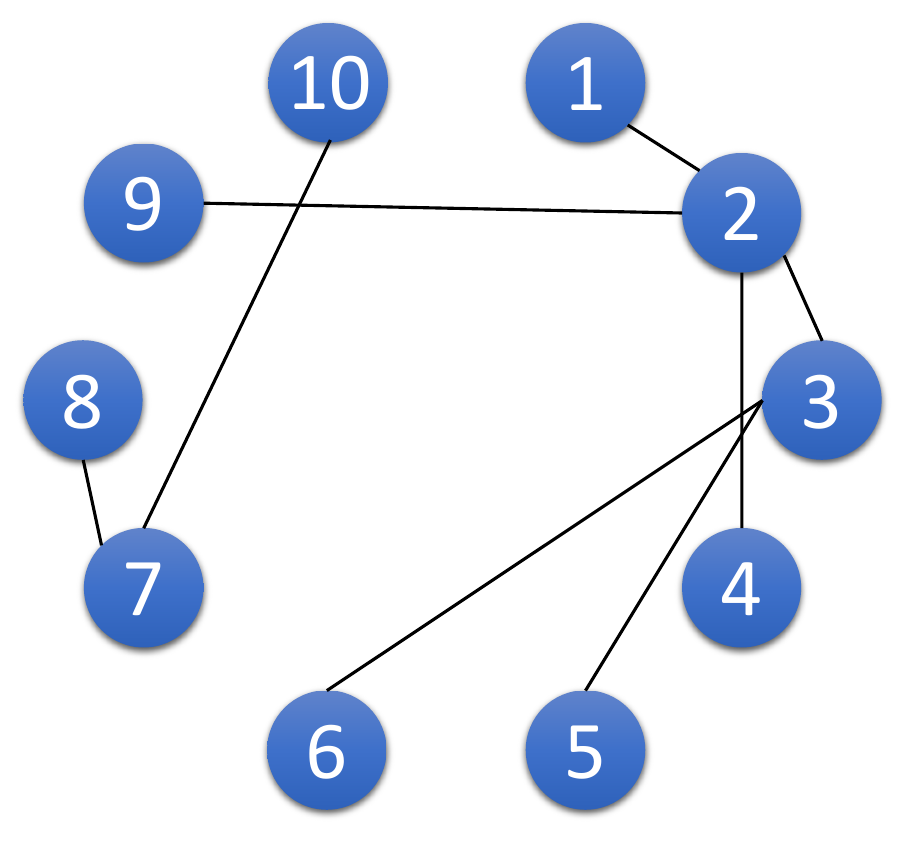}$G(3)$
		\\ \vspace*{0.3cm} \footnotesize Mean \\\includegraphics[scale=0.2]{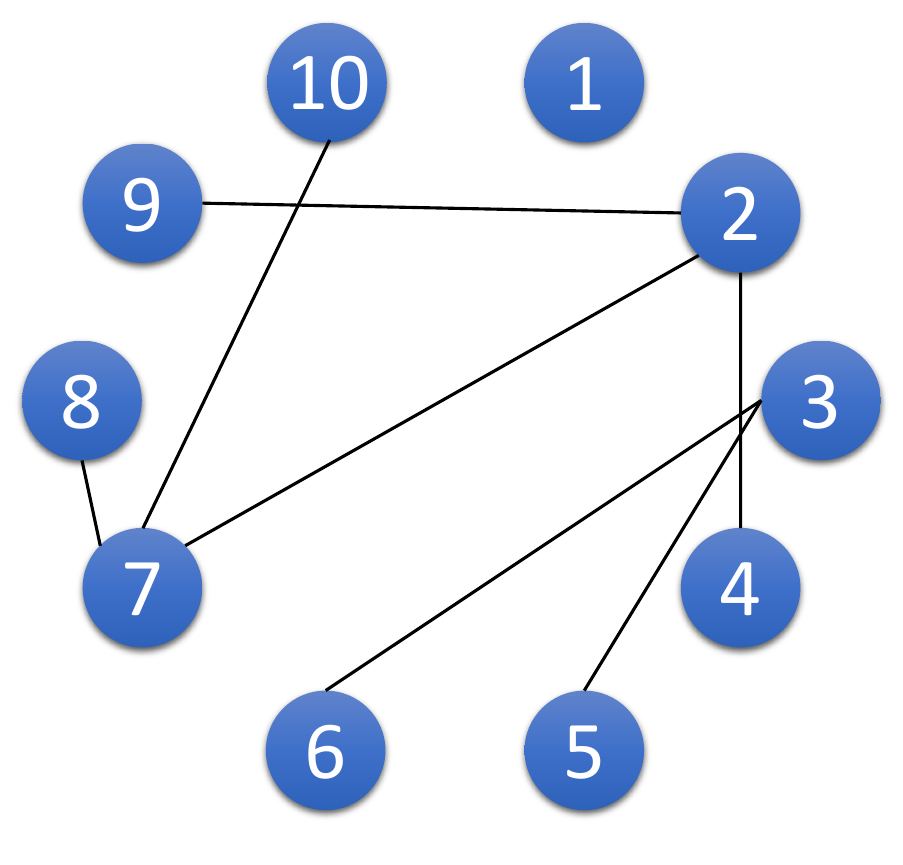}\tiny{$G(0)$}
		\includegraphics[scale=0.2]{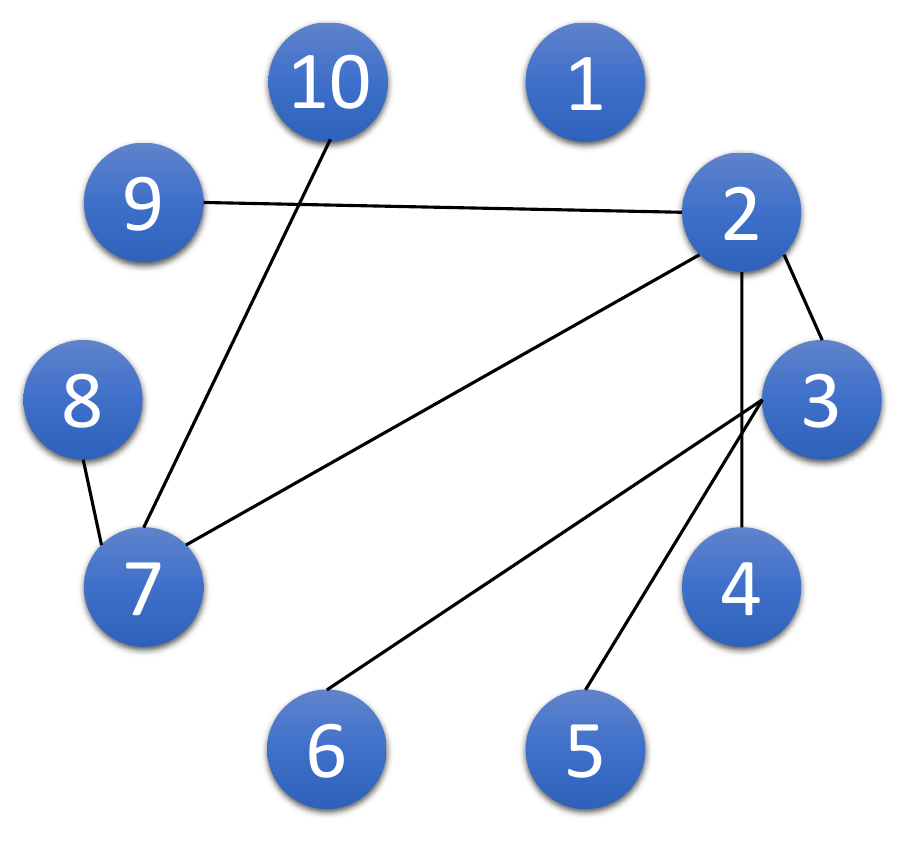}$G(1)$
		\includegraphics[scale=0.2]{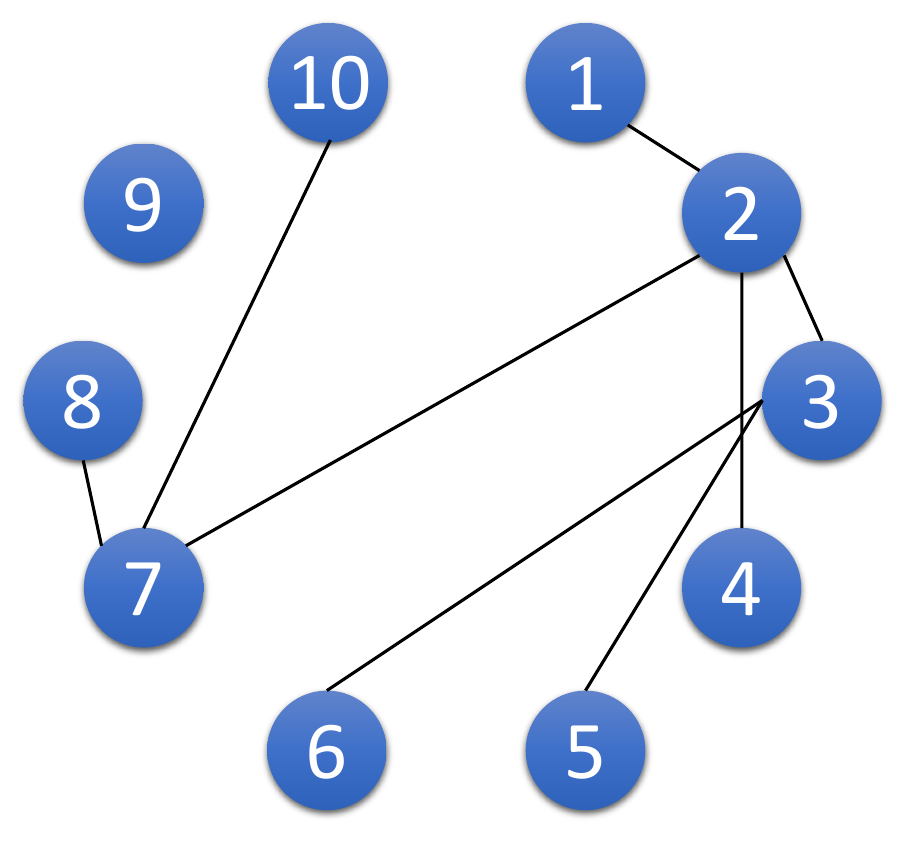}$G(2)$ 
		\includegraphics[scale=0.2]{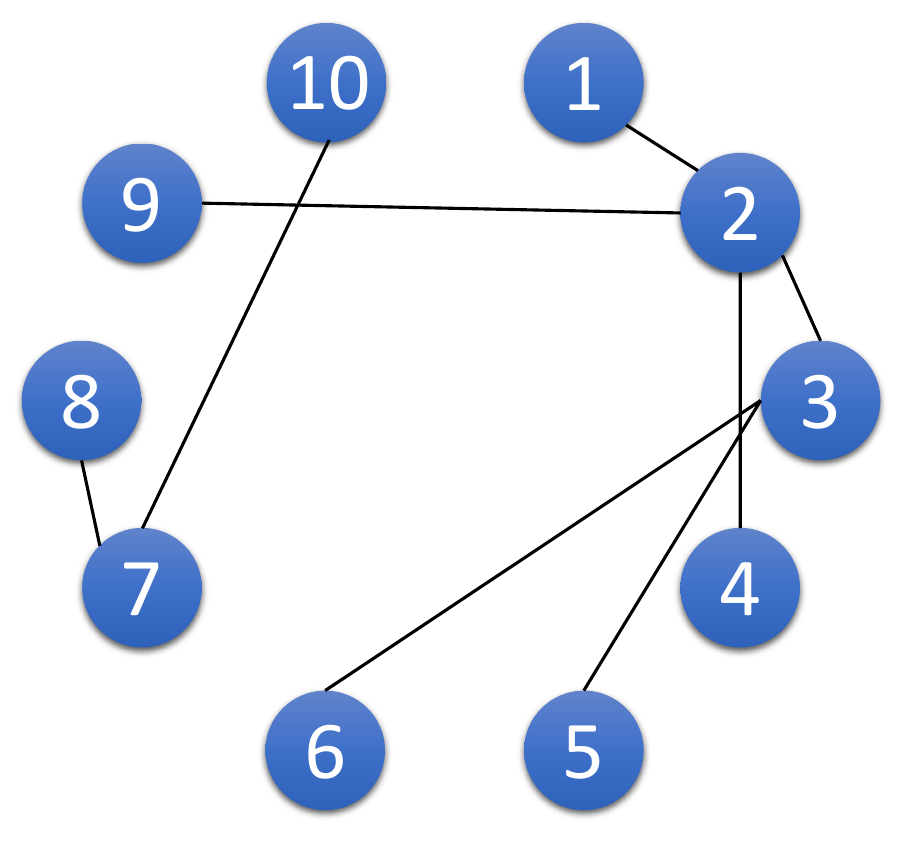}$G(3)$
			\\ \vspace*{0.3cm} 
   \footnotesize 25\% Quantile \\\includegraphics[scale=0.2]{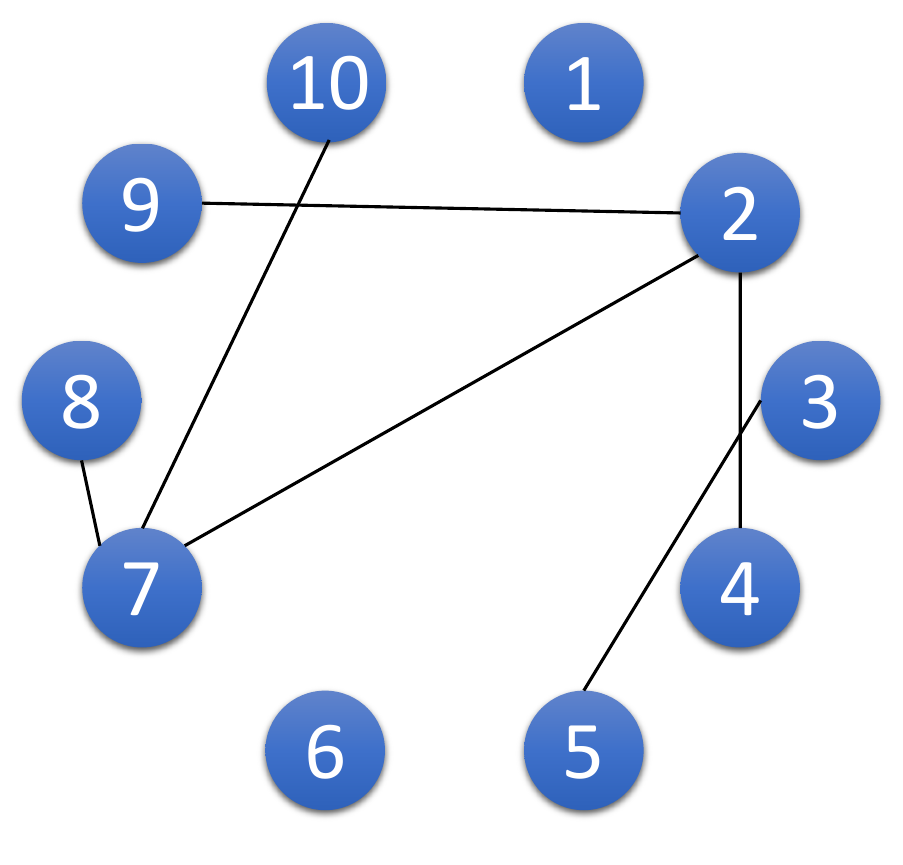}\tiny{$G(0)$}
		\includegraphics[scale=0.2]{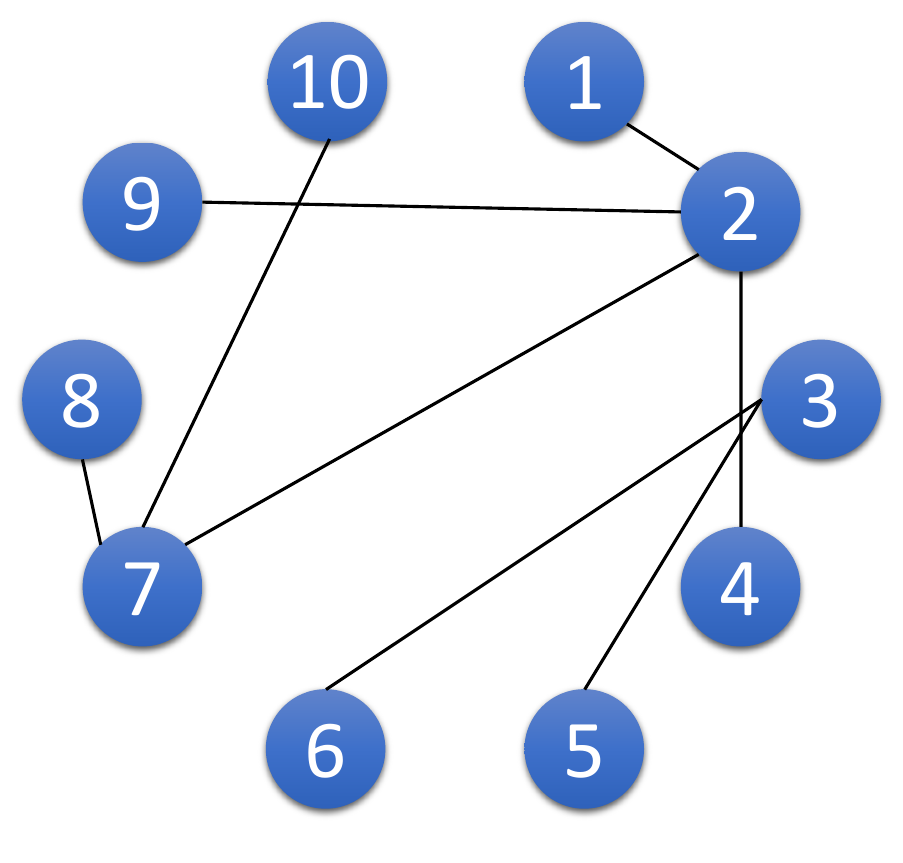}$G(1)$
		\includegraphics[scale=0.2]{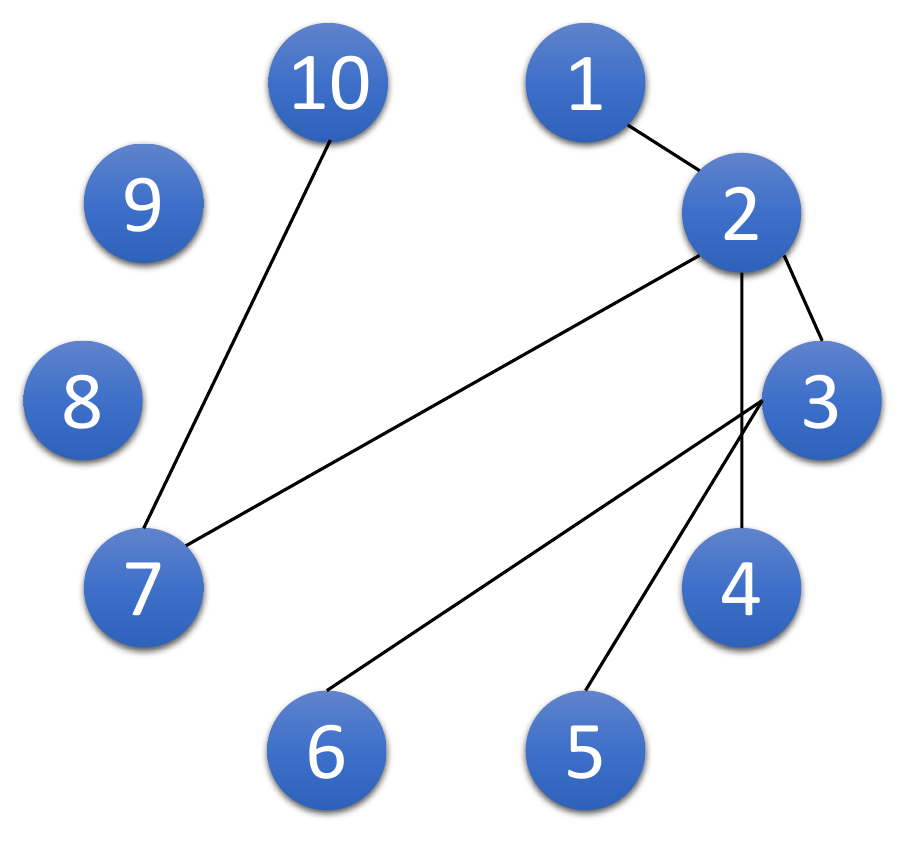}$G(2)$ 
		\includegraphics[scale=0.2]{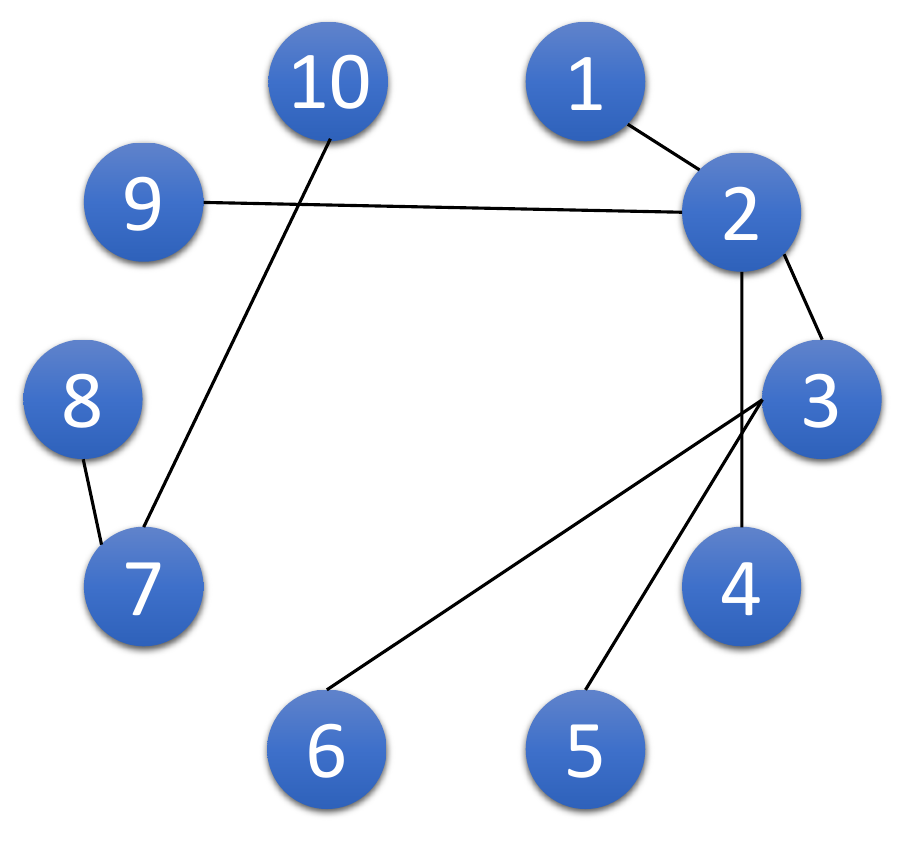}$G(3)$
  \caption{Graph structures obtained using the FB method in Scenario (A) in the low-dimensional setting at the 25\% quantile, mean, and 75\% quantile of the posterior distribution of edge inclusion probability, compared to the ground truth. Correctly identified edges are shown with solid lines, while spurious edges are represented by dotted lines.}
  \label{UQFB}
	\end{center}
\end{figure*}}

\small{\begin{figure*}[h!]
	\begin{center}
		 	\includegraphics[scale=0.4]{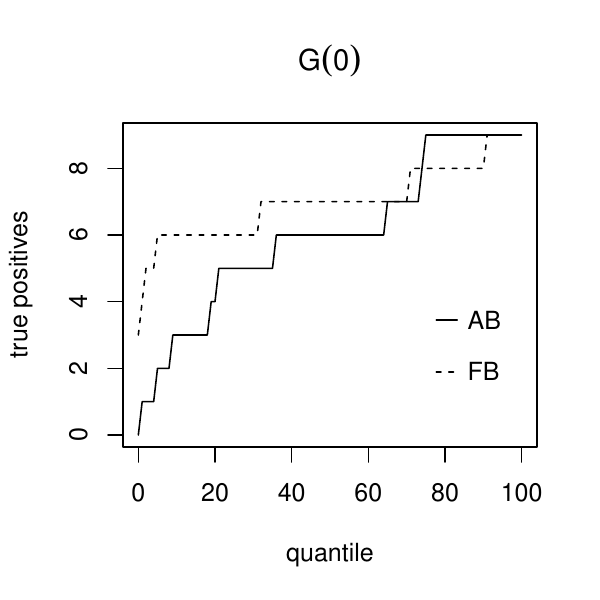}
		\includegraphics[scale=0.4]{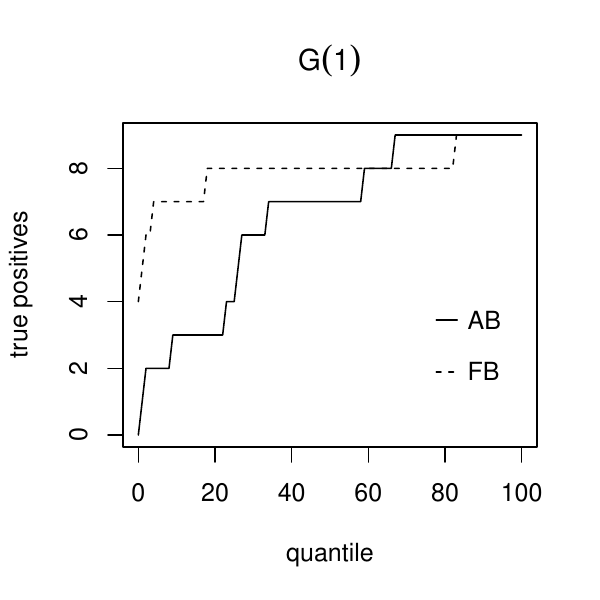}
		\includegraphics[scale=0.4]{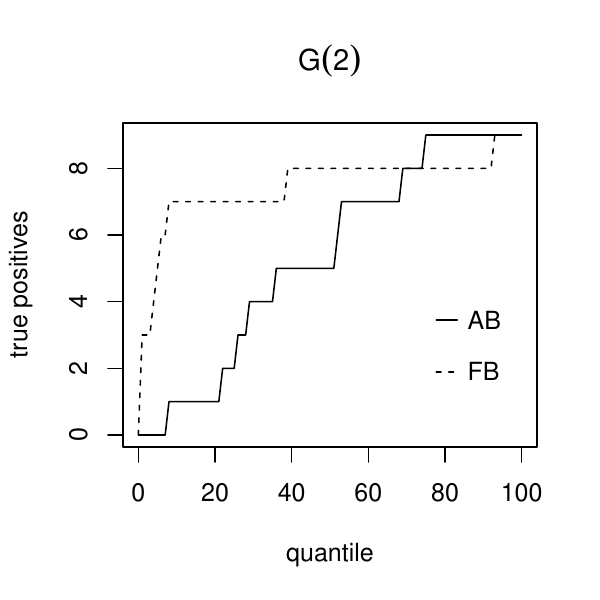}
		\includegraphics[scale=0.4]{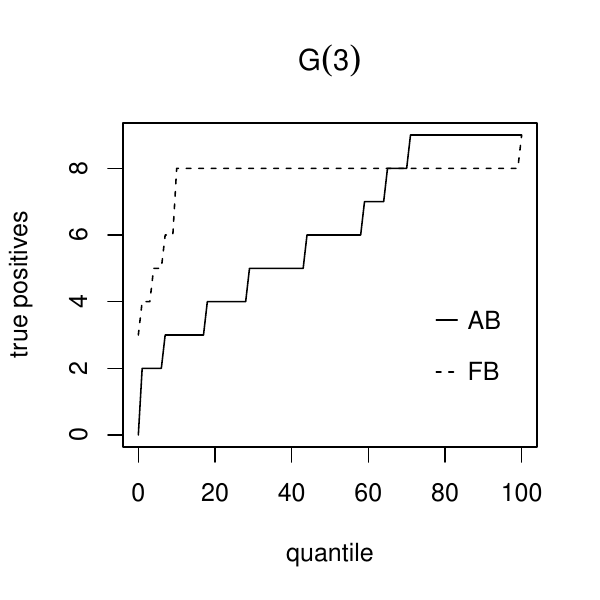} 
  \caption{Number of correctly identified edges selected in the graph at different quantiles for the FB method (dotted lines) and the AB method (solid lines) across four different levels in Scenario (A) in the low-dimensional case.}
  \label{UQEdges}
	\end{center}
\end{figure*}}

  \label{UQEdgesTh}

\section{Frequentist coverage}
\label{app:FC}

\begin{figure*}[h!]
	\begin{center}
		 	\includegraphics[scale=0.45]{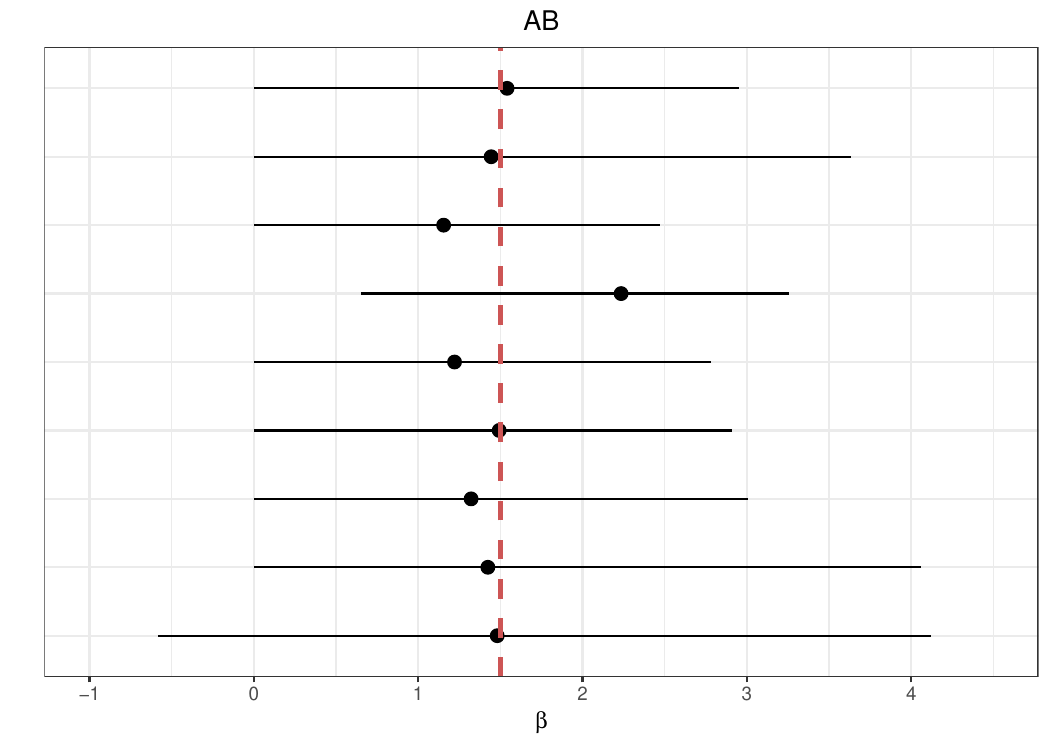}
		 	\includegraphics[scale=0.45]{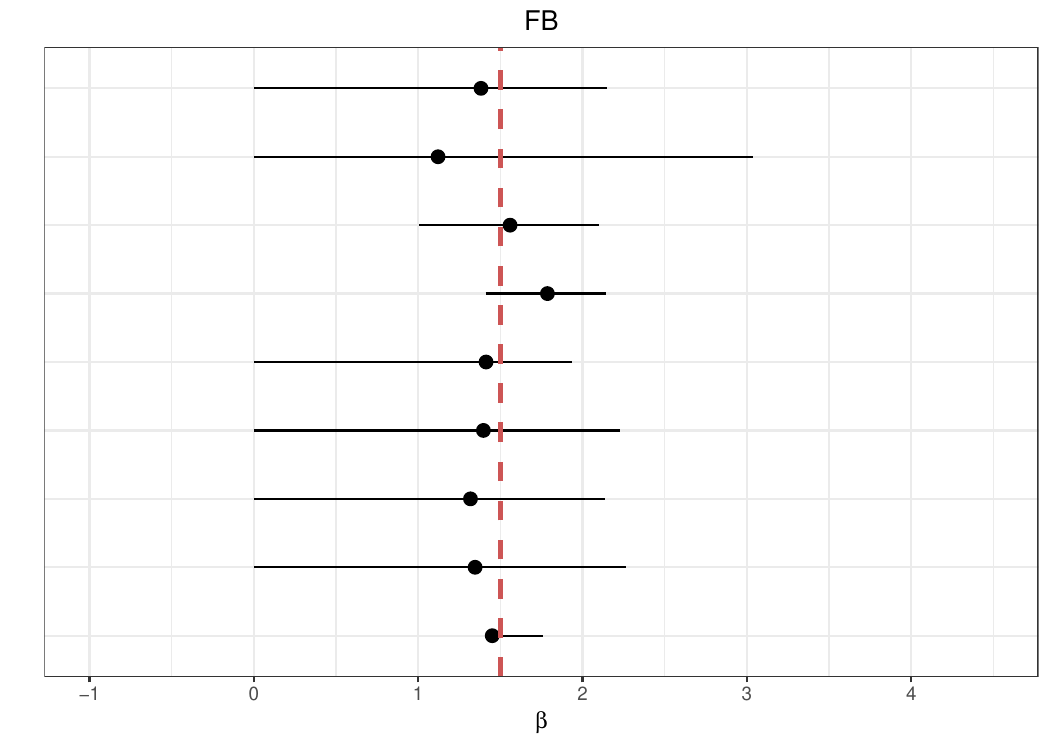}
  \caption{$90\%$ credible intervals for $\beta$ for Scenario (A) in the low-dimensional setting}
		\label{FCbetafp}
	\end{center}
\end{figure*}

\new{We investigate the frequentist coverage of the credible intervals associated with the key parameter in our data-generating mechanism. 
The simulated graphs used in this study are scale-free networks, generated via a preferential attachment process based on the Barab\'asi-Albert model. 
These networks were then employed to simulate binary data, where the probability was derived from a logistic regression model, using the scale-free graph structures and a coefficient of $\beta = 1.5$.

Figure \ref{FCbetafp} displays the 90\% credible intervals for $\beta$ for Scenario (A) in the low-dimensional setting. 
Our findings indicate that both of our proposed methods achieve good frequentist coverage, as all credible intervals contain the true value of $\beta = 1.5$. 
Moreover, the FB approach exhibits smaller variance compared to the AB method. 
This underscores the benefits of using the full likelihood in the FB method over the quasi-likelihood approach employed by the AB method, particularly in terms of uncertainty quantification and interval precision.}

\section{Other simulation scenarios}
\label{app:OtherSims}

\new{
We conducted additional simulation scenarios in both low-dimensional and high-dimensional settings. 
Following the same simulation specifications as described in the Simulation Studies section, we considered alternative undirected profiles $\G(A)$, $\G(B)$, $\G(C)$, and $\G(D)$, shown in Figure \ref{mulgraphsOld}. Tables \ref{T10old} and \ref{T50old} present the MCC, F1 score, and their standard errors across ten different simulations for both settings. 
The results demonstrated a robust performance of our methods, FB and AB, across these scenarios and those based on scale-free networks.
}

\small{\begin{figure}[h!]
	\begin{center}
		\footnotesize Scenario (A)\\
		 	\includegraphics[scale=0.12]{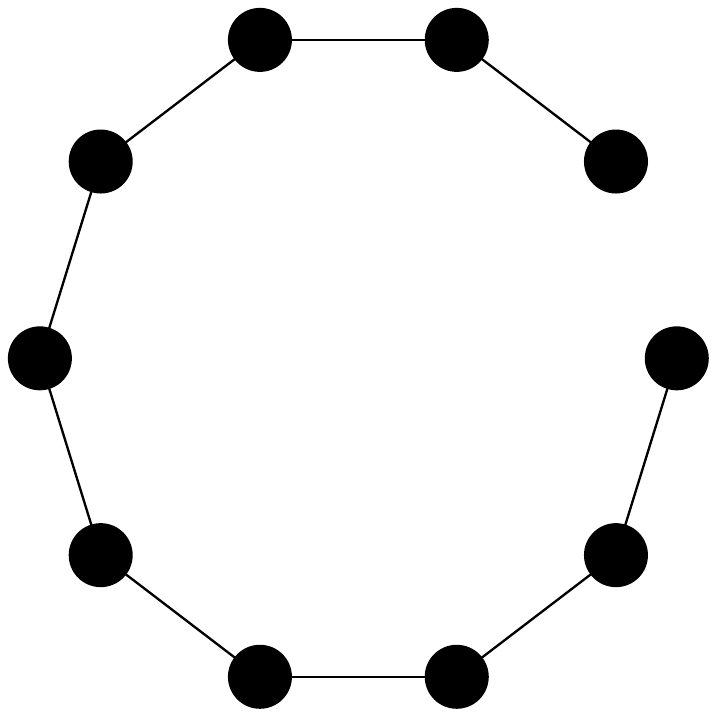}\tiny{$G(0)$}
		\includegraphics[scale=0.12]{figures/G0.pdf} $G(1)$ 
		\includegraphics[scale=0.12]{figures/G0.pdf} $G(2)$ 
		\includegraphics[scale=0.12]{figures/G0.pdf} $G(3)$ 
		\\ \vspace*{0.3cm} \footnotesize Scenario (B) \\\includegraphics[scale=0.12]{figures/G0.pdf}\tiny{$G(0)$}
		\includegraphics[scale=0.12]{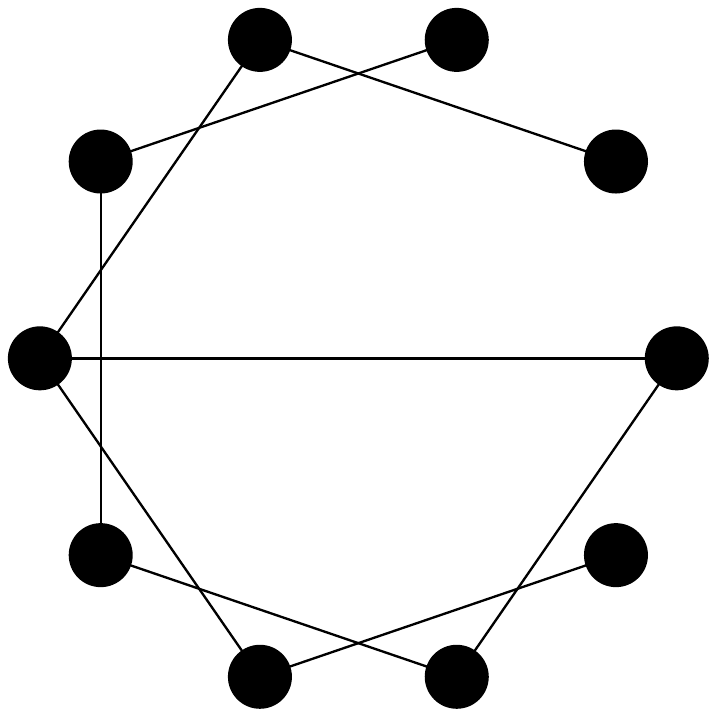}$G(1)$
		\includegraphics[scale=0.12]{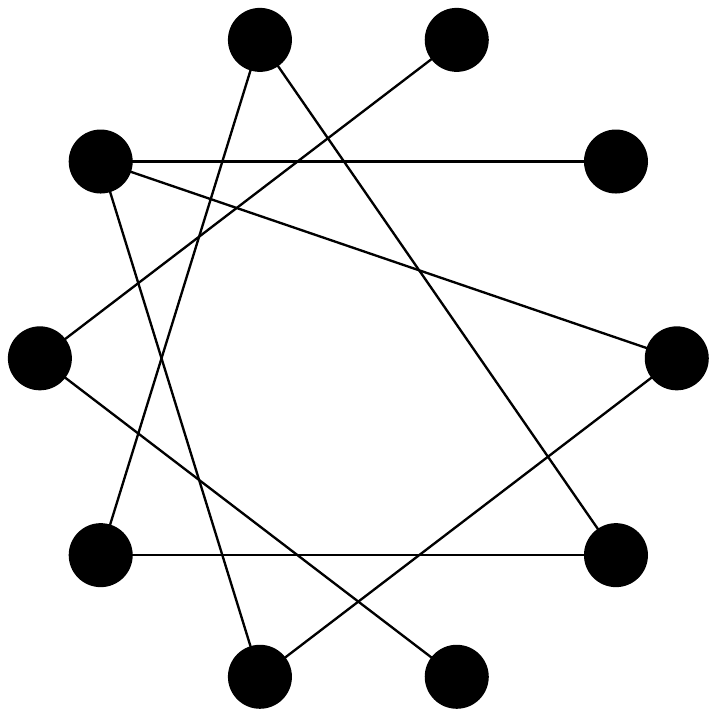}$G(2)$ 
		\includegraphics[scale=0.12]{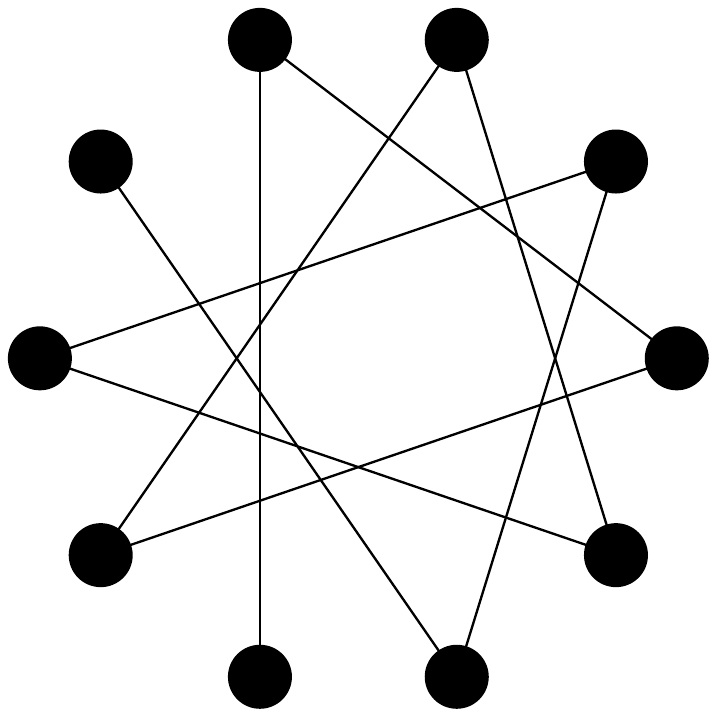}$G(3)$
		\\ \vspace*{0.3cm} \footnotesize Scenario (C) \\\includegraphics[scale=0.12]{figures/G0.pdf}\tiny{$G(0)$}
		\includegraphics[scale=0.12]{figures/G0.pdf}$G(1)$
		\includegraphics[scale=0.12]{figures/G1.pdf}$G(2)$ 
		\includegraphics[scale=0.12]{figures/G1.pdf}$G(3)$
			\\ \vspace*{0.3cm} \footnotesize Scenario (D) \\\includegraphics[scale=0.12]{figures/G0.pdf}\tiny{$G(0)$}
		\includegraphics[scale=0.12]{figures/G0.pdf}$G(1)$
		\includegraphics[scale=0.12]{figures/G0.pdf}$G(2)$ 
		\includegraphics[scale=0.12]{figures/G1.pdf}$G(3)$
  \caption{Multiple undirected graphs in the four different scenarios of the low-dimensional simulation study.}
		\label{mulgraphsOld}
	\end{center}
\end{figure}}

\begin{table*}[t]
\caption{MCC and F1 score and their standard error (SE) rates for all scenarios of the low-dimensional synthetic datasets.\label{T10old}}
\tabcolsep=0pt
\begin{tabular*}{\textwidth}{@{\extracolsep{\fill}}ccccc@{\extracolsep{\fill}}}
\toprule%
\multicolumn{5}{@{}c@{}}{MCC} \\
\hline
Model & Scenario (A) & Scenario (B) & Scenario (C) & Scenario (D) \\ 	\hline
SL     & 0.771(0.070) & 0.759(0.063) & 0.781(0.049) & 0.806(0.037) \\ 
DSSL   & \textbf{0.987}(0.025) & 0.581(0.144) & 0.578(0.079) & 0.726(0.038) \\
ABS & 0.775(0.072) & 0.753(0.049) & 0.803(0.061) & 0.827(0.038) \\
\cellcolor[HTML]{d9e6f2} AB    & \cellcolor[HTML]{d9e6f2} 0.831(0.055) &\cellcolor[HTML]{d9e6f2}  0.747(0.032) & \cellcolor[HTML]{d9e6f2} 0.807(0.061) & \cellcolor[HTML]{d9e6f2} 0.863(0.041)\\
FBS     & 0.828(0.050) & \textbf{0.811}(0.053) & 0.819(0.050) & 0.847(0.055)
		\\
  \cellcolor[HTML]{d9e6f2} FB    & \cellcolor[HTML]{d9e6f2} 0.909(0.047) & \cellcolor[HTML]{d9e6f2} 0.737(0.056) & \cellcolor[HTML]{d9e6f2} \textbf{0.831}(0.065) & \cellcolor[HTML]{d9e6f2} \textbf{0.875}(0.047)
		\\
		\hline \hline
  \multicolumn{5}{@{}c@{}}{F1} \\
\hline
Model & Scenario (A) & Scenario (B) & Scenario (C) & Scenario (D) \\ 	\hline
SL     & 0.812(0.057) & 0.803(0.051) & 0.816(0.044) & 0.763(0.047) \\ 
DSSL   & \textbf{0.989}(0.020) & 0.640(0.140) & 0.663(0.062) & 0.657(0.050) \\
ABS & 0.803(0.063) & 0.780(0.047) & 0.826(0.059) & 0.799(0.040) \\
\cellcolor[HTML]{d9e6f2} AB    & \cellcolor[HTML]{d9e6f2} 0.855(0.050) & \cellcolor[HTML]{d9e6f2} 0.785(0.030) & \cellcolor[HTML]{d9e6f2} 0.834(0.056) & \cellcolor[HTML]{d9e6f2} 0.836(0.046)\\
FBS     & 0.863(0.040) & \textbf{0.849}(0.042) & 0.854(0.041) & 0.810(0.068)\\
\rowcolor[HTML]{d9e6f2}
		FB    & 0.927(0.038) & 0.790(0.044) & \textbf{0.863}(0.053) & \textbf{0.844}(0.058)
		\\
		\hline
\end{tabular*}
\end{table*}

\begin{table*}[t]
\caption{MCC and F1 score and their standard error (SE) rates for all scenarios of the high-dimensional synthetic datasets. \label{T50old}}
\tabcolsep=0pt
\begin{tabular*}{\textwidth}{@{\extracolsep{\fill}}ccccc@{\extracolsep{\fill}}}
\toprule%
\multicolumn{5}{@{}c@{}}{MCC} \\
\hline
Model & Scenario (A) & Scenario (B) & Scenario (C) & Scenario (D) \\ 	\hline
SL & 0.911(0.006) & 0.923(0.012) & 0.916(0.011) & 0.922(0.011) \\ 
DSSL & \textbf{0.989}(0.011) & 0.830(0.015) & 0.723(0.015) & 0.788(0.013)\\
ABS & 0.934(0.008) & \textbf{0.928}(0.012) & 0.941(0.013) & 0.929(0.010) \\
\rowcolor[HTML]{d9e6f2}
AB & 0.958(0.009) & 0.925(0.012) & \textbf{0.945}(0.011) & \textbf{0.936}(0.011)
		\\
		\hline \hline
  \multicolumn{5}{@{}c@{}}{F1} \\
\hline
Model & Scenario (A) & Scenario (B) & Scenario (C) & Scenario (D) \\ 	\hline
SL & 0.914(0.006) & 0.925(0.011) & 0.918(0.011) & 0.924(0.011) \\ 
DSSL & \textbf{0.990}(0.010) & 0.829(0.016) & 0.710(0.016) & 0.781(0.013) \\
ABS & 0.935(0.008) & \textbf{0.929}(0.012) & 0.942(0.013) & 0.930(0.010) \\
\rowcolor[HTML]{d9e6f2}
		AB & 0.960(0.008) & 0.928(0.012) & \textbf{0.948}(0.010) & \textbf{0.940}(0.11)
		\\
		\hline
\end{tabular*}
\end{table*}

\section{Computational cost}
\label{app:CompCost}

We examine the computational cost associated with our proposed models as the complexity of the problem increases. 
Specifically, we analyze how the computational cost scales with an increase in the number of nodes  $p$ and the cardinality of the $X$. 
We aim to understand the efficiency and scalability of our models under different conditions and provide valuable insights into their practical applicability in large-scale settings.

Table \ref{tabTime} reports the averages and
standard deviations for the computation time across ten different simulations of Scenario (C) in the low-dimensional setting, varing the number of variables $p$. 
Given the computational resources available, we found performing the FB approach infeasible for scenarios with $p\geq 20$ variables and therefore, results are not
reported, as long vectors not supported in R (see caption of Table \ref{tabTime} for details).
This limitation further underscores the practical advantage of the AB model in high-dimensional settings.
    
Our study showed that the computational cost of the FB model is higher than that of the AB model. 
This is because the FB model requires estimating all $(p \times (p-1))/2$ interactions of the canonical parameter $\lambda_x$, which becomes increasingly complex as $p$ grows. 
Consequently, the benefit of using the likelihood approximation in the AB model is particularly appealing, especially when analyzing several high-dimensional datasets.

\begin{table}[ht]
\caption{ Average computation time in days, and standard error (SE), for ten simulation replicates. Note: we were unable to run FB for some scenarios due to computation limitations}
\label{tabTime}
\centering
\begin{tabular}{ccccc}
  \hline
 Model & p & Edges & Time(days) & SE \\ 
  \hline
  FB & 10 & 45 & 0.649 & 0.004 \\ 
  AB & 10 & 45 & 0.515 & 0.003 \\ 
  AB & 20 & 190 & 1.771 & 0.007 \\ 
  AB & 30 & 435 & 4.387 & 0.049 \\ 
  AB & 40 & 780 & 8.118 & 0.053 \\ 
  AB & 50 & 1,225 & 13.190 & 0.164 \\ 
  \hline
\end{tabular}
\end{table}


Table \ref{tabTimeX} presents the variation in computational cost as the cardinality of $X$ increases from 2 to 10. 
The table reports the average computation time, in seconds, for a single MCMC iteration across 100 iterations in the low-dimensional scenario C, where $p=10$ nodes are considered, and the cardinality of 
$X$ is varied. 
The results indicate that the FB model scales more efficiently than the AB model with respect to the cardinality of $X$. 
Specifically, the AB model only outperforms the FB model in terms of computation time when $X$ is small (2 and 3). However, while the FB model shows better scalability with $X$, it is important to note that this advantage does not extend to an increase in $p$. 
In fact, the FB model becomes infeasible to run in scenarios where 
$p\geq20$.

\begin{table}[t]
\caption{Average computation time in seconds, and standard error (SE) of one MCMC iteration for 100 iterations\label{tabTimeX}}
\tabcolsep=0pt
\centering
\begin{tabular}{c|cc|cc}
\hline
&\multicolumn{2}{c|}{AB}&\multicolumn{2}{c}{FB} \\
\hline
X & Time(seconds) & SE & Time(seconds) & SE \\ 	\hline
2 & 0.080 & 0.008 & 0.273 & 0.039 \\ 
3 & 0.237 & 0.021 & 0.406 & 0.048 \\ 
4 & 0.698 & 0.030 & 0.583 & 0.073 \\ 
5 & 2.117 & 0.068 & 1.007 & 0.121 \\ 
6 & 6.316 & 0.230 & 1.959 & 0.243 \\ 
7 & 17.079 & 0.367 & 4.435 & 0.562 \\ 
8 & 44.864 & 0.725 & 10.642 & 1.329 \\ 
9 & 117.972 & 2.689 & 26.754 & 2.628 \\ 
10 & 293.669 & 4.788 & 66.082 & 5.647 \\ 
		\hline
\end{tabular}
\end{table}


\clearpage
\newpage
\bibliography{reference}

\end{document}